\documentclass[12pt,reqno]{amsart}
\usepackage{comment}
\usepackage{MnSymbol}  
\usepackage{amsmath}
\usepackage{xcolor}
\usepackage{amsfonts}
\usepackage{bm}
\usepackage{array}  
\usepackage{hyperref}
\usepackage{enumerate}
\usepackage{tikz}
\usepackage{graphicx,comment}
\usepackage{tikz-cd}
\usetikzlibrary{calc, positioning,shapes,arrows,decorations.markings,patterns,angles,quotes}
\tikzset{Rightarrow/.style={double equal sign distance,={Implies},-},
	triple/.style={-,preaction={draw, Rightarrow}},
	quadruple/.style={preaction={draw,Rightarrow,shorten >=0pt},shorten >=1pt,-,double,double
		distance=0.2pt}}
\tikzstyle{line} = [draw, -latex']
\tikzset{mynode/.style={draw,circle, inner sep=2pt, outer sep=0pt}}
\tikzset{
wei/.style={circle, minimum size=0.4pt,inner sep=0.8pt},
}
\newcommand\bit{\begin{itemize}}
\newcommand\eit{\end{itemize}}
\newcommand\lan{\langle}
\newcommand\ran{\rangle}
\newcommand{\bp}{\begin{pmatrix}}
\newcommand{\ep}{\end{pmatrix}}
\newcommand\Pa{Painlev\'e }

\newcommand{\oc}[1]{{#1}^{\vee}}
\usepackage{setspace}
\usepackage{fullpage}
\usepackage{marginnote}
\definecolor{mycol}{rgb}{0,0.5,0}

\newcommand{\al}{\alpha}

\newcommand{\de}{\delta}

\newcommand{\De}{\Delta}
\newcommand{\be}{\beta}
\newcommand{\la}{\lambda}
\newcommand{\ga}{\gamma}
\newcommand{\Ga}{\Gamma}

\newcommand{\s}{\sigma}
\newcommand{\hs}{\hspace}
\newcommand{\dav}{V^{(1)*}}

\newcommand{\av}{V^{(1)}}
\theoremstyle{break}    \newtheorem{Cor}{Corollary}
\newtheorem{Rem}{Remark} \theoremstyle{marginbreak}

\newtheorem{Def}[Cor]{Definition}
\newtheorem{prop}{Proposition}

\newtheorem{eg}{Example}

\newcommand{\beq}{\begin{equation}}
\newcommand{\eeq}{\end{equation}}
\begin{document}
\title{Translations in affine Weyl groups and 
their applications in discrete integrable systems}
\maketitle
\author{Yang Shi}
\address{College of Science and Engineering, Flinders at Tonsley, Flinders University, SA 5042,
    Australia}
\email{yang.shi@flinders.edu.au}

\begin{abstract}
In this paper, we review the properties and representations of the Weyl groups
relevant in the study of discrete integrable systems.
Previously in \cite{jns4, Shi:19}, properties of Weyl groups of type $ADE$
(known as simply-laced) were shown to be useful in characterizing and establishing relations
between different integrable systems. 
Here we extend the formulations and discussions to include
non-simply-laced types, giving special attention to developing
formulas related to the translational elements of the affine
Weyl groups. As applications, we show how these are used to clarify
the natures of some integrable systems of type $E_8$ \cite{NN_elliptic}
and $F_4$ \cite{ahjn:16} appeared recently  in the literature.
\end{abstract}
\tableofcontents
\section{Introduction}\label{intro}
Many integrable equations and their discrete analogs admit 
Coxeter group or Weyl group symmetries.
Examples include Hirota's discrete analog of the KP equation  
on an octahedron \cite{hirota:81},
\begin{equation}\label{Hirota}
\al\, \tau_{1} \tau_{23}+\be\, \tau_{2} \tau_{13}+\ga\, \tau_{3} \tau_{12}=0,
\end{equation}
where $\tau=\tau(n, m, l)$ for $n, m, l \in \mathbb{Z}$,
and $\al, \be, \ga$ are constant parameters. 
We have used the convention
$\tau(n+1, m, l)=\tau_{1}$, and $\tau(n, m+1, l+1)=\tau_{23}$ etc.
Equation \eqref{Hirota} has Weyl symmetry of type $A_3$.

There is
the classification of Adler, Bobenko, and Suris  \cite{abs:03} for discrete analogs of KdV type equations on quadrilaterals, or ABS's {\it quad-equations}.
It contains Adler's discretization of the Krichever-Novikov equation \cite{adler98} as the Q4 equation, 
\begin{align}\label{Q4}
\rm{sn} (\alpha)(u u_{1}+u_{2} u_{12})&
-\rm{sn} (\be) (u u_{2}+u_{1} u_{12})\\\nonumber
&-\rm{sn}(\alpha-\beta)(u_{1} u_{2}+u u_{12}
      -\rm{sn} (\alpha) \rm{sn} (\beta)\left(1+k^2 u u_{1} u_{2} u_{12}\right))=0,
\end{align}
written in its elliptic form \cite{hie:05}.
We have $u=u(m, n), u_1=u(m+1, n)$, $u_2=u(m, n+1), u_{12}=u(m+1, n+1)$,  for $n, m \in \mathbb{Z}$,  $\alpha$ and $\beta$ are constant parameters, and $k$ is the modulus of the Jacobi \rm{sn} function. Quad-equations all have Weyl symmetry of type $B_2$.
Equations \eqref{Hirota} and \eqref{Q4} are examples of integrable partial difference equations (P$\Delta$Es). 

Following Okamoto's geometric description of 
the \Pa equations
\cite{oka:79}, 
Sakai gave a
classification of 22 types of discrete \Pa equations, these are 2nd-order integrable ordinary difference equations (O$\Delta$Es) \cite{sak:01}.
They are listed by their affine Weyl symmetry types in Figure \ref{Sakai}.
For a comprehensive 
overview of the current geometric theory of the discrete \Pa equations
se,e a recent survey \cite{KNY_review}.

The master equation of Sakai's list, 
in the sense that the other 21 types can be obtained from it in some degeneration limits,
is the elliptic \Pa equation of $E_8^{(1)}$ type, $e$-{\bf P}$(E_8^{(1)})$, shown in Equation \eqref{ePE8}.

It is given as a system of two 1st-order O$\Delta$Es
for $f(t)$, $g(t)$ in the variable $t$. $(f(t), g(t))$ is being considered as inhomogeneous coordinates of $\mathbb{P}^1 \times \mathbb{P}^1$ \cite{MSY:13}. The $2 \times 2$ $M$ matrices,
given in terms of Weierstrass's $\wp$ function, represent PGL(2)-action on $\mathbb{P}^1$, i.e., $w=\left(\begin{array}{ll}a & b \\ c & d\end{array}\right) z$ means $w=(a z+b) /(c z+d)$ and we have,
\begin{subequations}\label{ePE8}
\begin{align}
\overline{g}=g(t+\lambda)= & M\left(f, c_7, c_8, t-\frac{1}{4} \sum_{i=1}^6 c_i\right) M\left(f, c_5, c_6, t-\frac{1}{4} \sum_{i=1}^4 c_i\right) M\left(f, c_3, c_4, t-\frac{1}{4}\left(c_1+c_2\right)\right) \\\nonumber
& \times M\left(f, c_1, c_2, t\right) g, \\
\underline{f}=f(t-\lambda)= & M\left(g, d_7, d_8, t-\frac{1}{4} \sum_{i=1}^6 d_i\right) M\left(g, d_5, d_6, t-\frac{1}{4} \sum_{i=1}^4 d_i\right) M\left(g, d_3, d_4, t-\frac{1}{4}\left(d_1+d_2\right)\right) \\\nonumber
& \times M\left(g, d_1, d_2, t\right) f,
\end{align}  
where
\begin{align}\nonumber
 \quad \quad M\left(h, \kappa_1, \kappa_2, s\right)\\\label{Me8}
&\hs{-4em}
 =\left(\begin{array}{ll}-\wp\left(2 s-\frac{-\kappa_1+\kappa_2}{2}\right)&
 \wp\left(2 s-\frac{\kappa_1-\kappa_2}{2}\right)\\
 -1 & 1
 \end{array}\right)\\
&\hs{-3em} \times\left(\quad\begin{array}{ll}
\hspace{-1em}\left(h-\wp\left(\kappa_2\right)\right)\left(\wp(2 s)-\wp\left(2 s-\kappa_2\right)\right)\left(\wp\left(2 s-\frac{\kappa_1+\kappa_2}{2}\right)-\wp\left(2 s-\frac{\kappa_1-\kappa_2}{2}\right)\right)\quad\quad 0 \\\nonumber
0\quad\quad \left(h-\wp\left(\kappa_1\right)\right)\left(\wp(2 s)-\wp\left(2 s-\kappa_1\right)\right)\left(\wp\left(2 s-\frac{\kappa_1+\kappa_2}{2}\right)-\wp\left(2 s-\frac{-\kappa_1+\kappa_2}{2}\right)\right.
\end{array}\right) \\\nonumber
& \times\left(\begin{array}{cc}
1 & -\wp\left(2 s-\kappa_1\right) \\\nonumber
1 & -\wp\left(2 s-\kappa_2\right)
\end{array}\right) . 
\end{align}   

Here, 
$\lambda=\frac{1}{2} \sum_{i=1}^8 b_i,\; c_i=b_i+t,\; d_i=t-b_i$, 
and $b_i\; (1\leq i\leq 8)$ are constant parameters.
\end{subequations}

By saying that Equation \eqref{ePE8} has Weyl group symmetry of type
$E_8^{(1)}$, $W(E_8^{(1)})$,
we mean the equation is left invariant by
the transformations that form a birational representation
of $W(E_8^{(1)})$, given in Equation \eqref{msye8rep}. 

The element $T_{1}$ of $W(E_8^{(1)})$
that gives rise to the map in Equation \eqref{ePE8},
\begin{subequations}\label{T11}
\begin{equation}\label{T1act1}
  T_{1}:(b_i, t, f, g)\mapsto (b_i, t+\lambda, \overline{f}, \overline{g}),\quad 1\leq i\leq 8,
\end{equation}
acts on the simple roots $\{\al_j\mid 0\leq j\leq 8\}$ of $W(E_8^{(1)})$
by
\beq\label{TJ21}
T_{1}:\{\al_1, \al_3\}\mapsto
\{\al_1-2\de, \al_3+\de\},
\eeq
where only non-trivial actions are shown.
$T_{1}$ is a translation by $\al_1$
(with  $|\al_1|^2=2$, the shortest weight length) on the $E_8$ root/weight lattice, 
it has an expression as a product of 58 simple reflections of $W(E_8^{(1)})$,
 \begin{align}\nonumber
     T_{1}=
&s_3 s_4 s_2 s_5 s_4 s_3 s_6  s_5 s_4 s_2 s_7 s_6 s_5 
s_4 s_3 s_8 s_7 s_6 s_5 s_4 s_2 s_0 s_8 s_7 s_6 s_5 s_4 s_3s_1  \\\label{T1decomp1}
&s_3 s_4 s_5 s_6 s_7 s_8 s_0 s_2 s_4 s_5 s_6 s_7 s_8 s_3 
s_4 s_5 s_6 s_7 s_2 s_4 s_5 s_6 s_3 s_4 s_5 s_2 s_4 s_3s_1.
\end{align}
\end{subequations}
In fact, equations in Sakai's list \cite{sak:01}, referred to as the {\it canonical forms} of the discrete \Pa equations,
are all given by a translation 
of the shortest weight vector of its Weyl type, 
which we refer to as a {\it basic translation}.

The symmetry description has led to some natural generalisations of 
discrete integrable systems,
such as the \Pa equations of type $A_n^{(1)}$ \cite{NY:98}; 
and type $D_n^{(1)}$ generalisations \cite{sa:06, Masuda:15} of the $d$-{\bf P}$(D_4^{(1)})$ and $q$-{\bf P}$(D_5^{(1)})$ equations.
It also allows one to put different kinds of discrete systems 
on the same footing \cite{Doliwa:13, akt:15, jns4,noumi:18}. 
For example, the relation between a system of ABS's quad-equations on a $n$-cube (type
$B_n$) and a \Pa equation of type $A_{n-1}^{(1)}$ was found in \cite{jns2}
using the relation between the two weight lattices of these Weyl types.
Another example using the fact that 
the tau functions of the 
$e$-{\bf P}$(E_8^{(1)})$ equation
are indexed by $E_8$ weight vectors \cite{ORG:01}, Noumi \cite{noumi:18} showed that
Equation \eqref{ePE8} is equivalent to a system of $7560$ octahedron equations \eqref{Hirota}.

It is clear that symmetry plays an important role in the study of discrete integrable systems. Having integrable equations 
that comes as some birational representations of affine Weyl groups
such as those given in \cite{oka:79, sak:01, NY:98}
means some behaviours of the nonlinear systems
can be studied as certain properties of the Weyl group by using a linear representation, such as 
describing the discrete \Pa equations as
translations on the weight lattice.

A question that has been of considerable interest concerns
the nature of certain 2nd-order O$\Delta$Es 
which were shown to admit
symmetries that do not appear explicitly in Sakai's list
\cite{Takenawa:03, KNT:11, ahjn:16, Carsteaetal2017}. 
In our previous works \cite{jns4, Shi:19}, we discussed properties and representations of the affine Weyl group
useful for establishing
connections between such systems 
and Sakai's equations for the simply-laced Weyl groups. The formulas allowed us to reduce the problem of finding potentially complicated birational transformations
that relates the two O$\Delta$Es to some simple manipulations in linear algebra.
Moreover, we found that different kinds of equations amount to translations
of different lengths on the weight lattice. 

Here, we extend our previous discussions \cite{jns4, Shi:19}
to non-simply-laced types,
giving a detailed exposition on
the affine Weyl group, paying particular attention to describing its
elements of translation.
By employing a linear map $\pi$ from
the vector space $\av$ on which the Weyl groups act as groups of reflections to
a dual vector space $\dav$, we realise translational actions of the Weyl groups 
on an affine plane
of this dual space, where translations by vectors of different lengths
can be explicitly discussed.
Although
all formulas discussed here may be found one way or another in
classical texts such as Bourbaki \cite{Bbook} or Humphreys \cite{Hbook},
we believe having explicit formulas for
the actions of the affine Weyl group in some well-chosen representation  
can be useful in studying the integrable systems with Weyl symmetries in general.
 
As applications, two examples from the integrable system literature are chosen to illustrate how the formulas and properties of the Weyl
groups discussed here
can be used in this context.

First, we derive the relation between a ``new'' elliptic
difference equation with $W(E_8^{(1)})$ symmetry 
found recently in
\cite{NN_elliptic}
and Sakai's $e$-{\bf P}$(E_8^{(1)})$ equation \eqref{ePE8}.
As this particular equation
takes almost three pages to write down we refer the interested reader to
Equation (3.27) in \cite{NN_elliptic}. This equation
is said to be new
as the element $T_{J,1}\in W(E_8^{(1)})$ that gives rise to its discrete evolution
is a translation of squared length $4$ rather than
$2$ (as for $T_1$ of Sakai's $e$-{\bf P}$(E_8^{(1)})$ equation), that is not a basic translation in $W(E_8^{(1)})$, and
can not be conjugated to a basic translation under the actions of $W(E_8)$.
In particular, $T_{J,1}$ acts on the simple roots $\{\al_j\mid 0\leq j\leq 8\}$ of $W(E_8^{(1)})$
by
\beq\label{TJ11}
T_{J,1}:\{\al_1, \al_6\}\mapsto
\{\al_1-2\de, \al_6+\de\}.
\eeq
Using the dual representation of $W(E_8^{(1)})$,
we show that $T_{J,1}$ can be obtained by a composition of $T_1$ 
(Equation \eqref{T1decomp1})
and another
basic translation in $W(E_8^{(1)})$, thus clarify the relation
between these two elliptic difference equations of type $E_8^{(1)}$.

Our second example concerns a subsystem
of type $F_4^{(1)}$ of the $e$-{\bf P}$(E_8^{(1)})$ equation \eqref{ePE8}
discussed in \cite{ahjn:16}.
It is obtained from the Q4 equation \eqref{Q4} 
by imposing the condition $u_1=u_2$,
\begin{equation}\label{RCG}
\begin{aligned}
& \operatorname{cn}\left(\gamma_n\right) \operatorname{dn}\left(\gamma_n\right)\left(1-k^2 \operatorname{sn}^4\left(z_n\right)\right) u_n\left(u_{n+1}+u_{n-1}\right) \\
& \quad-\operatorname{cn}\left(z_n\right) \operatorname{dn}\left(z_n\right)\left(1-k^2 \operatorname{sn}^2\left(z_n\right) \operatorname{sn}^2\left(\gamma_n\right)\right)\left(u_{n+1} u_{n-1}+u_n^2\right) \\
& \quad+\left(\operatorname{cn}^2\left(z_n\right)-\operatorname{cn}^2\left(\gamma_n\right)\right) \operatorname{cn}\left(z_n\right) \operatorname{dn}\left(z_n\right)\left(1+k^2 u_n{ }^2 u_{n+1} u_{n-1}\right)=0,
\end{aligned}    
\end{equation}
where $n \in \mathbb{Z}$ now is the only independent variable, $u_n=u(n)$ is the dependent variable, $k$ is the modulus of the Jacobi \rm{sn} function, and $\gamma_e, \gamma_o$ are constant complex parameters with
$$
z_n=\left(\gamma_e+\gamma_o\right) n+z_0, \quad \gamma_n= \begin{cases}\gamma_e & \text { for } n=2 j \\ \gamma_o & \text { for } n=2 j+1 .\end{cases}
$$
The element $\varphi_a$ of $W(E_8^{(1)})$
that gives rise to the discrete evolution in Equation \eqref{RCG},
\begin{subequations}\label{psi1}
\begin{equation}
\varphi_a:n\mapsto n+1,   
\end{equation}
acts on the simple roots $\{\al_j\mid 0\leq j\leq 8\}$ of $W(E_8^{(1)})$
by
\begin{align}\label{psiact1}
    \varphi_a&:\{\al_1, \al_2, \al_3, \al_4, \al_5, \al_6, \al_7,\al_8, \al_0\}\\\nonumber
    &\mapsto
  \{\al_1+\de, -\al_2, \al_{23445}-\de, -\al_4, -\al_5, -\al_{12334456}+\de, -\al_7,-\al_8, -\al_0\}, 
    \end{align}
whereas we have,
\begin{align}\label{AHJNphi2}
    \varphi_a^2&:\{\al_1, \al_2, \al_3, \al_4, \al_5, \al_6, \al_7,\al_8, \al_0\}\\\nonumber
    &\mapsto
  \{\al_1+2\de, \al_2, \al_3-2\de, \al_4, \al_5, \al_6+\de, \al_7,\al_8, \al_0\}.
    \end{align}
That is, $\varphi_a$ is not a translation while $\varphi_a^2$ is.
We call elements such as $\varphi_a$ {\it quasi-translations}.
$\varphi_a$ has the following expression in terms of simple reflections of $W(E_8^{(1)})$,
 \begin{align}\label{psidecomp1}
     \varphi_a&=\varphi_s^2,\quad\mbox{where}\quad
\varphi_s=s_6s_5s_4s_2s_7s_6s_5s_4s_2s_8s_7s_6s_5s_4s_2s_0s_8s_7s_6s_5s_4s_1s_3s_2s_4s_5s_2s_4s_3.
\end{align}
\end{subequations}
It was verified (using MAGMA \cite{magma}) that $\varphi_a$ is an element
of a $F_4^{(1)}$ type subgroup of $W(E_8^{(1)})$.
Here, we show that this $F_4^{(1)}$ subgroup
in fact arises as part of a normalizer in $W(E_8^{(1)})$. 
Moreover, we give the sub-root system of $E_8^{(1)}$ 
on which the generators of this $F_4^{(1)}$ subgroup
(involutions in general)
can be realised as reflections, thereby allowing
translational (or quasi-translational)
type elements
to be constructed.
Finally, we show that $\varphi_a$
is an element of quasi-translation in the $F_4^{(1)}$ subgroup.

The paper is organised as follows.
We lay down some general facts about the finite Weyl group $W$ in Section \ref{Weyl}. In particular, we discuss in detail the root system of $W$--one of the most useful tools for studying the Weyl groups. Examples of root systems of
type $B_3$, $C_3$, $F_4$ and $G_2$ are used to highlight
the various properties missing from our previous discussions on simply-laced groups \cite{jns4, Shi:19}.
In Section \ref{AW}, after listing some general properties of the affine Weyl group $W^{(1)}$, 
a dual space $\dav$ is introduced along with
the coroots and fundamental weights.
Here, 
we show that $W^{(1)}$
contains a normal subgroup of translations on the root lattice $Q$,
$W^{(1)}=W\ltimes Q$, 
by studying the dual representation. Moreover, we write down explicitly
the elements of translation by both long and short coroots for non-simply-laced types.
In Section \ref{EAW}, we construct a certain extension of $W^{(1)}$, 
such
that it contains a normal subgroup of translations on the weight lattice $P$, that is $\widetilde{W}^{(1)}=W\ltimes P$. We write down explicitly
the expressions for translations on the weight lattice of
types $B_3$ and $C_3$.
In Section \ref{Dis} we discuss two examples from integrable systems
using the expressions derived earlier, and
finally give some concluding remarks and future prospects
in Section \ref{Con}.
\section{Weyl group}\label{Weyl}
Let $\Ga$ be a Dynkin diagram with $n$ nodes. We associate to it a {\it reflection group} or {\it Coxeter
group}, 
\begin{equation}\label{Cox}
 W=W(\Ga)=\langle s_i\mid s_i^2=1, (s_is_j)^{m_{ij}}=1, \; 1\leq i,j \leq n\rangle,   
\end{equation}
given by the {\it Coxeter presentation}, that is a generating set satisfying
some defining relations.
When  parameter $m_{ij}$ takes value in $\{2,3,4,6\}$,
known as the {\it crystallographic condition}, W is called a {\it Weyl group}.
Weyl groups are classified by the associated types of Dynkin diagrams (see Figure \ref{claW}), their defining relations encoded in the corresponding
Dynkin diagram $\Ga$ with the following rules. Each node of the diagram represents a generator $s_i$ (of order $2$) for $1\leq i\leq n$. The order of the product
of generators $s_i$ and $s_j$, that is
$m_{ij}$ takes the value of: $2, 3, 4$ or $6$ 
when two nodes labeled $i$ and $j$
are: disconnected, joined by a single, a double, or a triple edge, respectively.
As an example, see the fundamental relations \eqref{frB3} of $W(B_3)$ 
corresponding to $\Ga(B_3)$ in Figure \ref{B3Dyn}.
For each $\Ga$, there is a corresponding {\it Cartan matrix}, 
\beq
C(\Ga)=(a_{ij}),\quad {\rm where }\quad a_{ii}=2\quad\mbox{for all $1\leq i,j \leq n$.}
\eeq
The values of $a_{ij}$, for $i\neq j$, can be read off from $\Ga$
depending on the connection between nodes $i$ and $j$ following 
\begin{table}[ht]
\centering
\begin{tabular}{ |c|c|c|c| } 
 \hline
 $a_{ij}$&$a_{ji}$&$m_{ij}$ & $i$\quad\quad\quad $j$ \\ [0.5ex] 
 \hline
 0&  0&2 &\begin{tikzpicture}[scale=0.5]
			\node (a2) {$\circ$} ;
			\node [right=of a2](a3) {$\circ$} ;
\end{tikzpicture}\\ 
 -1&-1&3 & \begin{tikzpicture}[scale=0.5]
			\node (a2) {$\circ$} ;
			\node [right=of a2](a3) {$\circ$} ;
			\draw[-] (a2) -- node {} (a3);
\end{tikzpicture} \\ 
 -2&-1&4 & \begin{tikzpicture}[scale=0.5]
			\node (a2) {$\circ$} ;
			\node [right=of a2](a3) {$\circ$} ;
			\draw[double distance=1.5pt] (a2) -- node {} (a3);
			\node (b) at ($(a2)!0.5!(a3)$) {$>$};
\end{tikzpicture}\\ 
 -3&-1&6 & \begin{tikzpicture}[scale=0.5]
			\node (a2) {$\circ$} ;
			\node [right=of a2](a3) {$\circ$} ;
			\draw[triple] (a2) -- node {} (a3);
			\node (b) at ($(a2)!0.5!(a3)$) {$>$};
\end{tikzpicture} \\ 
 \hline
 \end{tabular}
 \vspace{2em}
\caption{For connections in the Dynkin
diagram between vertices
$i$ and $j$ in the last column, we give $a_{ij}$, entries of the Cartan matrix for $i\neq j$; $m_{ij}$, and the order of the product
$s_{i}s_{j}$.}\label{CD}
\end{table}

	\begin{figure}[h]
\centering
%
			\begin{tikzpicture}[scale=1]
			\centering
			\node  (a1) {$\circ$};
			\node [right=of a1](a2) {$\circ$} ;
			\node [right=of a2](a3) {$\circ$} ;
			\node [right=of a3](a4) {$\circ$} ;
			\node [left=of a1](an){};
			\draw (a1) node [anchor=north] {$1$} ;
			\draw (a2) node [anchor=north] {$2$} ;
			\draw (a3) node [anchor=north] {$n-1$} ;
			\draw (a4) node [anchor=north] {$n$} ;
			\draw (an) node {$A_n$} ;
			\draw[-] (a1) -- node {} (a2);
			\draw[dashed] (a2) --node {} (a3);
			\draw[-] (a3) --node {} (a4);
			\node (d) at ($(a1)!0.5!(a4)$) {};
			\node [below=of d](a0) {$\circ$} ;;
			\draw[-] (a4) -- (a0);
			\draw[-] (a1) -- (a0);
			\draw (a0) node [label=90:$0$]{};
			\path[use as bounding box] (-1.5,0) rectangle (0,0); 
			\end{tikzpicture}
			
			\begin{tikzpicture}[scale=1]
			\centering
			\node  (a1) {$\circ$};
			\node [right=of a1](a2) {$\circ$} ;
			\node [right=of a2](a3) {$\circ$} ;
			\node [right=of a3](a4) {$\circ$} ;
			\node [left=of a1](an){};
			\node [above=of a1](a5){$\circ$};
			
			\draw (a1) node [anchor=north] {$1$} ;
			\draw (a2) node [anchor=north] {$2$} ;
			\draw (a3) node [anchor=north] {$n-1$} ;
			\draw (a4) node [anchor=north] {$n$} ;
			\draw (a5) node [anchor=east] {$0$} ;
			\draw (an) node {$B_n$} ;
			\draw[-] (a1) -- node {} (a2);
			\draw[dashed] (a2) --node {} (a3);
			\draw[double distance=1.5pt] (a3) -- node {} (a4);
			\node (b) at ($(a3)!0.5!(a4)$) {$>$};
			\draw[-] (a5) -- node {} (a2);
			\path[use as bounding box] (-1.5,0) rectangle (0,0); 
			\end{tikzpicture}\\
	
				\vspace{7pt}	
			\begin{tikzpicture}[scale=1]
			\centering
			\node  (a1) {$\circ$};
			\node  [left=of a1](a0) {$\circ$};
			\node [right=of a1](a2) {$\circ$} ;
			\node [right=of a2](a3) {$\circ$} ;
			\node [right=of a3](a4) {$\circ$} ;
			\node [left=of a0](an){};
			\draw (a1) node [anchor=north] {$1$} ;
			\draw (a2) node [anchor=north] {$2$} ;
			\draw (a3) node [anchor=north] {$n-1$} ;
			\draw (a4) node [anchor=north] {$n$} ;
			\draw (an) node {$C_n$} ;
			\draw (a0) node [anchor=north] {$0$} ;
			\draw[-] (a1) -- node {} (a2);
			\draw[dashed] (a2) --node {} (a3);
			\draw[double distance=1.5pt] (a3) -- node {} (a4);
			\node (b) at ($(a4)!0.5!(a3)$) {$<$};
			\draw[double distance=1.5pt] (a1) -- node {} (a0);
			\node (b) at ($(a1)!0.5!(a0)$) {$>$};
			\path[use as bounding box] (-1.5,0) rectangle (0,0); 
			\end{tikzpicture}
				\vspace{20pt}
				
			\begin{tikzpicture}[scale=1]
			\centering
			\node  (a1) {};
			\node [above=of a1](a0) {$\circ$} ;
			\node [right=of a1](a2) {$\circ$} ;
			\node [right=of a2](a3) {} ;
			\node [right=of a3](a4) {$\circ$} ;
			\node [right=of a4](a5){};
			\node [above=of a5](a8) {$\circ$};
			\node [below=of a5](a9) {$\circ$};
			\node [below=of a1](a10) {$\circ$};
			\node [left=of a1](an){};
			\draw (a10) node [anchor=north] {$1$} ;
			\draw (a2) node [anchor=north] {$2$} ;
			\draw (a3) node [anchor=north] {} ;
			\draw (a4) node [anchor=west] {$n-2$} ;
			\draw (a8) node [anchor=west] {$n$} ;
			\draw (a9) node [anchor=west] {$n-1$} ;
			\draw (an) node {$D_n$} ;
			\draw (a0) node [anchor=east] {$0$} ;
			\draw[-] (a10) -- node {} (a2);
			\draw[dashed] (a2) --node {} (a3);
			\draw [dashed] (a3) --node {} (a4);
			\draw[-] (a4) -- (a8);
			\draw[-] (a4) -- (a9);
			\draw[-] (a0) -- node {} (a2);
			\path[use as bounding box] (-1.5,0) rectangle (0,0);
			\path[use as bounding box] (-1.5,0) rectangle (0,0); 
			\end{tikzpicture}
		\vspace{-22pt}
		
			\begin{tikzpicture}[scale=1]
			\node  (a1) {$\circ$};
			\node [right=of a1](a2){$\circ$};
			\node [right=of a2](a3) {$\circ$};
			\node [right=of a3](a4) {$\circ$};
			\node [right=of a4](a5){$\circ$};
			\node [above= of a3] (a6) {$\circ$};
			\node [left=of a1](an){};
			\node [above= of a6] (a0) {$\circ$};
			\draw (a1) node [anchor=north] {$2$} ;
			\draw (a2) node [anchor=north] {$3$} ;
			\draw (a3) node [anchor=north] {$4$} ;
			\draw (a4) node [anchor=north] {$5$} ;
			\draw (a5) node [anchor=north] {$6$} ;
			\draw (a6) node [anchor=east] {$1$} ;
			\draw (a0) node [anchor=east] {$0$} ;
			\draw (an) node {$E_6$} ;
			\draw[-] (a1) --  (a2);
			\draw[-] (a2) --  (a3);
			\draw[-] (a3) -- (a6);
			\draw[-] (a3) -- (a4);
			\draw[-] (a4) -- (a5);
			\draw[-] (a0) -- (a6);
			\path[use as bounding box] (-1.5,0) rectangle (0,0);
			\end{tikzpicture}\\
			\begin{tikzpicture}[scale=1]
			\node  (a1) {$\circ$};
			\node [left=of a1](a0){$\circ$};
			\node [right=of a1](a2){$\circ$};
			\node [right=of a2](a3) {$\circ$};
			\node [right=of a3](a4) {$\circ$};
			\node [right=of a4](a5){$\circ$};
			\node [right=of a5](a6) {$\circ$};
			\node [above= of a3] (a7) {$\circ$};
			\node [left=of a0](an){};
			\draw (a1) node [anchor=north] {$1$} ;
			\draw (a2) node [anchor=north] {$3$} ;
			\draw (a3) node [anchor=north] {$4$} ;
			\draw (a4) node [anchor=north] {$5$} ;
			\draw (a5) node [anchor=north] {$6$} ;
			\draw (a6) node [anchor=north] {$7$} ;
			\draw (a7) node [anchor=east] {$2$} ;
			\draw (a0) node [anchor=north] {$0$} ;
			\draw (an) node {$E_7$} ;
			\draw[-] (a1) --  (a2);
			\draw[-] (a2) --  (a3);
			\draw[-] (a3) -- (a7);
			\draw[-] (a3) -- (a4);
			\draw[-] (a4) -- (a5);
			\draw[-] (a5) -- (a6);
			\draw[-] (a1) --  (a0);
			\path[use as bounding box] (-1.5,0) rectangle (0,0);
			\end{tikzpicture}
			\vspace{20pt}
			\begin{tikzpicture}[scale=1]
			\node  (a1) {$\circ$};
			\node [right=of a1](a2){$\circ$};
			\node [right=of a2](a3) {$\circ$};
			\node [right=of a3](a4) {$\circ$};
			\node [right=of a4](a5){$\circ$};
			\node [right=of a5](a6) {$\circ$};
			\node [right=of a6](a7) {$\circ$};
			\node [above= of a3] (a8) {$\circ$};
			\node [left=of a1](an){};
			\node [right=of a7](a0) {$\circ$};
			\draw (a1) node [anchor=north] {$1$} ;
			\draw (a2) node [anchor=north] {$3$} ;
			\draw (a3) node [anchor=north] {$4$} ;
			\draw (a4) node [anchor=north] {$5$} ;
			\draw (a5) node [anchor=north] {$6$} ;
			\draw (a6) node [anchor=north] {$7$} ;
			\draw (a7) node [anchor=north] {$8$} ;
			\draw (a8) node [anchor=east] {$2$} ;
			\draw (an) node {$E_8$} ;
			\draw (a0) node [anchor=north] {$0$} ;
			\draw[-] (a1) --  (a2);
			\draw[-] (a2) --  (a3);
			\draw[-] (a3) -- (a8);
			\draw[-] (a3) -- (a4);
			\draw[-] (a4) -- (a5);
			\draw[-] (a5) -- (a6);
			\draw[-] (a6) -- (a7);
			\draw[-] (a0) -- (a7);
			\path[use as bounding box] (-1.5,0) rectangle (0,0);
			\end{tikzpicture}
			
				\begin{tikzpicture}[scale=1]
			\centering
			\node  (a1) {$\circ$};
			\node [left=of a1](a0) {$\circ$} ;
			\node [right=of a1](a2) {$\circ$} ;
			\node [right=of a2](a3) {$\circ$} ;
			\node [right=of a3](a4) {$\circ$} ;
			\node [left=of a0](an){};
			\draw (a1) node [anchor=north] {$1$} ;
			\draw (a2) node [anchor=north] {$2$} ;
			\draw (a3) node [anchor=north] {$3$} ;
			\draw (a4) node [anchor=north] {$4$} ;
			\draw (an) node {$F_4$} ;
			\draw (a0) node [anchor=north] {$0$} ;
			\draw[-] (a1) -- node {} (a2);
			\draw[-] (a3) --node {} (a4);
			\draw[double distance=1.5pt] (a2) -- node {} (a3);
			\node (b) at ($(a2)!0.5!(a3)$) {$>$};
			\draw[-] (a1) -- node {} (a0);
			\path[use as bounding box] (-1.5,0) rectangle (0,0); 
			\end{tikzpicture}\\
			\vspace{5pt}
				\begin{tikzpicture}
			\centering
			\node  (a1) {$\circ$};
			\node [right=of a1](a2) {$\circ$} ;
				\node [right=of a2](a0) {$\circ$} ;
			\node [left=of a1](an){};
			
			\draw (a1) node [anchor=north] {$1$} ;
			\draw (a2) node [anchor=north] {$2$} ;
			\draw (an) node {$G_2$} ;
			\draw (a0) node [anchor=north] {$0$} ;
			\draw[triple] (a1) -- node {} (a2);
			\node (b) at ($(a1)!0.5!(a2)$) {$<$};
			\draw[-] (a2) -- node {} (a0);
			\path[use as bounding box] (-1.5,0) rectangle (0,0);
\end{tikzpicture}
%

			\caption{Dynkin diagrams of affine Weyl groups.}\label{claW}
		\end{figure}

Diagrams of type $A_n$, $D_n$, $E_6$, $E_7$ and $E_8$ contain only
single edges are called {\it simply-laced}. Otherwise, they are
{\it non-simply-laced} types, that is $B_n$, $C_n$, $F_4$ and $G_2$.

Let $V$ be an $n$-dimensional real vector space.
Each node $i$ of $\Ga$ can be associated with
a vector $\al_i\in V$, called {\it a simple root}.
The set of simple roots $\De=\De(\Ga)=\{ \al_i\mid 1\leq i \leq n\}$, is the {\it simple system}, and it forms a basis for
$V$. The vector space $V$ is equipped with
a symmetric positive definite bilinear form given by
\begin{equation}\label{alaij}
\al_i\cdot\al_j=|\al_i||\al_j|\cos\left(\pi-\frac{\pi}{m_{ij}}\right)=\frac{|\al_j|^2a_{ij}}{2},
\quad\mbox{for all $1\leq i,j \leq n$,}
\end{equation}where
the quantity $\al_i\cdot\al_i=|\al_i|^2$ gives the usual interpretation of
 squared length of $\al_i$, its value depends on
 the type of Dynkin diagram. 
 
 The crystallographic condition of $W$ means that 
 the entries of $C(\Ga)$,
 \beq\label{aijmij}
 a_{ij}=\frac{2\al_i\cdot\al_j}{\al_j\cdot\al_j}
 \eeq
 take integer values. Moreover, from Equation \eqref{alaij}, we have
 \beq\label{rlC}
 \frac{|\al_i|^2}{|\al_j|^2}=\frac{a_{ij}}{a_{ji}},\quad\mbox{for all $1\leq i,j \leq n$.}
 \eeq

For simply-laced typed $\Ga$, $C(\Ga)$ is symmetric $a_{ij}=a_{ji}$ (see Table \ref{CD}).
Then by Equation \eqref{rlC}, there is only one root length, usually normalised to $|\al_i|^2=2$, for all $1\leq i \leq n$.

For non-simply-laced types, $C(\Ga)$ is is not symmetric 
and Equation \eqref{rlC} implies that we have roots of two different lengths, referred to as being {\it long} and {\it short}, respectively. The difference
in lengths of the simple roots is indicated 
in a non-simply-laced Dynkin diagram by
the arrow on a multiple edge: arrow points
from a long root to a short root (see Figure \ref{claW}). 

We give a chosen normalisation of root lengths in the
following remark.

{\Rem \label{rlRem}
\begin{subequations}
\begin{itemize}
\item{$B_n$}
The long roots are normalised to have squared length 2.
That is
\beq
|\al_j|^2=2, \quad 1\leq j \leq n-1\quad\mbox{which implies that}\quad
|\al_n|^2=1,
\eeq
since $\frac{|\al_{n-1}|^2}{|\al_n|^2}=\frac{a_{n-1\,n}}{a_{n\,n-1}}=2$.
\item{$C_n$}
The short roots are normalised to have squared length 2.
That is
\beq
|\al_j|^2=2,\quad 1\leq j \leq n-1\quad\mbox{which implies that}\quad
|\al_n|^2=4,
\eeq
since $\frac{|\al_{n-1}|^2}{|\al_n|^2}=\frac{a_{n-1\,n}}{a_{n\,n-1}}=\frac{1}{2}$.
\item{$F_4$}
The long roots are normalised to have squared length 2.
That is
\beq\label{F4rl}
|\al_1|^2=|\al_2|^2=2\quad\mbox{which implies that}\quad
|\al_3|^2=|\al_4|^2=1,
\eeq
since $\frac{|\al_{2}|^2}{|\al_3|^2}=\frac{a_{2\,3}}{a_{3\,2}}=2$.
\item{$G_2$}
The short roots are normalised to have squared length 2.
That is
\beq
|\al_1|^2=2,\quad\mbox{which implies that}\quad
|\al_2|^2=6,
\eeq
since $\frac{|\al_{2}|^2}{|\al_1|^2}=\frac{a_{2\,1}}{a_{1\,2}}=3$.\\
\end{itemize}
\end{subequations}
}

$W(\Ga)$ can be realised as a group of reflections in $V$ as follows.
The generator $s_j=s_{\al_j}$ of $W$ is realised
as the reflection along the hyperplane
orthogonal to the simple root $\al_j\in\De$. In particular, it acts on the $\De$
basis of $V$ by the usual formula of reflection,
\begin{equation}\label{sij}
s_j(\al_i)=\al_i-\frac{2\al_i\cdot\al_j}{\al_j\cdot\al_j}\al_j
=\al_i-a_{ij}\al_j\quad\mbox{for all}\quad i,j\in\{1,..., n\},
\end{equation}
where $a_{ij}$ is the $(i,j)$-entry
of $C(\Ga)$.
The element $s_j$ is called a {\it simple reflection}.

Define the {\it root system} of $W(\Ga)$ to be the set of vectors we get
by acting $W(\Ga)$ on $\De$,
\beq\label{frs}
\Phi=\Phi(\Ga)=W(\Ga)\De.
\eeq
{\Rem\label{abint}
From Equation \eqref{sij}, and the fact that $a_{ij}$ are integers, we
have that all elements of $\Phi$, known as {\it roots}, are
integer combinations of 
the simple system $\De$.
We have
\beq
\frac{2\al\cdot\be}{\be\cdot\be}\in \mathbb{Z}\quad \mbox{for all}\quad \al, \be\in \Phi.
\eeq
}

$\Phi$ is finite whenever $W$ is finite. Moreover, we have
\beq\label{prs1}
\Phi=\Phi_{+}\cup\Phi_{-},
\eeq
where
\beq\label{prs2}
\Phi_{+}=\{\al=\sum_{i=1}^{n} \la_i\al_i\mid\mathbb{Z}\ni\la_i\geq 0, \al_i\in \De\}\quad\text{and}
\quad\Phi_{-}=\{-\al\mid\al\in \Phi_{+}\}.
\eeq
$\sum_i \la_i$ is called the {\it height} of the root $\al$, and
$\Phi_{+}$ is the {\it positive root system}. 
We define $-\al$ to have the same height as  $\al$, for all $\al\in \Phi_{+}$.
Equations \eqref{prs1}
and \eqref{prs2} say that a root is 
some integer combinations of $\De$, with coefficients being either all positive or all negative, moreover $|\Phi|=2|\Phi_{+}|$. 
For each finite root system $\Phi$, there exists a {\it highest root}
\beq\label{lr}
\tilde{\al}=\sum_{i=1}^{n}c_i\al_i.
\eeq
We list the coefficients $c_i$ $(1\leq i \leq n)$ for all Weyl groups in Table \ref{basic} of Appendix \ref{list}
(see Figure \ref{claW} for the corresponding numbering of nodes
in the Dynkin diagrams).
For $\Phi$ of non-simply-laced type, where there are long and short roots, $\tilde{\al}$
is used to denote the highest long root, and we use $\tilde{\al}_s$ for the
short root in $\Phi^{+}$ with maximum height.

{\Rem\label{orbWQ}
The finite Weyl group $W$ acts transitively on roots of the same length.
That is, all long (respectively short) roots of $\Phi$ form a single $W$-orbit.
In general, vectors
in the same W-orbit will have the same length (note that the opposite is not true).
}
We list
$|\Phi_{+}|$ and $|W|$ for Weyl groups of all types 
in Table \ref{basic} of Appendix \ref{list}.

We now illustrate the properties of $W$ discussed so far
for the $B_3$, $C_3$, $F_4$ and $G_2$ types.
{\eg\label{B3}
Finite Weyl group of type $B_3$, $W(B_3)$. 
Given the Dynkin diagram 
of type $B_3$, $\Ga(B_3)$ in Figure \ref{B3Dyn},

\begin{figure}[ht]
\centering
\begin{tikzpicture}
		
			\node  (a1) {$\circ$};
			\node [right=of a1](a2) {$\circ$} ;
			\node [right=of a2](a3) {$\circ$} ;
			\draw (a1) node [anchor=north] {$1$} ;
			\draw (a2) node [anchor=north] {$2$} ;
			\draw (a3) node [anchor=north] {$3$} ;
			\draw[-] (a1) -- node {} (a2);
			\draw[double distance=1.5pt] (a2) -- node {} (a3);
			\node (b) at ($(a2)!0.5!(a3)$) {$>$};
			\path[use as bounding box] (-1.5,0) rectangle (0,0); 
\end{tikzpicture}
%
\caption{Dynkin diagram of finite type $B_3$, $\Ga(B_3)$.}\label{B3Dyn}
\end{figure}
the corresponding Cartan matrix, 
\beq\label{CarB3}
C(B_3)=(a_{ij})_{1\leq i,j\leq3}=\bp
2&-1&0\\
-1&2&-2\\
0&-1&2
\ep,
\eeq
and the defining relations for the Weyl group, 
\beq\label{frB3}
W(B_3)=\lan s_1,s_2,s_3\mid s_1^2=s_2^2=s_3^2=(s_1s_2)^3=(s_1s_3)^2=(s_2s_3)^4=1\ran,
\eeq
can be written down using $\Ga(B_3)$ with the rules
given in Table \ref{CD}.
The simple system $\De=\{\al_1,\al_2,\al_3\}$ is a basis of an $3$-dimensional real vector space $V$, equipped
with a symmetric bilinear form,
\beq\label{SymB3}
(\al_i\cdot\al_j)_{1\leq i,j\leq3}=\left(\frac{|\al_j|^2}{2}a_{ij}\right)_{1\leq i,j\leq3}
=\bp
2&-1&0\\
-1&2&-1\\
0&-1&1
\ep,
\eeq
on which $W(B_3)$ acts as a group of
reflections.

From Equation \eqref{SymB3} we have
\beq\label{laB3}
|\al_i|^2=2 \quad{\rm for}\quad(i=1,2)\quad {\rm and}\quad |\al_3|^2=1.
\eeq
That is, roots $\al_1$, $\al_2$ are long and $\al_3$ is short,
as indicated by how the arrow on the multiple edge points in $\Ga(B_3)$, see Figure \ref{B3Dyn}.

The simple reflection $s_j\in W(B_3)$ acts on the $\De$ basis of $V$
by Equation \eqref{sij} with $a_{ij}$ being the $(i,j)$-entry
of $C(B_3)$ from Equation \eqref{CarB3}.
The $B_3$ root system is given by $\Phi(B_3)=W(B_3)\De=\Phi^{+}\cup\Phi^{-}$, where
$\Phi^{+}$ has six long roots and three short roots. 
We list them below, starting with
the highest roots (long $\tilde{\al}$ and short $\tilde{\al}_s$, respectively).
 We give also
the corresponding element in $W$ takes the highest root to that root. 

\begin{align}\nonumber
&{\rm Long} & {\rm Short}&\\\nonumber
1:\quad&\al_1+2\al_2+2\al_3=\tilde{\al}, &&\\\nonumber
s_2:\quad&\al_1+\al_2+2\al_3, &&\\\nonumber
s_1s_2:\quad&\al_2+2\al_3, &&\\\label{rsB3}
s_3s_2:\quad&\al_1+\al_2, & 1:\quad&\al_1+\al_2+\al_3=\tilde{\al}_s,\\\nonumber
s_1s_3s_2:\quad&\al_2, & s_1:\quad&\al_2+\al_3,\\\nonumber
s_2s_3s_2:\quad&\al_1. & s_2s_1:\quad&\al_3.\\\nonumber
\end{align}
We see that, of the simple system $\De$, the long roots $\al_1$, $\al_2$ belong to the same $W$-orbit of $\tilde{\al}$, whereas the short simple root $\al_3$ is in the
$W$-orbit of $\tilde{\al}_s$.
{\Rem\label{Bnr}
 In general, for $B_n$ type root system, we have $n^2=n(n-1)+n$
positive roots in $\Phi^{+}$, with $n(n-1)$ long and $n$ short.
}
}
{\Rem\label{rsconv}
For simplicity,  
we sometimes adopt the notation: $\al_i+...+\al_j=\al_{i...j}$ to
express the sum of simple roots,
and $s_i...s_j=s_{i...j}$ for product of simple reflections.
For example, we now write 
$\tilde{\al}$ and $\tilde{\al}_s$ as $\al_{12233}$ and $\al_{123}$, respectively.
The element $s_2s_3s_2s_3\in W(B_3)$, can now be written as $s_{2323}$.
}
{\eg\label{C3}
Finite Weyl group of type $C_3$, $W(C_3)$. Given the Dynkin diagram $\Ga(C_3)$ 
 in Figure \ref{C3Dyn}.
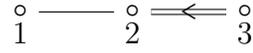
\begin{figure}[ht]
\centering
\begin{tikzpicture}
		
			\node  (a1) {$\circ$};
			\node [right=of a1](a2) {$\circ$} ;
			\node [right=of a2](a3) {$\circ$} ;
			\draw (a1) node [anchor=north] {$1$} ;
			\draw (a2) node [anchor=north] {$2$} ;
			\draw (a3) node [anchor=north] {$3$} ;
			\draw[-] (a1) -- node {} (a2);
			\draw[double distance=1.5pt] (a2) -- node {} (a3);
			\node (b) at ($(a2)!0.5!(a3)$) {$<$};
			\path[use as bounding box] (-1.5,0) rectangle (0,0); 
\end{tikzpicture}
%
\caption{Dynkin diagram of finite type $C_3$, $\Ga(C_3)$.}\label{C3Dyn}
\end{figure}

the corresponding Cartan matrix,
\beq\label{CarC3}
C(C_3)=(a_{ij})_{1\leq i,j\leq3}=\bp
2&-1&0\\
-1&2&-1\\
0&-2&2
\ep,
\eeq
and the defining relations for the Weyl group,

\beq\label{frC3}
W(C_3)=\lan s_1,s_2,s_3\mid s_1^2=s_2^2=s_3^2=(s_1s_2)^3=(s_1s_3)^2=(s_2s_3)^4=1\ran,
\eeq
can be written down using $\Ga(C_3)$ with the rules
given in Table \ref{CD}.

The simple system $\De=\{\al_1,\al_2,\al_3\}$ is a basis of an $3$-dimensional real vector space $V$, equipped
with a symmetric bilinear form given by
\beq\label{SymC3}
(\al_i\cdot\al_j)_{1\leq i,j\leq3}=\left(\frac{|\al_j|^2}{2}a_{ij}\right)_{1\leq i,j\leq3}
=\bp
2&-1&0\\
-1&2&-2\\
0&-2&4
\ep,
\eeq
on which $W(C_3)$ acts as a group of
reflections.

From Equation \eqref{SymC3} we have
\beq\label{laC3}
|\al_i|^2=2 \quad{\rm for}\quad(i=1,2)\quad {\rm and}\quad |\al_3|^2=4.
\eeq
That is, $\al_1$, $\al_2$ are short and $\al_3$ is long,
as indicated by the way the arrow is pointing in $\Ga(C_3)$, see Figure \ref{C3Dyn}.

Actions of $s_i\in W(C_3)$ on $\De$ are given by Equation \eqref{sij} with $a_{ij}$ being the $(i,j)$-entry
of $C(C_3)$ from Equation \eqref{CarC3}.
The $C_3$ root system is $\Phi=W(C_3)\De=\Phi^{+}\cup\Phi^{-}$.
In general, for $C_n$ type root system there are $n^2=n(n-1)+n$
positive roots in $\Phi^{+}$, with $n(n-1)$ short and $n$ long. Hence for $\Phi^{+}$ of $C_3$ type
we have the following six short roots and three long roots,
\begin{align}\nonumber
&{\rm short} & {\rm long}&\\\nonumber
1:\quad&\al_{1223}=\tilde{\al}_s, &&\\\nonumber
s_2:\quad&\al_{123}, &&\\\nonumber
s_1s_2:\quad&\al_{23}, &&\\\label{rsC3}
s_3s_2:\quad&\al_{12}, & 1:\quad&\al_{11223}=\tilde{\al},\\\nonumber
s_1s_3s_2:\quad&\al_2, & s_1:\quad&\al_{223},\\\nonumber
s_2s_3s_2:\quad&\al_1. & s_2s_1:\quad&\al_3.\\\nonumber
\end{align}
}
{\eg\label{F4}
Finite Weyl group of type $F_4$, $W(F_4)$. Given the Dynkin diagram $\Ga(F_4)$ in Figure \ref{F4Dyn}. 

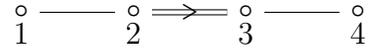
\begin{figure}[ht]
    \centering
\begin{tikzpicture}[scale=1]
			\node  (a1) {$\circ$};
			\node [right=of a1](a2) {$\circ$} ;
			\node [right=of a2](a3) {$\circ$} ;
			\node [right=of a3](a4) {$\circ$} ;
			\draw (a1) node [anchor=north] {$1$} ;
			\draw (a2) node [anchor=north] {$2$} ;
			\draw (a3) node [anchor=north] {$3$} ;
			\draw (a4) node [anchor=north] {$4$} ;
			\draw[-] (a1) -- node {} (a2);
			\draw[-] (a3) --node {} (a4);
			\draw[double distance=1.5pt] (a2) -- node {} (a3);
			\node (b) at ($(a2)!0.5!(a3)$) {$>$};
			\path[use as bounding box] (-1.5,0) rectangle (0,0); 
			\end{tikzpicture}
			\caption{Dynkin diagram of finite type $F_4$, $\Ga(F_4)$.}\label{F4Dyn}
\end{figure}
we have the corresponding Cartan matrix,
\beq\label{CarF4}
C(F_4)=(a_{ij})_{1\leq i,j\leq4}=\bp
2&-1&0&0\\
-1&2&-2&0\\
0&-1&2&-1\\
0&0&-1&2
\ep,
\eeq
and the defining relations for the Weyl group,
\begin{align}\nonumber
    W(F_4)&=\lan s_1,s_2,s_3,s_4\mid s_1^2=s_2^2=s_3^2=s_4^2=(s_1s_3)^2=(s_1s_4)^2=(s_2s_4)^2=1,\\\label{frF4}
    &(s_1s_2)^3=(s_2s_3)^4=(s_3s_4)^3=1\ran
\end{align}
using the rules
given in Table \ref{CD}.

The simple system $\De=\{\al_1,\al_2,\al_3,\al_4\}$ is a basis of an $4$-dimensional real vector space $V$, on which $W(F_4)$ acts as a group of
reflections. $V$ is equipped
with a symmetric bilinear form,
\beq\label{SymF4}
(\al_i\cdot\al_j)_{1\leq i,j\leq4}=\left(\frac{|\al_j|^2}{2}a_{ij}\right)_{1\leq i,j\leq4}
=\bp
2&-1&0&0\\
-1&2&-1&0\\
0&-1&1&-1/2\\
0&0&-1/2&1
\ep.
\eeq

From Equation \eqref{SymF4} we have
\beq\label{laF4}
|\al_i|^2=2 \quad{\rm for}\quad(i=1,2)\quad {\rm and}\quad |\al_j|^2=1, \quad j=3,4.
\eeq
That is, $\al_1$, $\al_2$ are long, while $\al_3$ and $\al_4$ are short,
as indicated by how the arrow points in $\Ga(F_4)$, see Figure \ref{F4Dyn}.
Actions of $s_i\in W(F_4)$ on $\De$ are given by Equation \eqref{sij} with $a_{ij}$ being the $(i,j)$-entry
of $C(F_4)$ from Equation \eqref{CarF4}.
The $F_4$ root system is $\Phi=W(F_4)\De=\Phi^{+}\cup\Phi^{-}$.
There are 24
positive roots in $\Phi^{+}$, with $12$ long and $12$ short. 
In particular, we have, 
\begin{equation}\label{hlrF4}
    \tilde{\al}=2\al_1+3\al_2+4\al_3+2\al_4,
\end{equation}
and
\begin{equation}\label{hsrF4}
    \tilde{\al}_s=\al_1+2\al_2+3\al_3+2\al_4.
\end{equation}
}
{\eg\label{G2}
Finite Weyl group of type $G_2$, $W(G_2)$. The Dynkin diagram 
of type $G_2$, $\Ga(G_2)$ is given in Figure \ref{G2Dyn}.

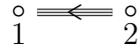
\begin{figure}[ht]
\centering
\begin{tikzpicture}
			\centering
			\node  (a1) {$\circ$};
			\node [right=of a1](a2) {$\circ$} ;
			
			\draw (a1) node [anchor=north] {$1$} ;
			\draw (a2) node [anchor=north] {$2$} ;
			\draw[triple] (a1) -- node {} (a2);
			\node (b) at ($(a1)!0.5!(a2)$) {$<$};
			\path[use as bounding box] (-1.5,0) rectangle (0,0);
			\end{tikzpicture}
%
			\caption{Dynkin diagram of finite type $G_2$, $\Ga(G_2)$.}\label{G2Dyn}
\end{figure}

The corresponding Cartan matrix,
\beq\label{CarG2}
C(G_2)=(a_{ij})_{1\leq i,j\leq2}=\bp
2&-1\\
-3&2\\
\ep,
\eeq
and the defining relations for the Weyl group,
\beq\label{frG2}
W(G_2)=\lan s_1,s_2\mid s_1^2=s_2^2=(s_1s_2)^6=1\ran,
\eeq
can be written down using $\Ga(G_2)$ with the rules
given in Table \ref{CD}.
The simple system $\De(G_2)=\De=\{\al_1,\al_2\}$ forms a basis of 
an $2$-dimensional real vector space $V$ on which $W(G_2)$ acts as a group of
reflections. 
The vector space $V$ is equipped
with a symmetric bilinear form,
\beq\label{SymG2}
(\al_i\cdot\al_j)_{1\leq i,j\leq2}=\left(\frac{|\al_j|^2}{2}a_{ij}\right)_{1\leq i,j\leq2}
=\bp
2&-3\\
-3&6\\
\ep.
\eeq

We have
\beq\label{laG2}
|\al_1|^2=2, \quad {\rm and}\quad |\al_2|^2=6.
\eeq
That is, roots $\al_1$ is short and $\al_2$ is long
as indicated by how the arrow on the multiple edge points in $\Ga(G_2)$, see Figure \ref{G2Dyn}.

The simple reflection $s_j\in W(G_2)$ acts on the $\De$ basis of $V$
by Equation \eqref{sij} with $a_{ij}$ being the $(i,j)$-entry
of $C(G_2)$ given by Equation \eqref{CarG2}.
The $G_2$ root system is $\Phi(G_2)=W(G_2)\De=\Phi^{+}\cup\Phi^{-}$.
$\Phi^{+}$ has three long roots and three short roots. 
We list them below, starting with
the highest roots (long $\tilde{\al}$ and short $\tilde{\al}_s$, respectively).
 We give also
the corresponding element in $W(G_2)$ that takes the highest long/short root to that root. 

\begin{align}\nonumber
&{\rm Long} & {\rm Short}&\\\nonumber
1:\quad&3\al_1+2\al_2=\tilde{\al}, &1:\quad&2\al_1+\al_2=\tilde{\al}_s,\\\label{rsG2}
s_2:\quad&3\al_1+\al_2, &s_1:\quad&\al_1+\al_2,\\\nonumber
s_1s_2:\quad&\al_2. &s_2s_1:\quad&\al_1.
\end{align}
}
\section{Affine Weyl group}\label{AW}
Let $\Ga^{(1)}$ be the affine extension of $\Ga$,  having an extra node labelled $0$,
associated with the affine reflection $s_0$ (or the affine simple root $\al_0$),
see diagrams in Figure \ref{claW}. The corresponding affine Weyl group,
\beq\label{aw}
W^{(1)}=\langle s_i\mid s_i^2=1, (s_is_j)^{m_{ij}}=1, \; 1\leq i,j \leq n\rangle,
\eeq
has its defining relations encoded by $\Ga^{(1)}$ with rules given in Table \ref{CD}. $W^{(1)}$ is a group of infinite order, to which $W=\langle s_i\mid \; 0\leq i\leq n\rangle$ is a finite standard parabolic
subgroup.
The corresponding Cartan matrix, $C(\Ga^{(1)})=(a_{ij})_{1\leq i,j\leq n,\, 0}$, 
the {\it generalised Cartan matrix}, is degenerate.  Here, we arrange the entries of $C(\Ga^{(1)})$
such that $(a_{i0})$ and $(a_{0j})$ are its last column and row, respectively. That is, we have $C(\Ga)=(C(\Ga^{(1)}))_{1\leq i,j\leq n}$.

The {\it simple affine system} 
\beq
\De^{(1)}
=\De\cup \{\al_0\}
=\{ \al_i\mid 0\leq i \leq n\}
\eeq forms a basis for an $n+1$-dimensional real vector space
$V^{(1)}$, on which one can define a semidefinite symmetric bilinear form using
$C(\Ga^{(1)})$,
\begin{equation}\label{alaij0}
\al_i\cdot\al_j=\frac{|\al_j|^2a_{ij}}{2},
\quad\mbox{for}\quad i,j\in\{0,1,..., n\}.
\end{equation}

The bilinear form on 
$V^{(1)}$ being semidefinite implies that there is a 1-dimensional subspace in $V^{(1)}$,
the {\it radical} of $V^{(1)}$: 
Rad$(V^{(1)})$,
is spanned by a vector $\de\in V^{(1)}$ called the {\it null root}, such that
\beq\label{aid}
\al_i\cdot\de=0\quad\mbox{for all}\quad 0\leq i \leq n.
\eeq
In particular, we have
\beq\label{d}
\de=\al_0+\tilde{\al}=\al_0+\sum_{i=1}^{n}c_i\al_i=\sum_{i=0}^{n}c_i\al_i,
\eeq
recall that $\tilde{\al}$ is the highest long root of $W$ defined in Equation \eqref{lr}.
By Equation \eqref{d}, we see that the set of vectors
$\{\al_1, \al_2, \ldots, \al_n, \de\}$
forms another basis of $V^{(1)}$.
Moreover, we see from the $\Ga^{(1)}$ diagrams  given in Figure \ref{claW}
that the root $\al_0\in \De^{(1)}$ is always a long root. That is,
\begin{equation}\label{a0ah}
|\al_0|^2=\al_0\cdot\al_0=(\de-\tilde{\al})\cdot(\de-\tilde{\al})=\tilde{\al}\cdot\tilde{\al}=|\tilde{\al}|^2.
\end{equation}
Recall that $W^{(1)}$ is a group of infinite order.
The corresponding affine root system
\beq
\Phi^{(1)}=\Phi(\Ga^{(1)})=W^{(1)}\De^{(1)},
\eeq
is also infinite. 
$W^{(1)}$ acts transitively on roots of $\Phi^{(1)}$ (of the same length).
Moreover, we have
\beq\label{prs1a}
\Phi^{(1)}=\Phi^{(1)}_{+}\cup\Phi^{(1)}_{-},
\eeq
where
\beq\label{prs2a}
\Phi^{(1)}_{+}=\{\al=\sum_{i=0}^n \la_i\al_i\mid\mathbb{Z}\ni\la_i\geq 0, \al_i\in \De^{(1)}\}\quad\text{and}
\quad\Phi^{(1)}_{-}=\{-\al\mid\al\in \Phi^{(1)}_{+}\}.
\eeq
The concept of the height of a root was discussed for
a finite Weyl group can be extended to the affine case.
{\Def\label{heiaw} Height of a root.
For a root in the positive system, $\al\in \Phi^{(1)}_{+}$, where $\al=\sum_{i=0}^n \la_i\al_i$, for $\al_i\in \De^{(1)}$ and $\mathbb{Z}\ni\la_i\geq 0$, the {\it height} of $\al$ is given by
$\sum_{i=0}^n \la_i$. 
We define $-\al$ to have the same height as  $\al$, for all $\al\in \Phi^{(1)}_{+}$.
}

Finally, it happens that all roots of the affine root system have the form
\beq\label{arsa}
\Phi^{(1)}=\{\al+m\de\mid\al\in\Phi,\; m\in\mathbb{Z}\},
\eeq
where $\Phi=W\De$ is the finite root system in a subspace of $V^{(1)}$, $V=\mbox{Span}(\De)$.

For $0\leq j\leq n$,  the generator $s_j\in W^{(1)}$ can
be realised as the reflection 
along the simple root $\al_j\in \De^{(1)}$, its action on $\De^{(1)}$ is
given by 
\begin{equation}\label{sij0}
s_j(\al_i)=s_{\al_j}(\al_i)
=\al_i-a_{ij}\al_j,\quad \mbox{for all}\quad i,j\in\{0,1,..., n\},
\end{equation}
where $a_{ij}$ is now the $(i,j)$-entry
of $C(\Ga^{(1)})$.

In general, the element of reflection along any root 
$\be\in\Phi^{(1)}$, $s_{\be}\in W^{(1)}$ 
acts on $V^{(1)}$ by,
\beq\label{sbv}
s_{\be}(v)=v-\frac{2v\cdot\be}{\be\cdot\be} \be,\quad\mbox{and
we have}\quad
s_{-\be}=s_{\be}
\eeq
for all $v\in V^{(1)}$.
Furthermore, $s_\be$ is related to a simple reflection $s_i$
by a conjugation,
\beq\label{sr}
s_{\be}=s_{w(\al_i)}=ws_iw^{-1},\quad\mbox{where}\quad w(\al_i)=\be,\quad w\in W^{(1)},\quad \al_i\in \De^{(1)},\quad \be\in\Phi^{(1)}.
\eeq
By Equations \eqref{sbv} and \eqref{aid} we have 
\beq\label{sid}
s_i(\de)=\de-\frac{2\de\cdot\al_i}{\al_i\cdot\al_i}\al_i=\de\quad\mbox{for all}\quad 0\leq i\leq n.
\eeq 
That is
\begin{equation}\label{AWde}
w(\de)=\de\quad\mbox{for all}\quad w\in W^{(1)}.  
\end{equation}

We now recall some useful properties and functions of 
the Weyl groups, which are also true for
Coxeter groups in general.
{\Def\label{Nuaw}
For each $w\in{W}^{(1)}$, define 
\beq
N(w)=\{\al\in \Phi^{(1)}_{+}\mid w(\al)\in \Phi^{(1)}_{-}\}.
\eeq
That is, $N(w)$ is the set of positive roots that
$w$ takes to some negative roots.
}
{\Def\label{law} Length of an element.
Each $w\in W^{(1)}$ can be expressed as a product of simple reflections. The least number of
simple reflections expression is called {\it reduced}, and this number is defined to be {\it the length of} $w$, $l(w)$.
Moreover, if $w_1$, $w_2\in W^{(1)}$ and $N(w_2)\subset N(w_1)$ then
\begin{equation}\label{dl}
l(w_1w_2^{-1})=l(w_1)-l(w_2).
\end{equation}
The general theory of Coxeter groups says that, for $w\in W^{(1)}$ and $\al\in\De^{(1)}$  we have,
\beq\label{lfun}
l(ws_{\al})=
 \begin{cases}
 l(w)+1, &\text{if}\quad w(\al)\in\Phi^{(1)}_{+}\\
 l(w)-1, &\text{if}\quad w(\al)\in\Phi^{(1)}_{-}.
 \end{cases}
 \eeq
 }
Then we have $|N(w)|=l(w)=k$, where $k$ is a non-negative integer. That is, 
Equation \eqref{lfun}
can be applied repeatedly until we can write $w$ as a product of $k$ simple reflections,
\begin{align}
ws_{l_1}...s_{l_k}&=1,\quad l_1, ..., l_k\in\{0,1,...,n\},\\\nonumber
\mbox{or}\quad\quad    w&=s_{l_k}...s_{l_1},
\end{align}
where we have used $s_j^2=1$ ($j\in\{0,1,...,n\}$).

Now as an example, we show how by knowing the actions 
of an element $T_1\in W(E_8^{(1)})$
on the $E_8^{(1)}$ simple system given in Equation \eqref{TJ21},
one can use Equation \eqref{lfun} to write $T_1$
as a product of simple reflections as given in Equation \eqref{T1decomp1}.
{\eg 
Let $\De^{(1)}=\{\al_j\mid 0\leq j\leq 8\}$ be the $E_8^{(1)}$ simple system,
we have $W(E_8^{(1)})=\lan s_i\mid 0\leq i \leq 8\ran$ and
$\Phi^{(1)}=W(E_8^{(1)})\De^{(1)}=\Phi^{(1)}_{+}\cup\Phi^{(1)}_{-}$ is the $E_8^{(1)}$ root system. The actions of $s_i\in W(E_8^{(1)})$ $(1\leq i \leq 8)$ on $\De^{(1)}$
are given by Equation \eqref{sij0} with $a_{ij}$
being the $(i,j)$-entry of $C(E_8^{(1)})$
given in Equation \eqref{CarE8a}, and
the $\de$ of $W(E_8^{(1)})$ is given in Equation \eqref{deE8}.

Given the action of an element $T_1$ of $W(E_8^{(1)})$
on its simple system $\De^{(1)}=\{\al_j\mid 0\leq j\leq 8\}$,
\beq\label{TJ233}
T_{1}:\{\al_1, \al_3\}\mapsto
\{\al_1-2\de, \al_3+\de\},
\eeq
where only non-trivial actions are shown after that.

We see
that $T_{1}(\al_i)\in \Phi^{(1)}_{-}$
only when $i=1$.
That is $l(T_1s_1)=l(T_1)-1$ by 
Equation \eqref{lfun}. Compute the action of $T_1s_1$ on $\De^{(1)}$ we have,
\beq
T_1s_1:\{\al_1, \al_3\}\mapsto
\{-\al_1+2\de, \al_3-\de\},
\eeq
that is $T_{1}s_1(\al_i)\in \Phi^{(1)}_{-}$
only when $i=3$,
so we have $l(T_1s_1s_3)=l(T_1s_1)-1$.
Calculate the action of $T_1s_1s_3$ on $\De^{(1)}$ we have,
\beq
T_1s_1s_3:\{\al_1, \al_3, \al_4\}\mapsto
\{\al_3+\de, -\al_1-\al_3+\de, \al_1+\al_3+\al_4-\de\},
\eeq
that is $T_{1}s_1s_3(\al_i)\in \Phi^{(1)}_{-}$ only when $i=4$,
that is $l(T_1s_1s_3s_4)=l(T_1s_1s_3)-1$.
Proceed in this fashion, we have  
$T_1s_1s_3s_4\cdots s_{l_k}=1$,  where $l_1, ..., l_k\in\{0,1,...,8\}$,
where we found $l_1=1$, $l_2=3$, $l_3=4$, ...,  $l_{58}=3$ and $k=58$. Rewriting this using $s_{j}^2=1$ (for $j\in\{0,1,...,8\}$) we have,
\begin{align}
T_1&=s_{l_k}\cdots s_{l_3}s_{l_2}s_{l_1}=s_{3}\cdots s_4s_3s_1,
\end{align}
which gives us the expression for $T_1$ as a product of 58 simple reflections
in Equation \eqref{T1decomp1}.
}
\subsection*{The longest element.}
For a finite Weyl group of rank $n$, $W=W(\Ga)$ with
the simple system $\De=\De(\Ga)$ and the corresponding finite positive root system $\Phi^{+}_{\Ga}=W\Ga$
 whereby slight abuse of notation we denote
both the Dynkin diagram and its type by $\Ga$), there exist
a unique element of maximal length $w_{\Ga}\in W$, its {\it longest element}, 
where $w_{\Ga}^2=1$ and
$N(w_{\Ga})=\Phi^{+}_{\Ga}$. 
Then we have $l(w_{\Ga})=\mid \Phi^{+}_{\Ga}\mid$.
Moreover, $
w_{\Ga}\De=-\De=\{-\al\mid\al\in \De\}.
$
In fact for a permutation $\sigma$ on the index set of $\Ga$ we have,
$
w_{\Ga}(\al_i)=-\al_{\sigma(i)}.
$
 For 
finite Weyl groups $\sigma$ is the identity for $A_1$, $B_n$, 
$C_n$, $D_n$ ($n$ even), $F_4$, $E_7$, $E_8$.
For the other types, $\sigma$ corresponds to the unique symmetry of order 2 of the Dynkin diagram.
{\eg
 Let $\De=\{\al_1, \al_2, \al_3\}$ be the simple system of type $A_3$,
and $W(A_3)=\lan s_1, s_2, s_3\ran$. From Table \ref{basic} we see that
$\mid\Phi^{+}_{A_3}\mid=6$, hence $l(w_{A_3})=6$. The longest element
of $W(A_3)$ is given by
\begin{equation}\label{lwa3}
w_{A_3}=s_2s_1s_3s_2s_1s_3,    
\end{equation}
it acts on the $A_3$ simple system by: 
\begin{equation}\label{lwa3act}
  w_{A_3}:\{\al_1,\al_2,\al_3\}\mapsto
\{-\al_3,-\al_2,-\al_1\}.  
\end{equation}
}
\subsection{Normalizer}\label{Norm} 
Let $W^{(1)}$ be an affine Weyl group with a simple system $\De^{(1)}$.
For a subset $J\subset \De^{(1)}$,
the group $W_J=\langle s_i\mid \al_i\in J\rangle$ 
is call the standard parabolic subgroup of $W^{(1)}$. 
The Normalizer of $W_J$ in $W^{(1)}$ is defined by 
$
N_W(W_J)=\{g\in W^{(1)}\mid g^{-1}W_Jg=W_J\}
=N_J\ltimes W_J.
$
In \cite{H, BH} it was shown that, 
$
N_J=\{w\in W^{(1)}\mid wJ=J\}.
$
That is, $N_J$ the set wise stabilizer of $J$--its element
either fixes or permutes the elements of $J$. 
The group $N_J$ is generated by 
{\it R-elements} and {\it M-elements}.
The R-elements, also known as {\it quasi-reflections}, are involutions 
 that act permutatively on the subset $J$, whereas 
the M-elements permute the R-elements.
Together, they generate a group of extended affine Weyl type
for which elements of translation (or quasi-translations in $W^{(1)}$) can be constructed.
Here we briefly explain how to find the R-elements when computing
a normalizer of $W_J$ in $W^{(1)}$. For
the full theory of normalizers in Coxeter groups see \cite{H, BH}.

Let $I$ and $J$ be disjoint subsets of $\De^{(1)}$, and let 
$L=I\cup J\subset \De^{(1)}$. Then there is a unique element $v[I,J]$
of $W_L$ given by,
\begin{equation}\label{vIJ}
  v[I, J]=w_{L}w_{J},   
\end{equation}
where $w_{L}$ and $w_{J}$ are the longest elements of 
the parabolic subgroups $W_L$ and $W_J$, respectively.
By Equation \eqref{dl} we have
\begin{equation}\label{lwLJ}
   l(v[I, J])=l(w_{L})-l(w_{J})=\mid \Phi^{+}_L \mid-\mid \Phi^{+}_J \mid. 
\end{equation}
In particular, when $|I|=1$,
elements $v[a, J]$ (for $a\in \De^{(1)}\setminus J$) which are involutions are called the {\it R-elements}.
It is also useful for us to consider the case when $|I|=2$,
that is elements of the type $v[\{a,b\},J]$
(for $a, b\in \De^{(1)}\setminus J$).
$v[\{a,b\},J]$ has exactly two standard expressions: 
\begin{equation}\label{vab}
 v[a_1,J_1]...v[a_nJ_n]=v[b_1,K_1]...v[b_nK_n],   
\end{equation}
where $a_n=a$, $b_n=b$, $J_n=K_n=J$. Furthermore, $a_i, b_j\in J\cup\{a,b\}$ and $J_i, K_j\subseteq J\cup\{a,b\}$ for all $i$, $j$.
Equation \eqref{vab} allows one to compute the orders for the pairwise products 
of the generators of $N_J$.
Let $V_J=\mbox{Span}\left(J\right)$, and $V_J^\perp$ be the orthogonal
complement of $V_J$ in $V^{(1)}$, that is, $V^{(1)}=V_J\bigoplus
V_J^\perp$.
Coxeter groups $W^{(1)}/W_J/{N}_J$ act on $V^{(1)}/V_J/V_{J^{\perp}}$ as reflection groups.
{\eg\label{2AinA3}
Given the finite Weyl group of type $A_3$,
$W(A_3)=\lan s_1, s_2, s_3\ran$ and its simple system $\De=\{\al_1, \al_2, \al_3\}$. 
Let $J=\{\al_1, \al_3\}\cong 2A_1$, 
then we have $W_J=\lan s_1, s_3\ran\cong W(2A_1)$.
The normalizer of ${W}_J$ in ${W}(A_3)$
is given by 
\begin{equation}\label{NWJa3}
N(W_J)=N_J\ltimes W_J=\lan s_{2312} \ran\ltimes
\lan s_1, s_2\ran
\cong W(A_1)\ltimes {W}(2A_1).
\end{equation}
\begin{proof}
To see this, recall that $N_J$ is generated by the R- and M-elements.
Since $J=\{\al_1, \al_3\}$ is the only type $2A_1$ subset of $\De$
there are no M-elements and we need only to find the R-elements, 
that is elements of the form $v[a, J]$ (for $a\in \De^{(1)}\setminus J$) which are involutions, and we see that there is only one possibility, $v[\al_2, J]$.
First, let us find the length of this element.
By Equation \eqref{lwLJ} we have,
\[
l(v[\al_2, J])=l(w_{J\cup\{\al_2\}}w_{J})
=l(w_{A_3}w_{2A_1})=l(w_{A_3})-l(w_{2A_1})
=\mid\Phi^{+}_{A_3}\mid-2\mid\Phi^{+}_{A_1}\mid=6-2=4.
\]
Moreover, since
$W_J\cong W(2A_1)$, we have $w_J=w_{2A_1}=s_1s_3$, and
\begin{equation}\label{lw2a1act}
  w_{2A_1}:\{\al_1,\al_2,\al_3\}\mapsto
\{-\al_1,\al_2,-\al_3\}.  
\end{equation}
Then by Equations \eqref{vIJ} and \eqref{lwa3} we have,
\begin{align}\nonumber
v[\al_2, J]&=w_{J\cup\{\al_2\}}w_{J},\\\nonumber
&=w_{A_3}w_{2A_1},\\\label{RA3}
&=s_2s_3s_1s_2s_1s_3s_1s_3,\\\nonumber
&=s_2s_3s_1s_2,\\\nonumber
&=s_{2312}.
\end{align}
Its action on $\De$ is obtained by composing
the actions of $w_{2A_1}$ and $w_{A_3}$ given in Equations \eqref{lw2a1act} and \eqref{lwa3act},
\[
v[\al_2, J]:\{\al_1,\al_2,\al_3\}\mapsto
\{\al_3,\al_2,\al_1\}.
\]
That is, $v[\al_2, J]$ permutes $\al_1$ and $\al_3$.  That is, it is an involution
hence an R-element.
\end{proof}
}
{\Rem For an arbitrary group $W$, the Normaliser of its subgroup usually
is no more than the subgroup itself. However, when $W$ is a Coxeter group and the subgroup a parabolic one, 
one can have some very non-trivial $N_J$.}

\subsection{A dual representation}
It is well-known that 
$W^{(1)}$ contains an abelian subgroup of translations
on the root lattice. To this end, it is useful to introduce a dual space $V^{(1)*}$ on which the translations are realised. 
{\Def\label{ds}
Let $V^{(1)*}$ be an $(n+1)$-dimensional real vector space, and 
$\langle {}\,, {} \rangle:$ $V^{(1)}\times V^{(1)\ast} \to \mathbb R$ be a bilinear
pairing between $V^{(1)}$ and $V^{(1)\ast}$.
Let $\{ h_1, \ldots, h_n, h_\de\} $  be  a basis of $ V^{(1)\ast}$   dual to
$\{\al_1, \al_2, \ldots, \al_n, \de\}\subset V^{(1)}$, that is we have,

\begin{subequations}
\begin{align}\label{ah1}
&\langle \al_i, h_j\rangle=\delta_{ij},\\\label{ah2}
&\langle \al_i, h_{\de}\rangle=
\langle \de, h_{j}\rangle=0,\quad \mbox{for}\quad 1\leq i, j\leq n,\\\label{ah3}
&\langle \de,h_\de \rangle=1.
\end{align}
\end{subequations}
The group
 $W^{(1)}$ acts on $V^{(1)\ast}$ via the contragredient action:
\beq\label{cona}
\lan w^{-1} v,  h\ran=\lan v, wh\ran, \quad\mbox{for}\quad v\in V^{(1)}, h\in V^{(1)\ast}, w\in W^{(1)}.
\eeq
}
A useful consequence of Definition \ref{ds} is
\begin{equation}\label{a0hj}
    \lan \al_0,h_j\ran=\lan \de-\sum_{i=1}^n c_i\al_i,h_j\ran=-c_j,\quad
    1\leq j\leq n.
\end{equation}
{\Def\label{dsX}
Define an $n$-dimensional hyperplane in $V^{(1)*}$,
\begin{equation}\label{Xk}
X_k=\{\,h\in V^{(1)\ast}\,|\, \langle \de, h\rangle=k\}.
\end{equation}
Since $w(\de)=\de$ for all $w\in W^{(1)}$, the set $X_k$ is preserved by $W^{(1)}$ for any $k\in\mathbb{R}$.
}
Whenever $k\neq 0$, the
set $X_k$ can be regarded as an $n$-dimensional affine space on which $W^{(1)}$ acts as affine transformations, such as translations.  
Moreover, we have $X_k=kh_\de+X_0$.

For $k=0$, $X_0$ is an $n$-dimensional subspace of $V^{(1)\ast}$. We see that the set $\{h_j\mid 1\leq j\leq n\}$ forms a basis for $X_0$ from 
the second equality in Equation \eqref{ah2}
of Definition \ref{ds}.
{\Def\label{wl}Vectors $h_j$ $( 1\leq j\leq n)$ are called {\it the fundamental weights} of $W^{(1)}$. 
The {\it weight lattice}  is the integer span of fundamental weights:
\begin{equation}
P=\bigoplus{}_{i=1}^{n}\mathbb{Z} h_i.
\end{equation}
}
\subsection*{A linear map from $V^{(1)}$ to $V^{(1)\ast}$}
We introduce a linear mapping $\pi$ from $V^{(1)}$ to its dual $V^{(1)\ast}$
which enables us to
quantitatively analyse different types of translational elements of $W^{(1)}$ and its  extensions. This is particularly relevant in the context of discrete integrable systems, since the dynamics of such systems arise from different types of translations in the Weyl group.
{\Def\label{p}
For each $u\in V^{(1)}$ there is a linear map $\pi(u):V^{(1)}\to\mathbb{R}$
defined by $v\mapsto v\cdot u$ for all $v\in V^{(1)}$. That is, $\pi(u)$ is
an element of $V^{(1)\ast}$ satisfying 
\beq\label{vpu}
v\cdot u=\lan v,\pi(u)\ran\quad\mbox{for all}\quad v\in V^{(1)}.
\eeq
}
For all $u,v\in V^{(1)}$, $w\in W^{(1)}$ we have
\[
\lan v,w\pi(u)\ran=\lan w^{-1}(v), \pi(u)\ran=w^{-1}(v)\cdot u=v\cdot w(u)=\lan v,\pi\left(w(u)\right)\ran,
\]
that is

\begin{equation}\label{wpi}
w\left(\pi(u)\right)=\pi\left(w(u)\right).
\end{equation}
Thus $\pi$ is a $W^{(1)}$-homomorphism from $V^{(1)}$ to $V^{(1)\ast}$.
We now discuss some useful properties of the map $\pi$.

Let $\al+m\de=\be\in \Phi^{(1)}$, for $\al\in\Phi$ and $0\neq m\in\mathbb{Z}$, we have
\[
\lan v, \pi(\be)\ran=v\cdot\be=v\cdot(\al+m\de)=v\cdot\al=\lan v, \pi(\al)\ran,
\]
that is
\beq\label{pba}
\pi(\be)=\pi(\al).
\eeq
So the kernel of $\pi$ is Rad$(V^{(1)})$ and its image is $X_0$.

For each $\be\in\Phi^{(1)}$, define
\beq\label{ocal}
\oc\be=\frac{2\be}{\be\cdot\be}=\frac{2\be}{|\be|^2},
\eeq
then by Equation \eqref{alaij0},  we have
\beq\label{aipj}
a_{ij}=\frac{2\al_i\cdot\al_j}{\al_j\cdot\al_j}=\al_i\cdot\oc\al_j=\lan \al_i,\pi(\oc\al_j)\ran,\quad \mbox{for}\quad 0\leq i, j\leq n,
\eeq
where $a_{ij}=\left(C(\Ga^{(1)})\right)_{ij}$.
{\prop\label{pahp}
$\{\pi(\oc\al_j)\mid 1\leq j\leq n\}$  is a basis of $X_0$. Moreover, we have,
\beq\label{pah}
\pi(\oc\al_j)=\sum_{k=1}^n a_{kj}h_k=\sum_{k=1}^n\left(C(\Ga)^T\right)_{jk}h_k,\quad
\mbox{for}\quad1\leq j\leq n.
\eeq
Or we have,
\beq\label{hpa}
h_i=\sum_{k=1}^n\left(C(\Ga)^T\right)^{-1}_{ik}\pi(\oc\al_k),\quad
\mbox{for}\quad1\leq i\leq n.
\eeq
}

\begin{proof}
Recall that $\{h_j\mid1\leq j\leq n\}$ is a basis of $X_0$.
Since for all $1\leq i\leq n$ we have,
\begin{align*}
    \lan \al_i,\pi(\oc\al_j)\ran
    &= \lan \al_i,\sum_{k=1}^n \lan \al_k, \pi(\oc\al_j)\ran h_k\ran,\\
    &=\lan \al_i,\sum_{k=1}^n a_{kj}h_k,\ran\\
    &=\lan \al_i,\sum_{k=1}^n\left(C(\Ga)^T\right)_{jk}h_k,\ran,
\end{align*}
where $1\leq j\leq n$, and we have used Equation \eqref{ah2} and Definition \ref{ds}.
That is, 
\[
\pi(\oc\al_j)=\sum_{k=1}^n a_{kj}h_k=\sum_{k=1}^n\left(C(\Ga)^T\right)_{jk}h_k,\quad
\mbox{for}\quad 1\leq j\leq n.
\]
As $C(\Ga)$ is nondegenerate, the set 
$\{\pi(\oc\al_j)\mid1\leq j\leq n\}$ is also a basis of $X_0$.
Equation \eqref{hpa} follows from Equation \eqref{pah}.
\end{proof}
{\Def\label{rl}
The {\it root lattice} of $W^{(1)}$ is the integer span of simple
coroots of the dual root system in $V^{(1)*}$:
\begin{equation}\label{rle}
Q=\bigoplus{}_{i=1}^{n}\mathbb{Z} \pi(\oc\al_i).
\end{equation}
}

We now look at the actions of $W^{(1)}$ on the $\{\pi(\oc\al_j)\mid1\leq j\leq n\}$ basis of $X_0$.
{\prop\label{skpajp} For $0\leq k, j\leq n$, we have
$s_k\in W^{(1)}$ acts on $\{\pi(\oc\al_j)\mid1\leq j\leq n\}$ by,
\begin{equation}\label{skpaj}
s_k(\pi(\oc\al_j))=\pi(\oc\al_j)-a_{kj}\pi(\oc\al_k)=\pi(\oc\al_j)-\left(C(\Ga^{(1)})^T
\right)_{jk}\pi(\oc\al_k).
\end{equation}
}
\begin{proof}
For any $0\leq i, k, j\leq n$ we have,
\begin{align*}
\lan \al_i,s_k(\pi(\oc\al_j))\ran&=\lan s_k(\al_i),\pi(\oc\al_j)\ran,\\
&=\lan\al_i-a_{ik}\al_k,\pi(\oc\al_j)\ran,\\
&=\lan \al_i,\pi(\oc\al_j)\ran-a_{ik}\lan \al_k,\pi(\oc\al_j)\ran,\\
&=\lan \al_i,\pi(\oc\al_j)\ran-\lan \al_i,\pi(\oc\al_k)\ran\lan \al_k,\pi(\oc\al_j)\ran,\\
&=\lan\al_i,\pi(\oc\al_j)-\lan \al_k,\pi(\oc\al_j)\ran\pi(\oc\al_k)\ran,\\
&=\lan\al_i,\pi(\oc\al_j)-a_{kj}\pi(\oc\al_k)\ran,
\end{align*}
where we have used Equations  \eqref{cona}, \eqref{sij0}, and \eqref{sbfx0ij}.
\end{proof}

{\Rem\label{CT}In fact, under the contragrdient actions of $W$,
the set $\{\pi(\oc\al_j)\mid1\leq j\leq n\}$
generates a dual root system $\oc\Phi$ in $X_0\subset V^{(1)\ast}$, its elements
are called {\it coroots}. 
When $\Ga$
is of a non-simply-laced type, the Dynkin diagram for the dual root system 
is 
obtained from that of $\Ga$ by reversing the direction of the arrow on the multiple edge of $\Ga$, which corresponds to the appearance of  $C(\Ga^{(1)})^T$ in Proposition \ref{skpajp}.
For simply-laced $ADE$ type systems, we have $C(\Ga^{(1)})^T=C(\Ga^{(1)})$.
}
{\prop\label{bfact}
In general, the element $s_{\be}\in W^{(1)}$ (for any 
$\be\in\Phi^{(1)}$) acts on $V^{(1)*}$ by 
\begin{equation}\label{sbf}
s_{\be}(f)=f-\lan\be,f\ran\pi(\oc\be),\quad\mbox{for any}\quad
f\in V^{(1)*}.
\end{equation}
}
\begin{proof}
By
Equations \eqref{sbv}, \eqref{vpu} and the contragredient action of $W^{(1)}$ we have,
\begin{align}\nonumber
\lan v,s_{\be}(f)\ran&=\lan s_\be(v),f\ran,\\\nonumber
&=\lan v-v\cdot\oc\be \be,f\ran,\\
&=\lan v,f\ran-v\cdot\oc\be\lan \be,f\ran,\\\nonumber
&=\lan v,f\ran-\lan v,\pi(\oc\be)\ran\lan \be,f\ran,\\\nonumber
&=\lan v,f-\lan\be,f\ran\pi(\oc\be)\ran.
\end{align}
\end{proof}

\subsection{Coroots}\label{Cr}
Proposition \ref{bfact} says that in order to understand how $W^{(1)}$ acts on the dual space $V^{(1)*}$
it is worth working out what the coroot of $\be$,
$\pi(\oc\be)$ is in $X_0$, for any $\be\in\Phi^{(1)}$. 
We establish some properties of $\pi(\oc\be)$
via the following six propositions in Section \ref{Cr}.
{\prop\label{pcba}
For any $\be=\al+m\de\in\Phi^{(1)}$, $\al\in\Phi$
and $m\in\mathbb{Z}$,
\begin{equation}\label{pcbae}
\pi(\oc\be)=\pi(\oc\al).
\end{equation}
}
\begin{proof}
We know that the image of the
function $\pi$ is in $X_0$.
Moreover, for any $\be=\al+m\de\in\Phi^{(1)}$, $\al\in\Phi$
and $m\in\mathbb{Z}$, we have
\begin{equation}\label{mhd}
\lan\be,h_\de\ran=m,\quad\mbox{and}\quad\be\cdot\be=(\al+m\de).(\al+m\de)=\al\cdot\al,
\end{equation} 
that is
\begin{equation}\label{pcbap}
\pi(\oc\be)=\pi(\frac{2\be}{\be\cdot\be})=\frac{2\pi(\be)}{\be\cdot\be}
=\frac{2\pi(\al)}{\al\cdot\al}=\pi(\frac{2\al}{\al\cdot\al})=\pi(\oc\al),
\end{equation}
where have used Equation \eqref{pba}.
\end{proof}

For example, if we let $\be=\al_0=-\tilde\al+\de$, that is, $\al=-\tilde\al$ and $m=1$. Then we have the coroot of $\al_0$ is,
\begin{equation}\label{pa0ah}
\pi(\oc\al_0)=\pi(\oc\be)=\pi(\oc\al)=\pi(-\oc{\tilde\al})=-\pi(\oc{\tilde\al}).
\end{equation}

{\prop
Let $\al\in\Phi$, then $\pi(\oc \al)$ belongs to the dual root system $\oc\Phi$
generated by $\{\pi(\oc\al_j)\mid1\leq j\leq n\}$ under the contragrdient action of $W$. That is, for $\al\in \Phi$ where we have
\beq
w(\al_j)=\al\quad\mbox{for some}\quad w\in W, \;\al_j\in\De,
\eeq
then, 
\beq
w\left(\pi(\oc\al_j)\right)=\pi(\oc \al).
\eeq 
}
\begin{proof}
Let $\al\in \Phi$, then we have
$w(\al_j)=\al$ for some $w\in W, \;\al_j\in\De$, moreover
we have $|\al_j|=|\al|$. Then
\[
w\left(\pi(\oc\al_j)\right)=w\left(\pi(\frac{2\al_j}{|\al_j|^2})\right)
=\frac{2w\left(\pi(\al_j)\right)}{|\al_j|^2}=\frac{2\pi\left(w(\al_j)\right)}{|\al|^2}
=\frac{2\pi(\al)}{|\al|^2}=\pi(\oc \al).
\]
\end{proof}
This means that $\pi(\oc \al)$ can be written as some integer combinations
of $\pi(\oc\al_j)\,(1\leq j\leq n)$, just as $\al\in\Phi$
can be written as some integer combinations
of $\al_j\,(1\leq j\leq n)$.
The integer coefficients of $\al_j$ in $\al\in\Phi$ and those of $\pi(\oc\al_j)$ in $\pi(\oc \al)\in \oc\Phi$
are related by the following proposition.
{\prop\label{miki} If $\al\in \Phi$, that is $\al=\sum_{i=1}^n m_i\al_i$, for some $m_i\in \mathbb{Z}$ and $\al_i\in\De$,
then the coroot of $\al$, $\pi(\oc \al)$ is an element in $\oc\Phi$, 
\beq\label{mikieq}
\pi(\oc\al)=\sum_{i=1}^n k_i\pi(\oc\al_i),\quad\mbox{where}\quad k_i=m_i\frac{|\al_i|^2}{|\al|^2}\quad
\mbox{or}\quad m_i=\frac{|\al|^2}{|\al_i|^2}k_i, \quad\mbox{for}\quad 1\leq i\leq n.
\eeq
}
\begin{proof}We have

\begin{align*}
\pi(\oc\al)&=\pi\left(\frac{2}{|\al|^2}\sum_{i=1}^n m_i\al_i\right)\\
&=\sum_{i=1}^n m_i\pi\left(\frac{2\al_i|\al_i|^2}{|\al|^2|\al_i|^2}\right)\\
&=\sum_{i=1}^n m_i\frac{|\al_i|^2}{|\al|^2}\pi(\oc\al_i)\\
&=\sum_{i=1}^n k_i\pi(\oc\al_i).
\end{align*}
\end{proof}
{\Rem
Proposition \ref{miki} says
that for simply-laced type root
systems, where $|\al_i|^2=|\al|^2=2$ (for all $\al_i,\al\in\Phi$), we have $m_i=k_i$
($1\leq i\leq n$). For these types we can identify the simple coroots with the simple roots $\pi(\oc\al_j)=\al_j\,(1\leq j\leq n)$.
For non-simply-laced-types,
$m_i$ and $k_i$ are not the same for some values of $i$,  $\pi(\oc\al_i)$ is identified with $\frac{2\al_i}{|\al_i|^2}$. The 
values of the $k_i$'s in $\pi(\oc{\tilde{\al}})$ for non-simply-laced-types
are listed in Table \ref{basic} of Appendix \ref{list}.}

Now we show that $\pi(\oc\be)\in X_0$ for any $\be\in \Phi^{(1)}$
is an element of the weight lattice $P$.
{\prop\label{muihp}
For any $\be\in \Phi^{(1)}$, let 
$b_k=\lan\al_k,\pi(\oc\be)\ran$ for $ 0\leq k\leq n$.
We have
\begin{equation}\label{muih}
\pi(\oc\be)=\sum_{j=1}^{n}b_jh_j=\sum_{j=1}^{n}\lan\al_j,\pi(\oc\be)\ran h_j.
\end{equation}
Moreover,
\begin{equation}\label{muicon}
\sum_{i=0}^{n}c_ib_i=0,\quad {or}\quad b_0=-\sum_{i=1}^{n}c_ib_i,
\end{equation}
where $c_i$ is the coefficient of $\al_i$ in $\de$. In particular, the coefficients
$b_i$ (for $0\leq i\leq n$) are all integral. That is, $\pi(\oc\be)\in X_0$
is an element of the weight lattice $P=\bigoplus{}_{i=1}^{n}\mathbb{Z} h_i$.
}
\begin{proof} Recall that $\{h_j\mid1\leq j\leq n\}$ 
is a basis of $X_0$, and $\pi(\oc\be)\in X_0$ for any $\be\in \Phi^{(1)}$.
For any $0\leq i\leq n$ we have,
\begin{align*}
&\lan \al_i,\pi(\oc\be)\ran,\\
&=\lan \al_i,\sum_{i=1}^{n}\lan \al_j,\pi(\oc\be)\ran h_j\ran,\\
&=\lan \al_i,\sum_{i=1}^{n}b_j h_j\ran,
\end{align*}
that is, 
\[
\pi(\oc\be)=\sum_{j=1}^{n}b_jh_j=\sum_{j=1}^{n}\lan\al_j,\pi(\oc\be)\ran h_j.
\]

To show that Equation \eqref{muicon} is true, observe that
\begin{equation}\label{b0bi}
0=\lan\de,\pi(\oc\be)\ran=\lan\sum_{i=0}^{n}c_i\al_i,\pi(\oc\be)\ran
=\sum_{i=0}^{n}c_ib_i,\quad \mbox{or}\quad b_0=-\sum_{i=1}^{n}c_ib_i,
\end{equation}
since $c_0=1$ by Equation \eqref{d}.

To show that $b_i$ for $0\leq i\leq n$ are all integers, first
for $1\leq i\leq n$ we have,
\[
b_i=\lan\al_i,\pi(\oc\be)\ran=\lan\al_i,\pi(\oc\al)\ran=
\al_i\cdot\oc\al=\frac{2\al_i\cdot\al}{\al\cdot\al},
\] which are integers for all $\al_i, \al\in \De$ by Remark \ref{abint}.
By Equation \eqref{b0bi} and the fact that $b_i$ and $c_i$ ($1\leq i\leq n$) are all integers we have $b_0$ is also an integer.
\end{proof}
{\prop\label{pa0h}
For each $0\leq j\leq n$, $\pi(\oc\al_j)$ can be expressed as a linear
combination of $\{h_k\mid 1\leq k\leq n\}$ with coefficients from the $j$-th
column of $C(\Ga^{(1)})$,
\beq
\pi(\oc\al_j)
=\sum_{k=1}^{n}\left(C(\Ga^{(1)})\right)_{kj}h_k
=\sum_{k=1}^{n}a_{kj}h_k.
\eeq

Moreover, for any $0\leq j\leq n$ we have
\beq\label{aijcon}
\sum_{i=0}^{n}c_ia_{ij}=0,
\eeq
where $c_i$ is the coefficient of $\al_i$ in $\de$, for $0\leq i\leq n$.
}
\begin{proof}
In Proposition \ref{muihp}, 
let $\be=\al_j$, for $0\leq j\leq n$  we have,
\beq
b_k=\lan\al_k,\pi(\oc\al_j)\ran=a_{kj},\quad 0\leq k\leq n,
\eeq
and
\beq
\pi(\oc\al_j)=\sum_{k=1}^{n}b_kh_k=\sum_{k=1}^{n}\lan\al_k,\pi(\oc\al_j)\ran h_k
=\sum_{k=1}^{n}a_{kj}h_k
=\sum_{k=1}^{n}\left(C(\Ga^{(1)})\right)_{kj}h_k.
\eeq

For $1\leq j\leq n$ we recover Equation \eqref{pah}. When $j=0$, we have,
\begin{equation}\label{prhh}
\pi(\oc{\al_0})=\sum_{i=1}^{n}\lan\al_i,\pi(\oc\al_0)\ran h_i=\sum_{i=1}^{n}a_{i0} h_i
\end{equation}
where $a_{i0}$ ($1\leq i\leq n$) are the first $n$ entries of the last column of $C(\Ga^{(1)})$.

Equation \eqref{aijcon} is just a special case of Equation \eqref{muicon}
on letting $\be=\al_j$ ($0\leq j\leq n$).
\end{proof}

We give $\pi(\oc{\tilde{\al}})$ and $\pi(\oc{\tilde{\al}_s})$ for all Weyl group types in terms of the fundamental weights
in Table \ref{basic} of Appendix \ref{list}.

Now, introducing a bilinear form on $X_0\subset V^{(1)*}$, we discuss the lengths of coroots and fundamental weights in $X_0$.
 
{\Def\label{sbfX0}
Let $({}\,,{})$: $X_0\times X_0\to \mathbb{R}$ 
be a symmetric positive definite bilinear form on $X_0\subset V^{(1)*}$
such that,
\beq\label{paipj}
\left(\pi(\oc\al_i),\pi(\oc\al_j)\right)
=\frac{2}{|\al_i|^2}a_{ij}, \quad 1\leq i, j\leq n.
\eeq
}
Using $|\pi(\be)|$ to denote the length of the vector $\pi(\be)\in X_0$ for any 
$\be\in V^{(1)}$, from Equation \eqref{paipj} we have,
\begin{equation}\label{lpor}
|\pi(\oc\al_i)|^2=\left(\pi(\oc\al_i),\pi(\oc\al_i)\right)=\frac{2}{|\al_i|^2}a_{ii}=\frac{4}{|\al_i|^2},\quad  1\leq i\leq n.
\end{equation}
Hence for any $\al\in\Phi$ we have,
\beq\label{picl}
|\pi(\oc\al)|^2=
 \begin{cases}
 4, &\text{for}\quad |\al|^2=1,\\
 2, &\text{for}\quad |\al|^2=2,\\
  1, &\text{for}\quad |\al|^2=4,\\
   \frac{2}{3}, &\text{for}\quad |\al|^2=6.\\
 \end{cases}
 \eeq

{\prop\label{sbfx0ij} The bilinear form $({}\,,{})$
given in Definition \ref{sbfX0} on $X_0$ is related to a restriction of
the bilinear pairing between $V^{(1)}$ and $V^{(1)*}$ 
$\langle {}\,, {} \rangle$  given in Definition \ref{ds} by
\beq\label{paipjdef}
\left(\pi(\al_i),\pi(\oc\al_j)\right)=a_{ij}=\lan \al_i,\pi(\oc\al_j)\ran,\quad
1\leq i, j\leq n.
\eeq
}
\begin{proof} For $1\leq i, j\leq n$ we have
\begin{align*}
\left(\pi(\al_i),\pi(\oc\al_j)\right)&=    \left(\frac{|\al_i|^2\pi(\oc\al_i)}{2},\pi(\oc\al_j)\right),\\
&=\frac{|\al_i|^2}{2}\left(\pi(\oc\al_i),\pi(\oc\al_j)\right),\\
&=a_{ij},\\
&=\al_i\cdot\oc\al_j,\\
&=\lan \al_i,\pi(\oc\al_j)\ran,
\end{align*}
where we have used Definition \ref{sbfX0} and Equation \eqref{aipj}.
\end{proof}

{\prop\label{blh}
The bilinear form $({}\,,{})$ in the $\{h_j\mid1\leq j\leq n\}$ basis of $X_0$ is
\beq\label{hij}
(h_i, h_j)=\sum_{k=1}^n\left(C(\Ga)^T\right)^{-1}_{ik}\frac{2}{|\al_k|^2}\de_{kj},\quad
1\leq i, j\leq n.
\eeq
}

\begin{proof}
From Equation \eqref{hpa} we have
\begin{align*}
(h_i, h_j)&=\left(\sum_{k=1}^n\left(C(\Ga)^T\right)^{-1}_{ik}\pi(\oc\al_k),h_j\right),\\
&=\sum_{k=1}^n\left(\left(C(\Ga)^T\right)^{-1}_{ik}\frac{2}{|\al_k|^2}\pi(\al_k),h_j\right),\\
&= \sum_{k=1}^n\left(C(\Ga)^T\right)^{-1}_{ik}\frac{2}{|\al_k|^2}\lan\al_k,h_j\ran,\\
&= \sum_{k=1}^n\left(C(\Ga)^T\right)^{-1}_{ik}\frac{2}{|\al_k|^2}\de_{kj},\\
\end{align*}
where we have used Equation \eqref{ocal}, Proposition \ref{sbfx0ij} and Equation \eqref{ah2}.
\end{proof}
{\Rem\label{hl}
In Proposition \ref{blh}, observe that  for simply-laced type ($ADE$)
root systems, where $|\al_k|^2=2$ $(1\leq k\leq n)$ and $C(\Ga)$ is symmetric,
$|h_j|^2$ $(1\leq j\leq n)$
are just the diagonal entries of $C(\Ga)^{-1}$. We list 
$|h_j|^2$ $(1\leq j\leq n)$ for all Weyl groups in Table \ref{basic} of Appendix \ref{list}.
}

We now illustrate the properties of $W^{(1)}$ discussed so far
for $B_3$, $C_3$, $F_4$ and $G_2$ type systems.
{\eg \label{pB3}Affine Weyl group of type $B_3$, $W(B_3^{(1)})$.
The Dynkin diagram of type $B_3^{(1)}$, $\Ga(B_3^{(1)})$ is given 
in Figure \ref{rsB3a}.
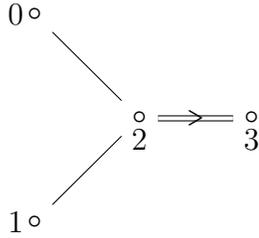
\begin{figure}[h!]
\centering
\begin{tikzpicture}[scale=1]
\node  (a1) {};
\node [right=of a1](a2){$\circ$};
\node [right=of a2](a5){$\circ$};

\node [above=of a1](a10) {$\circ$};
\node [below=of a1](a11) {$\circ$};
\node [left=of a1](an){};
\draw (a5) node [anchor=north] {$3$};
\draw (a2) node [anchor=north] {$2$} ;
\draw (a10) node [anchor=east] {$0$} ;
\draw (a11) node [anchor=east] {$1$} ;
\draw[double distance=1.5pt] (a2) -- node {} (a5);
\draw[-] (a2) --  (a10);
\draw[-] (a2) -- (a11);
\node (b) at ($(a2)!0.5!(a5)$) {$>$};
\end{tikzpicture}
%
\caption{Dynkin diagram of affine $B_3$ type, $\Ga(B_3^{(1)})$.}\label{rsB3a}
\end{figure}
The corresponding generalized Cartan matrix of type $B_3^{(1)}$,
\beq\label{CarB3a}
C(B_3^{(1)})=(a_{ij})_{1\leq i,j\leq 3,0}=(\al_i\cdot\oc\al_j)_{1\leq i,j\leq 3,0}=\bp
2&-1&0&0\\
-1&2&-2&-1\\
0&-1&2&0\\
0&-1&0&2
\ep,
\eeq
and the defining relations for the Weyl group $W(B_3^{(1)})=\lan s_i\mid 0\leq i \leq 3\ran$,
\begin{align}\nonumber
&s_1^2=s_2^2=s_3^2=1,\;\;(s_1s_{2})^3=1, \quad (s_{1}s_3)^2=1,\quad (s_{2}s_3)^4=1,\\\label{funWB3a}
&s_0^2=1,\;\;(s_0s_{2})^3=1, \quad (s_{0}s_3)^2=(s_{0}s_1)^2=1,
\end{align}
can be written down using $\Ga(B_3^{(1)})$ with the rules
given in Table \ref{CD}.
The $B_3^{(1)}$ simple system $\De^{(1)}=\{\al_1,\al_2,\al_3,\al_0\}$ forms a basis for an $4$-dimensional real vector space $V^{(1)}$.
Generators $s_j$ act on $V^{(1)}$ by Equation \eqref{sij0}, with $a_{ij}$ being the $(i,j)$-entry
of $C(B_3^{(1)})$ given in Equation \eqref{CarB3a}.
The dual space $ V^{(1)*}$ and its hyperplanes $X_k$ are given by Definitions \ref{ds}
and \ref{dsX}, respectively.
The set of simple coroots of $B_3$, $\{\pi(\oc\al_1), \pi(\oc\al_2),\pi(\oc\al_3)\}$
form a dual system of $C_3$ type (see Figure \ref{rsB3d}) and
is a basis of $X_0\subset V^{(1)*}$.

\begin{figure}[ht]
\centering
\begin{tikzpicture}
		
			\node  (a1) {$\circ$};
			\node [right=of a1](a2) {$\circ$} ;
			\node [right=of a2](a3) {$\circ$} ;
			\draw (a1) node [anchor=north] {$\pi(\oc\al_1)$} ;
			\draw (a2) node [anchor=north] {$\pi(\oc\al_2)$} ;
			\draw (a3) node [anchor=north] {$\pi(\oc\al_3)$} ;
			\draw[-] (a1) -- node {} (a2);
			\draw[double distance=1.5pt] (a2) -- node {} (a3);
			\node (b) at ($(a2)!0.5!(a3)$) {$<$};
			\path[use as bounding box] (-1.5,0) rectangle (0,0); 
\end{tikzpicture}
%
\caption{Dynkin diagram for the dual system.}\label{rsB3d}
\end{figure}
The group $W(B_3^{(1)})$ acts on $\{\pi(\oc\al_j) \mid 1\leq j\leq 3\}$ by Proposition \ref{skpajp}.
By Equation \eqref{pah}, $\{\pi(\oc\al_j) \mid 1\leq j\leq 3\}$ can be expressed in terms of the fundamental weights $\{h_j \mid 1\leq j\leq 3\}$ by
\beq\label{pahB3}
\bp
\pi(\oc\al_1)\\
\pi(\oc\al_2)\\
\pi(\oc\al_3)
\ep=C(B_3)^T\bp
h_1\\
h_2\\
h_3
\ep=\bp
2&-1&0\\
-1&2&-1\\
0&-2&2
\ep\bp
h_1\\
h_2\\
h_3
\ep,
\eeq
where $C(B_3)$ is given by Equation \eqref{CarB3},
or we have
\beq\label{hpaB3}
\bp
h_1\\
h_2\\
h_3
\ep=
\left(C(B_3)^T\right)^{-1}
\bp
\pi(\oc\al_1)\\
\pi(\oc\al_2)\\
\pi(\oc\al_3)
\ep=\bp
1&1&\frac{1}{2}\\
1&2&1\\
1&2&\frac{3}{2}
\ep\bp
\pi(\oc\al_1)\\
\pi(\oc\al_2)\\
\pi(\oc\al_3)
\ep.
\eeq

The matrix of symmetric bilinear form $(\;,\;)$ on subspace $X_0$ in $\{\pi(\oc\al_1), \pi(\oc\al_2),\pi(\oc\al_3)\}$ basis  
is given by Equation \eqref{paipj}: 
\beq\label{SymB3d}
\left(\left(\pi(\oc\al_i),\pi(\oc\al_j)\right)\right)_{1\leq i,j\leq3}
=\left(\frac{2}{|\al_i|^2}a_{ij}\right)_{1\leq i,j\leq3}
=\bp
1&0&0\\
0&1&0\\
0&0&2
\ep\bp
2&-1&0\\
-1&2&-2\\
0&-1&2
\ep
=\bp
2&-1&0\\
-1&2&-2\\
0&-2&4
\ep.
\eeq
The diagonal entries of the last matrix in Equation \eqref{SymB3d} tell us that 
$|\pi(\oc\al_i)|^2$ for $1\leq j\leq 3$ are
2, 2, and 4, respectively. 
The bilinear form in $\{h_j \mid 1\leq j\leq 3\}$ basis of $X_0$ is given by Equation
\eqref{hij}:
\begin{align}\label{hijB3}
\left((h_i, h_j)\right)_{1\leq i,j\leq3}
&=\left(C(B_3)^T\right)^{-1}\left(\frac{2}{|\al_k|^2}\de_{kj}\right)_{1\leq k,j\leq3},\\\nonumber
&=\left(C(B_3)^T\right)^{-1}
\bp
1&0&0\\
0&1&0\\
0&0&2
\ep
=\bp
1&1&\frac{1}{2}\\
1&2&1\\
1&2&\frac{3}{2}
\ep\bp
1&0&0\\
0&1&0\\
0&0&2
\ep
=\bp
1&1&1\\
1&2&2\\
1&2&3
\ep.    
\end{align}
The diagonal entries of the last matrix in Equation \eqref{hijB3}
tell us that $|h_j|^2$ for $1\leq j\leq 3$ are 1, 2 and 3, respectively. That is, $h_1$ 
with $|h_1|^2=1$ is the shortest of the fundamental weights of the $B_3$ system. 

Now let us consider the coroots of some non-simple roots of the finite $B_3$ root system (given in Equation \eqref{rsB3}) using Proposition \ref{miki}.
In particular, consider the highest short and long roots of the $B_3$ system:
$\tilde{\al}_s=\al_{123}$ and $\tilde{\al}=\al_{12233}$.\\
For
$\tilde{\al}_s=\al_{123}$, we have $m_i=1$ for $1\leq i\leq 3$.  Moreover, we have 
$|\al_{1}|^2=|\al_{2}|^2=2$, and $|\al_{3}|^2=|\tilde{\al}_s|^2=1$ from Equations \eqref{SymB3} and \eqref{rsB3}. Then by Proposition \ref{miki} we have,
\begin{equation}\label{psB3}
\pi(\oc{\tilde{\al}_s})=\pi(\oc\al_{123})=2\pi(\oc\al_1)+2\pi(\oc\al_2)+\pi(\oc\al_3)=2h_1,
\end{equation} 
where $|\pi(\oc{\tilde{\al}_s})|^2=4|h_1|^2=4$. That is $\pi(\oc{\tilde{\al}_s})$
is a long root in the dual $C_3$ root system in $X_0$.

For $\tilde{\al}=\al_{12233}$, we have $m_1=1$, $m_2=m_3=2$, and $|\al_{12233}|^2=2$. By Proposition \ref{miki}
we have
\beq\label{plB3}
\pi(\oc{\tilde{\al}})=\pi(\oc\al_{12233})
=\frac{|\al_{1}|^2}{2}\pi(\oc\al_1)+2\frac{|\al_{2}|^2}{2}\pi(\oc\al_2)+
2\frac{|\al_{3}|^2}{2}\pi(\oc\al_3)
=\pi(\oc\al_1)+2\pi(\oc\al_2)+\pi(\oc\al_3)=h_2,
\eeq
where $|\pi(\oc{\tilde{\al}})|^2=|h_2|^2=2$. That is, $\pi(\oc{\tilde{\al}})$
is a short root in the dual $C_3$ system in $X_0$.
To express $\pi(\oc{\tilde{\al}_s})$ and $\pi(\oc{\tilde{\al}})$ in terms of the fundamental weights we have used Equation \eqref{hpaB3}.
The last expression of Equation \eqref{plB3}
can also be obtained using Proposition \ref{pa0h} and Equation \eqref{pa0ah},
\begin{equation}\label{p0h}
\pi(\oc{\al_0})=\sum_{k=1}^3 \left(C(B_3^{(1)})\right)_{k0}h_k=
0.h_1+(-1).h_2+0.h_3=-h_2,
\end{equation}
with $C(B_3^{(1)})$ given by Equation \eqref{CarB3a}.
}
{\Rem\label{pBC3}
Comparing the expressions for $\pi(\oc{\tilde{\al}_s})$ and $\pi(\oc{\tilde{\al}})$
given in Equations \eqref{psB3} and \eqref{plB3} with roots of the finite $C_3$
system given in Equation \eqref{rsC3},
we see that the map $\pi$ takes $\oc{\tilde{\al}_s}$ and $\oc{\tilde{\al}}$ of $B_3$ type to the highest long and short root 
of the $C_3$ type
dual system generated by $\{\pi(\oc\al_i)\mid1\leq j\leq 3\}$, respectively.}

{\eg \label{pC3}Affine Weyl group of type $C_3$, $W(C_3^{(1)})$.
The Dynkin diagram of type $C_3^{(1)}$, $\Ga(C_3^{(1)})$ is given in Figure \ref{rsC3a}.

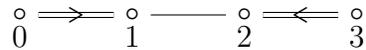
\begin{figure}[ht]
\centering
\begin{tikzpicture}[scale=1]
			\centering
			\node  (a1) {$\circ$};
			\node  [left=of a1](a0) {$\circ$};
			\node [right=of a1](a2) {$\circ$} ;
			\node [right=of a2](a4) {$\circ$} ;
			\draw (a1) node [anchor=north] {$1$} ;
			\draw (a2) node [anchor=north] {$2$} ;
			\draw (a4) node [anchor=north] {$3$} ;
			\draw (a0) node [anchor=north] {$0$} ;
			\draw[-] (a1) -- node {} (a2);
			\draw[double distance=1.5pt] (a2) -- node {} (a4);
			\node (b) at ($(a4)!0.5!(a2)$) {$<$};
			\draw[double distance=1.5pt] (a1) -- node {} (a0);
			\node (b) at ($(a1)!0.5!(a0)$) {$>$};
			\path[use as bounding box] (-1.5,0) rectangle (0,0); 
			\end{tikzpicture}
			\caption{Dynkin diagram of affine $C_3$ type, $\Ga(C_3^{(1)})$.}\label{rsC3a}
\end{figure}
The corresponding generalized Cartan matrix of type $C_3^{(1)}$, 
\beq\label{CarC3a}
C(C_3^{(1)})=(a_{ij})_{1\leq i,j\leq 3,0}=(\al_i\cdot\oc\al_j)_{1\leq i,j\leq 3,0}=\bp
2&-1&0&-1\\
-1&2&-1&0\\
0&-2&2&0\\
-2&0&0&2
\ep,
\eeq
and the defining relations for the Weyl group $W(C_3^{(1)})=\lan s_i\mid 0\leq i \leq 3\ran$,
\begin{align}\label{funWC3}
&s_1^2=s_2^2=s_3^2=1,\;\;(s_1s_{2})^3=1, \quad (s_{1}s_3)^2=1,\quad (s_{2}s_3)^4=1,\\\label{funWC3a}
&s_0^2=1,\;\;(s_0s_{1})^4=1, \quad (s_{0}s_2)^2=(s_{0}s_3)^2=1,
\end{align}
can be written down using $\Ga(C_3^{(1)})$ with the rules
given in Table \ref{CD}.
The $C_3^{(1)}$ simple system $\De^{(1)}=\{\al_1,\al_2,\al_3,\al_0\}$ forms a basis for an $4$-dimensional real vector space $V^{(1)}$.
Generators $s_j$ act on $V^{(1)}$ by Equation \eqref{sij0}, where $a_{ij}$ is the $(i,j)$-entry
of $C(C_3^{(1)})$ given in Equation \eqref{CarC3a}.
The dual space $ V^{(1)*}$ and its hyperplanes $X_k$ are given by Definitions \ref{ds}
and \ref{dsX}, respectively.
The set of simple coroots of $C_3$, $\{\pi(\oc\al_1), \pi(\oc\al_2),\pi(\oc\al_3)\}$
form a dual system of $B_3$ type in $X_0\subset V^{(1)*}$ (see Figure \ref{rsC3d}).
\begin{figure}[ht]
\centering
\begin{tikzpicture}
		
			\node  (a1) {$\circ$};
			\node [right=of a1](a2) {$\circ$} ;
			\node [right=of a2](a3) {$\circ$} ;
			\node [left=of a0](an){};
			\draw (a1) node [anchor=north] {$\pi(\oc\al_1)$} ;
			\draw (a2) node [anchor=north] {$\pi(\oc\al_2)$} ;
			\draw (a3) node [anchor=north] {$\pi(\oc\al_3)$} ;
			\draw[-] (a1) -- node {} (a2);
			\draw[double distance=1.5pt] (a2) -- node {} (a3);
			\node (b) at ($(a2)!0.5!(a3)$) {$>$};
			\path[use as bounding box] (-1.5,0) rectangle (0,0); 
\end{tikzpicture}
%
\caption{Dynkin diagram for the dual system.}\label{rsC3d}
\end{figure}
The group $W(C_3^{(1)})$ acts on $\{\pi(\oc\al_j) \mid 1\leq j\leq 3\}$ by Proposition \ref{skpajp}.
By Equation \eqref{pah}, $\{\pi(\oc\al_j) \mid 1\leq j\leq 3\}$ can be expressed in terms of 
the fundamental weights by
\beq\label{pahC3}
\bp
\pi(\oc\al_1)\\
\pi(\oc\al_2)\\
\pi(\oc\al_3)
\ep=C(C_3)^T\bp
h_1\\
h_2\\
h_3
\ep=\bp
2&-1&0\\
-1&2&-2\\
0&-1&2
\ep\bp
h_1\\
h_2\\
h_3
\ep,
\eeq
where $C(C_3)$ is given by Equation \eqref{CarC3}. Moreover, we have
\beq\label{hpaC3}
\bp
h_1\\
h_2\\
h_3
\ep=
\left(C(C_3)^T\right)^{-1}
\bp
\pi(\oc\al_1)\\
\pi(\oc\al_2)\\
\pi(\oc\al_3)
\ep=
\bp
1&1&1\\
1&2&2\\
\frac{1}{2}&1&\frac{3}{2}
\ep
\bp
\pi(\oc\al_1)\\
\pi(\oc\al_2)\\
\pi(\oc\al_3)
\ep.
\eeq

The matrix of symmetric bilinear form on the subspace $X_0\subset V^{(1)*}$ in $\{\pi(\oc\al_1), \pi(\oc\al_2),\pi(\oc\al_3)\}$ basis  
is given by Equation \eqref{paipj}: 
\beq\label{SymC3d}
\left(\left(\pi(\oc\al_i),\pi(\oc\al_j)\right)\right)_{1\leq i,j\leq3}
=\left(\frac{2}{|\al_i|^2}a_{ij}\right)_{1\leq i,j\leq3}
=\bp
1&0&0\\
0&1&0\\
0&0&\frac{1}{2}
\ep\bp
2&-1&0\\
-1&2&-1\\
0&-2&2
\ep
=\bp
2&-1&0\\
-1&2&-1\\
0&-1&1
\ep.
\eeq
The diagonal entries of the last matrix in Equation \eqref{SymC3d} tell us that 
$|\pi(\oc\al_i)|^2$ for $1\leq j\leq 3$ are
2, 2, and 1, respectively. 
The bilinear form in the $\{h_j\mid 1\leq j\leq 3\}$ basis of $X_0$ is given by Equation
\eqref{hij}:
\begin{align}\label{hijC3}
\left((h_i, h_j)\right)_{1\leq i,j\leq3}
&=\left(C(C_3)^T\right)^{-1}\left(\frac{2}{|\al_k|^2}\de_{kj}\right)_{1\leq k,j\leq3},\\
\nonumber
&=\left(C(C_3)^T\right)^{-1}
\bp
1&0&0\\
0&1&0\\
0&0&\frac{1}{2}
\ep
=\bp
1&1&1\\
1&2&2\\
\frac{1}{2}&1&\frac{3}{2}
\ep\bp
1&0&0\\
0&1&0\\
0&0&\frac{1}{2}
\ep
=\bp
1&1&\frac{1}{2}\\
1&2&1\\
\frac{1}{2}&1&\frac{3}{4}
\ep.
\end{align}
The diagonal entries of the matrix on the right in Equation \eqref{hijC3}
tell us that $|h_j|^2$ for $1\leq j\leq 3$ are $1$, $2$ and $\frac{3}{4}$, respectively. Hence $h_3$ with $|h_3|^2=\frac{3}{4}$ is the shortest of the fundamental weights of 
a $C_3$ system. 

Using Proposition \ref{miki}, we now compute the coroots of
the highest short and long roots of the finite $C_3$ system,
$\tilde{\al}_s=\al_{1223}$ and $\tilde{\al}=\al_{11223}$, respectively.
For
$\tilde{\al}_s=\al_{1223}$, we have $m_1=1$, $m_2=2$ and $m_3=1$. From Equations \eqref{laC3},
and \eqref{rsC3} we know that
$|\al_{1}|^2=|\al_{2}|^2=2$, $|\al_{3}|^2=4$ and $|\al_{1223}|^2=2$. Then by Proposition \ref{miki} we have,
\begin{equation}\label{hrC3}
\pi(\oc{\tilde{\al}_s})=\pi(\oc\al_{1223})=\pi(\oc\al_1)+2\pi(\oc\al_2)+2\pi(\oc\al_3)=h_2,
\end{equation} 
where $|\pi(\oc{\tilde{\al}_s})|^2=|h_2|^2=2$.  That is $\pi(\oc{\tilde{\al}_s})$
is a long root in the dual $C_3$ system in $X_0$.
For $\tilde{\al}=\al_{11223}$, we have $m_1=m_2=2$, $m_3=1$ and $|\al_{11223}|^2=4$. Then 
by Proposition \ref{miki} we have
\beq\label{hsrC3}
\pi(\oc{\tilde{\al}})=\pi(\oc\al_{11223})
=2\frac{|\al_{1}|^2}{4}\pi(\oc\al_1)+2\frac{|\al_{2}|^2}{4}\pi(\oc\al_2)+
\frac{|\al_{3}|^2}{4}\pi(\oc\al_3)
=\pi(\oc\al_1)+\pi(\oc\al_2)+\pi(\oc\al_3)=h_1,
\eeq
where $|\pi(\oc{\tilde{\al}})|^2=|h_1|^2=1$. That is $\pi(\oc{\tilde{\al}})$
is a short root in the dual $B_3$ system in $X_0$.
To express $\pi(\oc{\tilde{\al}_s})$ and $\pi(\oc{\tilde{\al}})$ in terms of the fundamental weight we have used Equation \eqref{hpaC3}.
The last expression on the right of Equation \eqref{hsrC3}
can also be obtained using Proposition \ref{pa0h} and Equation \eqref{pa0ah},
\begin{equation}\label{p0Ch}
\pi(\oc{\al_0})=\sum_{k=1}^3 \left(C(C_3^{(1)})\right)_{k0}h_k=(-1).h_1+0.h_2+0.h_3=-h_1,
\end{equation}
with $C(C_3^{(1)})$ given by Equation \eqref{CarC3a}.
}
{\eg \label{pF4}Affine Weyl group of type $F_4$, $W(F_4^{(1)})$.
The Dynkin diagram of type $F_4^{(1)}$, $\Ga(F_4^{(1)})$ is given in Figure \ref{claW}.
The corresponding generalized Cartan matrix of type $F_4^{(1)}$,
\beq\label{CarF4a}
C(F_4^{(1)})=(a_{ij})_{1\leq i,j\leq 4,0}=(\al_i\cdot\oc\al_j)_{1\leq i,j\leq 4,0}=\bp
2&-1&0&0&-1\\
-1&2&-2&0&0\\
0&-1&2&-1&0\\
0&0&-1&2&0\\
-1&0&0&0&2
\ep,
\eeq
and the defining relations for the Weyl group $W(F_4^{(1)})=\lan s_i\mid 0\leq i \leq 4\ran$,
\begin{align}\nonumber
&s_1^2=s_2^2=s_3^2=s_4^2=(s_1s_3)^2=(s_1s_4)^2=(s_2s_4)^2=1,\quad (s_1s_2)^3=(s_2s_3)^4=(s_3s_4)^3=1\\\label{funWF4a}
&s_0^2=1,\;\;(s_0s_{1})^3=1, \quad (s_{0}s_2)^2=(s_{0}s_3)^2=(s_{0}s_4)^2=1,
\end{align}
can be written down using $\Ga(F_4^{(1)})$ with the rules
given in Table \ref{CD}.
The $F_4^{(1)}$ simple system $\De^{(1)}=\{\al_1,\al_2,\al_3,\al_4,\al_0\}$ forms a basis for an $5$-dimensional real vector space
$V^{(1)}$.
Generators $s_j$ act on $V^{(1)}$ by Equation \eqref{sij0}, where $a_{ij}$ is the $(i,j)$-entry
of $C(F_4^{(1)})$ given in Equation \eqref{CarF4a}.
The dual space $ V^{(1)*}$ and its hyperplanes $X_k$ are given by Definitions \ref{ds}
and \ref{dsX}, respectively.
The set of simple coroots of $F_4$, $\{\pi(\oc\al_1), \pi(\oc\al_2),\pi(\oc\al_3),\pi(\oc\al_4)\}$
form a dual system again of $F_4$ type (see Figure \ref{rsF4d})
and is a basis of $X_0\subset V^{(1)*}$.
\begin{figure}[ht]
\centering
\begin{tikzpicture}
		
			\node  (a1) {$\circ$};
			\node [right=of a1](a2) {$\circ$} ;
			\node [right=of a2](a3) {$\circ$} ;
			\node [right=of a3](a4) {$\circ$} ;
			\node [left=of a0](an){};
			\draw (a1) node [anchor=north] {$\pi(\oc\al_1)$} ;
			\draw (a2) node [anchor=north] {$\pi(\oc\al_2)$} ;
			\draw (a3) node [anchor=north] {$\pi(\oc\al_3)$} ;
			\draw (a4) node [anchor=north] {$\pi(\oc\al_4)$} ;
			\draw[-] (a1) -- node {} (a2);
			\draw[-] (a3) --node {} (a4);
			\draw[double distance=1.5pt] (a2) -- node {} (a3);
			\node (b) at ($(a2)!0.5!(a3)$) {$<$};
			\path[use as bounding box] (-1.5,0) rectangle (0,0); 
\end{tikzpicture}
%
\caption{Dynkin diagram for the dual system.}\label{rsF4d}
\end{figure}
The group $W(F_4^{(1)})$ acting on $\{\pi(\oc\al_j) \mid 1\leq j\leq 4\}$ given by Proposition \ref{skpajp}.
By Equation \eqref{pah}, $\{\pi(\oc\al_j) \mid 1\leq j\leq 4\}$ can be expressed in terms of the fundamental weights $\{h_j \mid 1\leq j\leq 4\}$ by
\beq\label{pahF4}
\bp
\pi(\oc\al_1)\\
\pi(\oc\al_2)\\
\pi(\oc\al_3)\\
\pi(\oc\al_4)
\ep=C(F_4)^T\bp
h_1\\
h_2\\
h_3\\
h_3
\ep=\bp
2&-1&0&0\\
-1&2&-1&0\\
0&-2&2&-1\\
0&0&-1&2
\ep\bp
h_1\\
h_2\\
h_3\\
h_4
\ep,
\eeq
where $C(F_4)$ is given by Equation \eqref{CarF4},
or we have
\beq\label{hpaF4}
\bp
h_1\\
h_2\\
h_3\\
h_4
\ep=
\left(C(F_4)^T\right)^{-1}
\bp
\pi(\oc\al_1)\\
\pi(\oc\al_2)\\
\pi(\oc\al_3)\\
\pi(\oc\al_4)
\ep=\bp
2 & 3 & 2 & 1 \\
 3 & 6 & 4 & 2 \\
 4 & 8 & 6 & 3 \\
 2 & 4 & 3 & 2
\ep\bp
\pi(\oc\al_1)\\
\pi(\oc\al_2)\\
\pi(\oc\al_3)\\
\pi(\oc\al_4)
\ep.
\eeq

The matrix of symmetric bilinear form $(\;,\;)$ on subspace $X_0$ in $\{\pi(\oc\al_1), \pi(\oc\al_2),\pi(\oc\al_3),\pi(\oc\al_4)\}$ basis  
is given by Equation \eqref{paipj}: 
\begin{align}\label{SymF4d}
\left(\left(\pi(\oc\al_i),\pi(\oc\al_j)\right)\right)_{1\leq i,j\leq4}
&=\left(\frac{2}{|\al_i|^2}a_{ij}\right)_{1\leq i,j\leq4},\\\nonumber
&=\bp
1&0&0&0\\
0&1&0&0\\
0&0&2&0\\
0&0&0&2
\ep\bp
2&-1&0&0\\
-1&2&-2&0\\
0&-1&2&-1\\
0&0&-1&2
\ep
=\bp
2&-1&0&0\\
-1&2&-2&0\\
0&-2&4&-2\\
0&0&-2&4
\ep.
\end{align}

The diagonal entries of the last matrix in Equation \eqref{SymF4d} tell us that 
$|\pi(\oc\al_i)|^2$ (for $1\leq j\leq 4$) are
2, 2, 4 and 4, respectively. That is, $\pi(\oc\al_3)$ and $\pi(\oc\al_4)$ are long while
$\pi(\oc\al_1)$ and $\pi(\oc\al_2)$ are short as indicated by the Dynkin diagram in Figure \ref{rsF4d}.

The bilinear form in $\{h_j \mid 1\leq j\leq 4\}$ basis of $X_0$ is given by Equation
\eqref{hij}:
\begin{align}\label{hijF4}
\left((h_i, h_j)\right)_{1\leq i,j\leq4}
&=\left(C(F_4)^T\right)^{-1}\left(\frac{2}{|\al_k|^2}\de_{kj}\right)_{1\leq k,j\leq4},\\\nonumber
&=\left(C(F_4)^T\right)^{-1}
\bp
1&0&0&0\\
0&1&0&0\\
0&0&2&0\\
0&0&0&2
\ep=\bp
2 & 3 & 2 & 1 \\
 3 & 6 & 4 & 2 \\
 4 & 8 & 6 & 3 \\
 2 & 4 & 3 & 2
\ep\bp
1&0&0&0\\
0&1&0&0\\
0&0&2&0\\
0&0&0&2
\ep
=\bp
2 & 3 & 4 & 2 \\
 3 & 6 & 8 & 4 \\
 4 & 8 & 12 & 6 \\
 2 & 4 & 6 & 4
\ep.    
\end{align}

The diagonal entries of the last matrix in Equation \eqref{hijF4}
tell us that $|h_j|^2$ for $1\leq j\leq 4$ are 2,6,12 and 4, respectively. 
That is, $h_1$ with $|h_1|^2=2$ is the shortest of the fundamental weights of the $F_4$ system. 

Now let us consider the coroots of some non-simple roots of the finite $F_4$ root system using Proposition \ref{miki}.
In particular, consider the highest short and long roots of the $F_4$ system:
$\tilde{\al}_s=\al_{12233344}$ and $\tilde{\al}=\al_{11222333344}$.\\
For
$\tilde{\al}_s=\al_{12233344}$, we have $m_i$ for $1\leq i \leq 4$ are $1,2,3,2$, respectively.  Moreover, we have 
$|\al_{1}|^2=|\al_{2}|^2=2$, $|\al_{3}|^2=|\al_{4}|^2=|\tilde{\al}_s|^2=1$ from Equation \eqref{SymF4}. Then by Proposition \ref{miki}, we have
\begin{align}\nonumber
\pi(\oc{\tilde{\al}_s})&=\pi(\oc\al_{12233344})
=(1\cdot2)\pi(\oc\al_1)+(2\cdot2)\pi(\oc\al_2)+(3\cdot 1)\pi(\oc\al_3)+(2\cdot 1)\pi(\oc\al_4)\\
\label{psF4}
&=2\pi(\oc\al_1)+4\pi(\oc\al_2)+3\pi(\oc\al_3)+2\pi(\oc\al_4)=h_4.
\end{align}
Hence we have $|\pi(\oc{\tilde{\al}_s})|^2=|h_4|^2=4$, that is $\pi(\oc{\tilde{\al}_s})$
is a long root in the dual $F_4$ system in $X_0$.

For $\tilde{\al}=\al_{11222333344}$, we have $|\tilde{\al}|^2=2$, and $m_i$ for $1\leq i \leq 4$ are $2,3,4,2$, respectively. By Proposition \ref{miki}
we have
\begin{align}\nonumber
\pi(\oc{\tilde{\al}})&=\pi(\oc\al_{11222333344})
=(2\cdot1)\pi(\oc\al_1)+(3\cdot1)\pi(\oc\al_2)+(4\cdot \frac{1}{2})\pi(\oc\al_3)+(2\cdot \frac{1}{2})\pi(\oc\al_4)\\
\label{plF4}
&=2\pi(\oc\al_1)+3\pi(\oc\al_2)+2\pi(\oc\al_3)+\pi(\oc\al_4)=h_1,
\end{align}
and $|\pi(\oc{\tilde{\al}})|^2=|h_1|^2=2$. That is, $\pi(\oc{\tilde{\al}})$
is a short root in the dual $F_4$ system in $X_0$.
To express $\pi(\oc{\tilde{\al}_s})$ and $\pi(\oc{\tilde{\al}})$ in terms of the fundamental weights we have used Equation \eqref{hpaF4}.
The last expression of Equation \eqref{plF4}
can also be obtained using Proposition \ref{pa0h} and Equation \eqref{pa0ah},
\begin{equation}\label{p0Fh}
\pi(\oc{\al_0})=\sum_{k=1}^4 \left(C(F_4^{(1)})\right)_{k0}h_k=
(-1).h_1+0.h_2+0.h_3+0.h_4=-h_1,
\end{equation}
with $C(F_4^{(1)})$ given in Equation \eqref{CarF4a}.
}
{\Rem\label{pF4r}
Comparing the expressions for $\pi(\oc{\tilde{\al}_s})$ and $\pi(\oc{\tilde{\al}})$
given in Equations \eqref{psF4} and \eqref{plF4} with roots of the finite $F_4$
system in Example \ref{F4},
we see that the map $\pi$ takes $\oc{\tilde{\al}_s}$ and $\oc{\tilde{\al}}$ of $F_4$ type to the highest long and short root 
of the $F_4$ type
dual system generated by $\{\pi(\oc\al_i)\mid1\leq j\leq 4\}$, respectively.}

{\eg \label{pG2}Affine Weyl group of type $G_2$, $W(G_2^{(1)})$.
The Dynkin diagram of type $G_2^{(1)}$, $\Ga(G_2^{(1)})$ is given in Figure \ref{claW}.
The corresponding generalized Cartan matrix of type $G_2^{(1)}$, 
\beq\label{CarG2a}
C(G_2^{(1)})=(a_{ij})_{1\leq i,j\leq 2,0}=(\al_i\cdot\oc\al_j)_{1\leq i,j\leq 2,0}=\bp
2&-1&0\\
-3&2&-1\\
0&-1&2
\ep,
\eeq
and the defining relations for the Weyl group $W(G_2^{(1)})=\lan s_i\mid 0\leq i \leq 2\ran$,
\begin{align}\label{funWG2}
&s_1^2=s_2^2=1,\;\;(s_1s_{2})^6=1, \\\label{funWG2a}
&s_0^2=1,\;\;(s_0s_{1})^2=1, \quad (s_{0}s_2)^3=1,
\end{align}
can be read off from $\Ga(G_2^{(1)})$ with the rules
given in Table \ref{CD}.
The $G_2^{(1)}$ simple system $\De^{(1)}=\{\al_1,\al_2,\al_0\}$ forms a basis for an $3$-dimensional real vector space $V^{(1)}$.
Generators $s_j$ act on $V^{(1)}$ by Equation \eqref{sij0}, where $a_{ij}$ is the $(i,j)$-entry
of $C(G_2^{(1)})$ given in Equation \eqref{CarG2a}.
The dual space $ V^{(1)*}$ and its hyperplanes $X_k$ are given by Definitions \ref{ds}
and \ref{dsX}, respectively.
The set of simple coroots of $G_2$, $\{\pi(\oc\al_1), \pi(\oc\al_2)\}$
form a dual system also of $G_2$ type (see Figure \ref{rsG2d})
and is a basis of $X_0\subset V^{(1)*}$.
\begin{figure}[ht]
\centering
\begin{tikzpicture}
		\hs{-5em}
			\node  (a1) {};
			\node [right=of a1](a2) {$\circ$} ;
			\node [right=of a2](a3) {$\circ$} ;
			\draw (a2) node [anchor=north] {$\pi(\oc\al_1)$} ;
			\draw (a3) node [anchor=north] {$\pi(\oc\al_2)$} ;
			\draw[triple] (a2) -- node {} (a3);
			\node (b) at ($(a2)!0.5!(a3)$) {$>$};
			\path[use as bounding box] (-1.5,0) rectangle (0,0); 
\end{tikzpicture}
\caption{Dynkin diagram for the dual system.}\label{rsG2d}
\end{figure}
The group $W(G_2^{(1)})$ acts on $\{\pi(\oc\al_1), \pi(\oc\al_2)\}$ by Proposition \ref{skpajp}.
By Equation \eqref{pah}, $\{\pi(\oc\al_1), \pi(\oc\al_2)\}$ can be expressed in terms of 
the fundamental weights by
\beq\label{pahG2}
\bp
\pi(\oc\al_1)\\
\pi(\oc\al_2)
\ep=C(G_2)^T\bp
h_1\\
h_2
\ep=\bp
2&-3\\
-1&2
\ep\bp
h_1\\
h_2
\ep,
\eeq
where $C(G_2)$ is given by Equation \eqref{CarG2}. Moreover, we have
\beq\label{hpaG2}
\bp
h_1\\
h_2
\ep=
\left(C(G_2)^T\right)^{-1}
\bp
\pi(\oc\al_1)\\
\pi(\oc\al_2)
\ep=
\bp
2&3\\
1&2
\ep
\bp
\pi(\oc\al_1)\\
\pi(\oc\al_2)
\ep.
\eeq

The matrix of symmetric bilinear form on the subspace $X_0\subset V^{(1)*}$ in $\{\pi(\oc\al_1), \pi(\oc\al_2)\}$ basis  
is given by Equation \eqref{paipj}: 
\beq\label{SymG2d}
\left(\left(\pi(\oc\al_i),\pi(\oc\al_j)\right)\right)_{1\leq i,j\leq2}
=\left(\frac{2}{|\al_i|^2}a_{ij}\right)_{1\leq i,j\leq2}
=\bp
1&0\\
0&\frac{1}{3}
\ep\bp
2&-1\\
-3&2
\ep
=\bp
2&-1\\
-1&\frac{2}{3}
\ep.
\eeq
The diagonal entries of the last matrix in Equation \eqref{SymG2d} tell us that 
$|\pi(\oc\al_1)|^2=2$ and $|\pi(\oc\al_2)|^2=2/3$.
The bilinear form in the $\{h_1, h_2\}$ basis of $X_0$ is given by Equation
\eqref{hij}:
\beq\label{hijG2}
\left((h_i, h_j)\right)_{1\leq i,j\leq2}
=\left(C(G_2)^T\right)^{-1}
\bp
1&0\\
0&\frac{1}{3}
\ep
=\bp
2&3\\
1&2
\ep\bp
1&0\\
0&\frac{1}{3}
\ep
=\bp
2&1\\
1&\frac{2}{3}
\ep.
\eeq
The diagonal entries of the matrix on the right in Equation \eqref{hijG2}
tell us that $|h_1|^2=2$ and $|h_2|^2=2/3$. 
Hence $h_2$ is the shortest fundamental weight of the $G_2$ system. 

Using Proposition \ref{miki}, we compute the coroots of
the highest short root $\tilde{\al}_s=\al_{112}$, and long root  $\tilde{\al}=\al_{11122}$,
of the finite $G_2$ system given in Equation \eqref{rsG2}.

For
$\tilde{\al}_s=\al_{112}$, we have $m_1=2$ and $m_2=1$. From Example \ref{G2} we know that
$|\al_{1}|^2=2$, $|\al_{2}|^2=6$ and $|\al_{112}|^2=2$. Then by Proposition \ref{miki} we have
\begin{equation}\label{hsrG2}
\pi(\oc{\tilde{\al}_s})=\pi(\oc\al_{112})=(2\cdot 1) \pi(\oc\al_1)+(1\cdot 3) \pi(\oc\al_2)
=2\pi(\oc\al_1)+3\pi(\oc\al_2)=h_1,
\end{equation} 
so $|\pi(\oc{\tilde{\al}_s})|^2=|h_1|^2=2$.  That is $\pi(\oc{\tilde{\al}_s})$
is a long root in the dual $G_2$ system in $X_0$.
For $\tilde{\al}=\al_{11122}$, we have $m_1=3$, $m_2=2$ and $|\al_{11122}|^2=6$. Then 
by Proposition \ref{miki} we have
\beq\label{hrG2}
\pi(\oc{\tilde{\al}})=\pi(\oc\al_{11122})
=(3\cdot \frac{1}{3})\pi(\oc\al_1)+(2\cdot 1)\pi(\oc\al_2)
=\pi(\oc\al_1)+2\pi(\oc\al_2)=h_2,
\eeq
where $|\pi(\oc{\tilde{\al}})|^2=|h_2|^2=2/3$. That is $\pi(\oc{\tilde{\al}})$
is a short root in the dual $G_2$ system in $X_0$.
To express $\pi(\oc{\tilde{\al}_s})$ and $\pi(\oc{\tilde{\al}})$ in terms of the fundamental weight we have used Equation \eqref{hpaG2}.
The last expression on the right of Equation \eqref{hsrG2}
can also be obtained using Proposition \ref{pa0h} and Equation \eqref{pa0ah},
\begin{equation}\label{p0Gh}
\pi(\oc{\al_0})=\sum_{k=1}^2 \left(C(G_2^{(1)})\right)_{k0}h_k=0.h_1+(-1).h_2=-h_2,
\end{equation}
with $C(G_2^{(1)})$ given by Equation \eqref{CarG2a}.
}
\subsection{Translations}\label{TX1}
We are now ready to investigate actions of $W^{(1)}$ on the dual space $V^{(1)*}$. In particular, we construct certain elements of $W^{(1)}$ which act as translations on the affine subspace $X_1\subset V^{(1)*}$.
{\prop\label{tbX0} 
For any root of the affine root system: $\be=\al+m\de\in\Phi^{(1)}$, where $0\neq m\in\mathbb{Z}$, 
and $\al\in\Phi$, we form the element $t_\be=s_\al s_\be\in W^{(1)}$. We have
\beq\label{tbh}
t_\be(h)=h,
\eeq
for all $h\in X_0$. That is, $t_\be\in W^{(1)}$
fixes all elements of $X_0$.
}
\begin{proof}
Let $t_\be=s_\al s_\be$ for any $\be=\al+m\de\in\Phi^{(1)}$ ($0\neq m\in\mathbb{Z}$, 
and $\al\in\Phi$). 
First, observe that for $\al\in\Phi$ we have
\begin{equation}\label{sah}
s_\al(h)\in X_0, \quad h\in X_0,
\end{equation}
since
\begin{equation}
\lan \de,s_\al(h)\ran=\lan s_\al(\de),h\ran=\lan \de,h\ran=0,
\end{equation}
that is $s_\al\in W$ preserves $X_0$.

Now by Equation \eqref{sbf}, for $h\in X_0$ we have
\begin{equation}\label{sbh}
s_\be(h)=h-\lan \al+m\de,h\ran\pi(\oc\be)=h-\lan \al+m\de,h\ran\pi(\oc\al)=h-\lan \al,h\ran\pi(\oc\al)
=s_\al(h).
\end{equation}

Then we have,
\beq\label{tbhp}
t_\be(h)=s_\al s_\be(h)=s_\al s_\al(h)=h,
\eeq
for all $h\in X_0$. 
\end{proof}

{\prop\label{tbhdp}For any $\be=\al+m\de\in\Phi^{(1)}$ ($0\neq m\in\mathbb{Z}$, 
and $\al\in\Phi$),
the element $t_\be\in W^{(1)}$ acts on $h_\de\in V^{(1)*}$ by
\begin{equation}\label{tbhde}
t_\be(h_\de)=h_\de+\lan\be,h_\de\ran\pi(\oc\be).
\end{equation}
}
\begin{proof}
By Equation \eqref{bfact} we have,
\begin{equation}\label{sahd}
s_\al(h_\de)=h_\de-\lan\al,h_\de\ran\pi(\oc\be)=h_\de,
\end{equation}
where $\lan \al,h_\de\ran=0$ by Definition \ref{ds}, since $\al\in\Phi$ is a linear combination of 
$\{\al_i|1\leq i \leq n\}$.
Moreover, 
\begin{equation}\label{sbhde}
s_\be(h_\de)=h_\de-\lan\be,h_\de\ran\pi(\oc\be).
\end{equation}

By Equation \eqref{sbf} we have
\begin{align*}
s_\al(\pi(\oc\be))&=\pi(\oc\be)-\lan \al,\pi(\oc\be)\ran\pi(\oc\be)\\
&=\pi(\oc\be)-\lan \al,\pi(\oc\al)\ran\pi(\oc\be)\\
&=\pi(\oc\be)-2\pi(\oc\be)\\
&=-\pi(\oc\be),
\end{align*}
where we have used $\lan \al,\pi(\oc\al)\ran=\al\cdot\oc\al=\frac{2\al\cdot\al}{\al\cdot\al}=2$.
Hence we found that,
\begin{equation}\label{tbhd}
t_\be(h_\de)=(s_\al s_\be)(h_\de)=s_\al (h_\de-\lan\be,h_\de\ran\pi(\oc\be))
=h_\de-\lan\be,h_\de\ran s_\al(\pi(\oc\be))=h_\de+\lan\be,h_\de\ran\pi(\oc\be).
\end{equation}
\end{proof}

The dual space $V^{(1)*}$ is the direct sum 
$\mathbb{R}h_\de\oplus X_0$, actions of $t_\be$ on $V^{(1)*}$ and $V^{(1)}$ (by the
contragredient action) in general are given by the following Proposition.
{\prop\label{tbfv}
The element $t_\be=s_\al s_\be\in W^{(1)}$ for any $\be=\al+m\de\in\Phi^{(1)}$ $(0\neq m\in\mathbb{Z}$, and $\al\in\Phi)$ acts on $V^{(1)*}$ and $V^{(1)}$
by
\begin{equation}\label{tbf}
t_\be(f)=f+\lan\de,f\ran\lan\be,h_\de\ran\pi(\oc\be)
=f+\lan\de,f\ran m \pi(\oc\al),\quad\mbox{for all}\quad f\in V^{(1)*},
\end{equation}
and
\begin{equation}\label{tbv}
t_\be(v)=v-\lan\be,h_\de\ran\lan v,\pi(\oc\be)\ran\de
=v-m\lan v,\pi(\oc\al)\ran\de,\quad\mbox{for all}\quad v\in V^{(1)},
\end{equation}
respectively.
}
\begin{proof}
Let $f$ be an arbitrary element of $V^{(1)*}$ then
$
f_0=f-\lan \de,f\ran h_\de 
$ is an element of the subspace $X_0$ since
$
\lan\de,f_0\ran=\lan\de,f\ran-\lan\de,f\ran\lan\de,h_\de\ran=0.
$
It follows from Propositions \ref{tbX0} and \ref{tbhdp} that for $\be=\al+m\de\in\Phi^{(1)}$ ($0\neq m\in\mathbb{Z}$, and $\al\in\Phi$), 
\begin{align}\label{tbfp}\nonumber
t_\be(f)&=t_\be(f_0)+\lan\de,f\ran t_\be(h_\de),\\\nonumber
&=f_0+\lan\de,f\ran\left(h_\de+\lan\be,h_\de\ran\pi(\oc\be)\right),\\\nonumber
&=f+\lan\de,f\ran\lan\be,h_\de\ran\pi(\oc\be),\\
&=f+\lan\de,f\ran m \pi(\oc\al),
\end{align}
for all $f\in V^{(1)*}$.

Action of $t_\be$ on $v\in V^{(1)}$ can be obtained using Equation \eqref{tbfp} and
the contragredient action
of $W^{(1)}$. That is,
\begin{align*}
\lan t_\be^{-1}(v),f\ran&=\lan v,t_\be (f)\ran,\\
&=\lan v,f+\lan\de,f\ran m \pi(\oc\al)\ran,\\
&=\lan v,f\ran+\lan\de,f\ran m \lan v,\pi(\oc\al)\ran,\\
&=\lan v+m\lan v,\pi(\oc\al)\ran\de,f\ran,
\end{align*}
so $t_\be^{-1}(v)=v+m\lan v,\pi(\oc\al)\ran\de$ or
\begin{equation}
t_\be(v)=v-m\lan v,\pi(\oc\al)\ran\de,
\end{equation}
for all $v\in V^{(1)}$.
\end{proof}
Recall that we have
$n$-dimensional affine spaces $X_k=kh_\de+X_0$ in $V^{(1)*}$ on which $t_\be$ acts as  a translation
by Equation \eqref{tbf}.  
Since for every non-zero value of $k$, $X_k$ behaves like every other, we henceforth consider only $X_1$. 
Moreover, it is useful to have the actions of $t_\be$ on the $\{\al_1, \al_2, ..., \al_0\}$ basis of $V^{(1)}$.
Hence we have the following.

{\prop\label{tbX1} The element $t_\be=s_\al s_\be\in W^{(1)}$ for any $\be=\al+m\de\in\Phi^{(1)}$ $(0\neq m\in\mathbb{Z}$, and $\al\in\Phi)$ acts on $f\in X_1$ 
by
\begin{equation}\label{tbh1}
t_\be(f)=f+m\sum_{j=1}^{n}b_jh_j,
\end{equation} and
on the $\{\al_1, \al_2, ..., \al_0\}$ basis
of $V^{(1)}$ by
\begin{equation}\label{tbal}
t_{\be}(\al_i)=\al_i-mb_i\de,
\end{equation}
where $b_i=\lan\al_i,\pi(\oc\be)\ran$ for $0\leq i\leq n$.
}
\begin{proof}
By Proposition \ref{tbfv}, having $\lan f,\de\ran=1$ for all $f\in X_1$, gives us 
\begin{equation}
t_\be(f)=f+\lan\be,h_\de\ran\pi(\oc\be)
=f+m\sum_{j=1}^{n}b_jh_j,\quad b_j=\lan\al_j,\pi(\oc\be)\ran,
\end{equation} where we have used Proposition \ref{muihp}.
The actions of $t_\be$ on the $\{\al_1, \al_2, ..., \al_0\}$ basis
are given by Equation \eqref{tbv},
\begin{equation}
t_\be(\al_i)
=\al_i-m\lan \al_i,\sum_{j=1}^{n}b_jh_j\ran\de
=\al_i-m b_i\de,\quad\mbox{for all}\quad 0\leq i\leq n.
\end{equation}
\end{proof}
\subsubsection{Translations by simple coroots.}
Here we apply Proposition \ref{tbfv} to write down some explicit examples of translations in $W^{(1)}$.
{\prop\label{tj}
Let 
\beq
t_j=t_{\be_j}=s_{\al_j}s_{\be_j},\quad\mbox{where}\quad\be_j=\al_j+\de\quad\mbox{for}\quad 1\leq j\leq n, \quad\mbox{and}\quad
\be_0=\al_0.
\eeq
Element $t_j$ ($0\leq j\leq n$)
acts on $f\in X_1$ by
\begin{equation}\label{tjf}
t_j(f)=f+\pi(\oc\al_j)=f+\sum_{k=1}^{n}a_{kj}h_k,
\end{equation}
and on $\{\al_1, \al_2, ..., \al_0\}$ basis
of $V^{(1)}$ by
\begin{equation}\label{tjal}
t_{j}(\al_i)=\al_i-a_{ij}\de, 
\end{equation}
where $0\leq i,j\leq n$, and $a_{ij}$ is the $(i,j)$-entry of $C(\Ga^{(1)})$.
}
\begin{proof}
In Proposition \ref{tbX1}, for $\be_j$ ($1\leq j\leq n$) we have $\be=\al_j+\de$,
that is, $\al=\al_j$ and $m=1$,
and $t_j=t_{\be_j}=t_{\al_j+\de}=t_{\be}=s_{\al}s_{\be}=s_{\al_j}s_{\be_j}$.

For $\be_0=\al_0$, we have $\be=\al_0=-\tilde{\al}+\de$,
that is, $\al=-\tilde{\al}$, $m=1$ and
\begin{align}\label{t0}
t_{0}=t_{\al_0}=t_{\be}&=s_{\al}s_{\be}=s_{-\tilde{\al}}s_{\al_0}=s_{\tilde{\al}}s_{\al_0}=s_{\tilde{\al}}s_{0}.
\end{align}
Element $t_j$ ($0\leq j\leq n$) acts on $f\in X_1$ by
\beq
t_j(f)=f+\pi(\oc\al_j)=f+\sum_{k=1}^{n}a_{kj}h_k,
\eeq
where we have used Proposition \ref{pa0h}.
Its action on $\{\al_1, \al_2, ..., \al_0\}$ basis
of $V^{(1)}$ is given by
Equation \eqref{tbal}, we have
\beq
t_{j}(\al_i)=\al_i-\lan \al_i,\pi(\oc\al_j)\ran\de=\al_i-a_{ij}\de, \quad\mbox{for all}\quad 0\leq i,j\leq n.
\eeq
\end{proof}

{\Rem\label{TransQ}
Recall that the group $W^{(1)}$ acts transitively on roots in $\Phi^{(1)}$ (of the same length) and hence also on the coroots  (of the same length) via the contragredient action.
That is, it is enough
to write down an element of translation by a simple coroot of the 
simply-laced system, and two simple coroots of different lengths for 
the non-simply-laced system.
In particular, elements of translations are related by the formula
given in Equation \eqref{Fwh}.}

We end this section with
a discussion on the expression of $T_1\in W(E_8^{(1)})$ given earlier in Equation \eqref{T1decomp1} for Sakai's $e$-{\bf P}$(E_8^{(1)})$ equation employing the formulas discussed in Equation \eqref{TX1}.

{\eg \label{T1form}
Let $\De^{(1)}=\{\al_j\mid 0\leq j\leq 8\}$
be the $E_8^{(1)}$ simple system with the numbering on $\Ga(E_8^{(1)})$
in Figure \ref{claW}, we have $W(E_8^{(1)})=\lan s_i\mid 0\leq i \leq 8\ran$ and
$\Phi^{(1)}$ is the $E_8^{(1)}$ root system.
An element $T_1$ of $W(E_8^{(1)})$ given as a product of 58 simple reflections,
\begin{align}\nonumber
     T_{1}=
&s_3 s_4 s_2 s_5 s_4 s_3 s_6  s_5 s_4 s_2 s_7 s_6 s_5 
s_4 s_3 s_8 s_7 s_6 s_5 s_4 s_2 s_0 s_8 s_7 s_6 s_5 s_4 s_3s_1  \\\label{T1decomp}
&s_3 s_4 s_5 s_6 s_7 s_8 s_0 s_2 s_4 s_5 s_6 s_7 s_8 s_3 
s_4 s_5 s_6 s_7 s_2 s_4 s_5 s_6 s_3 s_4 s_5 s_2 s_4 s_3s_1,
\end{align}
is said to be a translation by $\al_1$
in $W(E_8^{(1)})$. 
The actions of $T_1$
on the simple roots $\{\al_j\mid 0\leq j\leq 8\}$
can be computed using Equation \eqref{T1decomp} by composing the actions of
simple reflections given in Equation \eqref{sij0} with $C(E_8^{(1)})$ from Equation \eqref{CarE8a},
where the compositions are taken from right to left,
\beq\label{TJ23}
T_{1}:\{\al_1, \al_3\}\mapsto
\{\al_1-2\de, \al_3+\de\},
\eeq
and they 
coincide with
those given in Equation  \eqref{TJ21}.
On the other hand, by Proposition \ref{tj}, for $j=1$ we have translation of $\al_1$
in $W(E_8^{(1)})$ given by,
\begin{equation}\label{t1e8}
    t_1=t_{\al_1+\de}=s_{\al_1}s_{\al_1+\de},
\end{equation}
where $\de$ is given by Equation \eqref{deE8}.
The action of $t_1$ on $\De^{(1)}$ is given by Equation \eqref{tjal} with 
the $a_{ij}$'s from
$C(E_8^{(1)})$.
Since the only non-zero entries in the first column of $C(E_8^{(1)})$
are $a_{11}=2$ and $a_{31}=-1$, we have
\begin{equation}
    t_1(\al_1)=\al_1-2\de, \quad t_1(\al_3)=\al_3+\de,\quad \mbox{and}
    \quad t_1(\al_i)=\al_i,\quad \mbox{for}\quad i\neq 1,3,
\end{equation}
which agrees with the action computed in Equation \eqref{TJ23}.

The expression for $T_1$ in Equation \eqref{T1decomp} is given as a product of
simple reflections of the form: $58=28+1+28+1$.
On the other hand, our formula for $t_1$ in Equation \eqref{t1e8} is given as a product of two reflections, along the two roots $\al_1$, and $\al_1+\de\in \Phi^{(1)}$. 
The element $t_1$ given in Equation \eqref{t1e8} can
be rewritten into the form given in Equation \eqref{T1decomp} following a series of observations.
First, we have
\begin{equation}
 t_1=t_{\al_1+\de}=t_{\al_1-\de}^{-1}=(s_{\al_1}s_{\al_1-\de})^{-1}=s_{\al_1-\de}
 s_{\al_1}=s_{\de-\al_1}s_{\al_1},    
\end{equation}
where we have used Equation \eqref{tbfv}. Next, we find an expression for $s_{\de-\al_1}$.
Recall that $W^{(1)}$
acts transitively on the roots of $\Phi^{(1)}$, that is
there is an $w\in W(E_8^{(1)})$ such that
\begin{equation}
w(\al_1)=\de-\al_1.
\end{equation}
By \cite[Lemma~4.4]{Brink:98}, we have the length of $w$ is equal to
the height of $\de-\al_1$ minus 1.
Recall that the height of a root is given earlier in Definition \ref{heiaw}.
That is, the height of $\de-\al_1$ is given by,
\begin{equation}
    1+\sum_{i=1}^{8}c_i-1=1+29-1=29.
\end{equation}
So we have,
\begin{equation}
    l(w)=29-1=28.
\end{equation}
An algorithm for finding an expression of $w$ is also
given in \cite{Brink:98} and we have,
\begin{equation}\label{wa1da1}
 w=s_3 s_4 s_2 s_5 s_4 s_3 s_6  s_5 s_4 s_2 s_7 s_6 s_5 
s_4 s_3 s_8 s_7 s_6 s_5 s_4 s_2 s_0 s_8 s_7 s_6 s_5 s_4 s_3,
\end{equation}
where $l(w)=28$ and $w^2=1$.
Then by Equation \eqref{sr} we have $s_{\de-\al_1}=ws_1w^{-1}$, and finally,
\begin{equation}
    t_1=s_{\de-\al_1}s_{\al_1}=ws_1w^{-1}s_1=ws_1ws_1,
\end{equation}
the second expression from the right coincides with the expression 
given in Equation \eqref{T1decomp}.
}

{\Rem
Note that a different, geometrically motivated expression for 
$t_1\in W(E_8^{(1)})$ is given in \cite{KMNOY:03}.
}
{\Rem
The relation between the height of a root $\be\in\Phi^{(1)}_{+}$ and the
length of the element $w$ such that $w(\al_i)=\be$ (where $\al_i$ is a simple root)
is given for general Coxeter groups and their root systems in \cite{Brink:98},
where the concept of the height of a root is generalised to the {\it depth} of a root.
}

\subsection{A normal subgroup of translations, ${W}^{(1)}=W\ltimes Q$}\label{TG}
In the previous section, we constructed elements of translation $t_\be$ (for any $\be\in\Phi^{(1)}$) and looked at in particular translations by simple coroots of $W^{(1)}$.

Now, using
a representation $R$ of $W^{(1)}$ on the dual vector space $V^{(1)*}$ 
we investigate the properties of set of all translations in $W^{(1)}$ and show that it forms
an abelian normal subgroup of $W^{(1)}$.

{\Def\label{Rd}
Let $R$ be a representation of $W^{(1)}$ on $V^{(1)*}$.
That is, if $w\in W^{(1)}$ then $R(w)$ is the linear transformation of $V^{(1)*}$
given by 
\begin{equation}\label{R}
\left(R(w)\right)(f)=w(f)\quad\mbox{for all}\quad f\in V^{(1)*}.
\end{equation}
The translational transformations of $R$, $F_h:V^{(1)*}\to V^{(1)*}$ are given by  
\begin{equation}\label{RF}
F_h(f)=f+\lan\de,f\ran h\quad\mbox{for each}\quad h\in X_0
\quad\mbox{and for all}\quad f\in V^{(1)*}.
\end{equation}
}

Comparing Equation \eqref{RF} with the action of $t_\be$ on $f\in V^{(1)*}$
given in Equation \eqref{tbf} we see that
\begin{equation}\label{RFh}
R(t_\be)=F_h, \quad\mbox{with}\quad h=\lan\be,h_\de\ran\pi(\oc\be)=m \pi(\oc\al),
\end{equation} 
for $\be=\al+m\de\in\Phi^{(1)}, \al\in \Phi, m\in\mathbb{Z}$.

{\prop\label{Tnormsub}
Let
\begin{equation}\label{Tg}
T=\{t_\be\mid\be\in\Phi^{(1)}\} .
\end{equation}
$T$ is a finitely generated abelian normal subgroup of translations
on the root lattice of $W^{(1)}$, 
\begin{equation}\label{Tns}
W^{(1)}=W\ltimes T=W\ltimes\langle t_{j} \mid 1\leq j\leq n\rangle
=W\ltimes Q,			
\end{equation}where $t_j$ are given by Proposition \ref{tj}.
}
\begin{proof}
Let $h,k\in X_0$ and $f\in V^{(1)*}$ then
\begin{align*}
(F_hF_k)(f)&=F_h(f+\lan \de, f\ran k)\\
&=(f+\lan \de, f\ran k)+\lan \de,f+\lan \de, f\ran k\ran h\\
&=f+\lan \de, f\ran k+\lan \de, f\ran h\\
&=F_{h+k}(f),
\end{align*}where we used $\lan \de, k\ran=0$, since $k\in X_0$. That is,
\begin{equation}\label{Tabi}
F_hF_k=F_{h+k}=F_kF_h.
\end{equation}
Moreover, $F_h$ is the identity if and only if $h=0$ and
\begin{equation}\label{Fnh}
F_{nh}=(F_h)^n\quad \mbox{for all}\quad  n\in \mathbb{Z}.
\end{equation}

If $w\in W^{(1)}$ and $h\in X_0$ then for all $v\in V^{(1)*}$ and $f\in V^{(1)*}$
we have
\begin{align*}
\lan v,w(F_h(w^{-1}(f)))\ran&=\lan w^{-1}v, F_h(w^{-1}(f))\ran\\
&=\lan w^{-1}(v),w^{-1}(f)+\lan\de,w^{-1}(f)\ran h\ran\\
&=\lan w^{-1}(v),w^{-1}(f)\ran+\lan\de,w^{-1}(f)\ran\lan w^{-1}(v),h\ran\\
&=\lan v,f\ran+\lan w(\de),f\ran\lan v,w(h)\ran\\
&=\lan v, f+\lan \de,f\ran w(h)\ran\\
&=\lan v, F_{w(h)}(f)\ran,
\end{align*} where we have used $w(\de)=\de$.
So we have, 
\begin{equation}\label{Fwh}
w(F_h(w^{-1}(f)))=F_{w(h)}(f),
\end{equation}
that is, a translation by $w(h)$ (for $h\in X_0, w\in W^{(1)}$)  on $X_1$ is related
to a translation by $h$ on $X_1$ by a conjugation of $w$.
In particular, if we have some $w\in W^{(1)}$ such that
$w(\al_j+\de)=\be$, then we have
\begin{equation}\label{Fwtj}
  t_{\be}=wt_jw^{-1},  
\end{equation}
where $t_j$ is the translation on $X_1$ by $\pi(\oc\al_j)$.
Since
\begin{equation}
R(t_{\be})=R(wt_jw^{-1})=F_{\pi\left(\oc {w(\al_j)}\right)}=F_{w\left(\pi(\oc\al_j)\right)},\quad 1\leq j\leq n,\quad w\in W^{(1)},
\end{equation}
where we have used Equation \eqref{wpi}.

So far, we have shown that $T$ is an abelian normal subgroup of $W^{(1)}$. To prove that it is finitely
generated let us define
\begin{equation}\label{TjRtj}
T_j=F_{\pi(\oc{\al_j})}=R(t_{\be_j})=R(t_j),\quad\text{where}\quad \be_j=\al_j+\de\quad\mbox{for each}\quad 1\leq j\leq n.
\end{equation}
Then for any $\be\in\Phi^{(1)}$: $\be=\al+m\de=\sum_{i=1}^n m_i\al_i+m\de$, ($m,\,m_i\in \mathbb{Z}, \al\in\Phi, \al_i\in\De)$
we have
\begin{equation}
R(t_\be)=F_{m\pi(\oc\al)}=F_{m\pi(\oc\al)}
=F_{m\sum_{i=1}^n k_i\pi(\oc\al_i)}
=T_1^{mk_1}T_2^{mk_2}\cdots T_n^{mk_n}
=R(\prod_{i=1}^n t_i^{m k_i}),
\end{equation}
where $
\mathbb{Z}\ni k_i=m_i\frac{|\al_i|^2}{|\al|^2}$, and we have used Proposition \ref{miki}, Equations \eqref{RFh}, \eqref{Tabi} and \eqref{Fnh}.
That is, $T$ is finitely generated by $\{t_j\, \mid 1\leq j\leq n\}$.
Finally, the root lattice $Q$ (given by Definition \eqref{rl}) is isomorphic to $T$ 
by 
\[
\sum_{i=1}^n k_i\pi(\oc\al_i)\mapsto \prod_{i=1}^n t_i^{k_i},\quad k_i\in\mathbb{Z}\quad\mbox{for}\quad 1\leq i\leq n.
\]
\end{proof}

We illustrate properties of $W^{(1)}$ discussed in Sections \ref{TX1} and \ref{TG}
for $B_3$, $C_3$, $F_4$ and $G_2$ type systems.
\subsection{Translations in $W(B_3^{(1)})$.}\label{TaB3}
Recall that in $\De(B_3^{(1)})$, $\al_i$ for $i=0,1,2$ are long while $\al_3$ is short.
Moreover,  
$\de=\al_0+\tilde{\al}=\al_{012233}$ 
and $\tilde{\al}_s=\al_{123}$.
\begin{enumerate}
\item{Translations associated with long roots of $\Phi(B_3^{(1)})$.}
We first write down an explicit expression for
$t_0=t_{\al_0}$ in terms of simple reflections of $W(B_3^{(1)})$.
Expressions for $t_1=t_{\al_1+\de}$ and $t_2=t_{\al_2+\de}$ or any other long roots of $\Phi(B_3^{(1)})$ can
be obtained by using either Proposition \ref{tj} or Equation \eqref{Fwh} by
the fact that root $\al_0$ is a long root of $\Phi(B_3^{(1)})$ and is in the
same orbit of any long root.

From Equation \eqref{t0}
we have $t_{\al_0}=s_{\tilde{\al}}s_{0}$.
By Equation \eqref{rsB3},
we know that 
\beq
s_{232}(\tilde{\al})=\al_1\quad\mbox{or}\quad s_{232}(\al_1)=\tilde{\al}.
\eeq
That is,
\beq
s_{\tilde{\al}}=s_{s_{232}(\al_1)}=s_{232}s_1s_{232}=s_{2321232},
\eeq
where we have used Equation \eqref{sr}. We have then,
\beq\label{t0B3}
t_{\al_0}=s_{\tilde{\al}}s_{0}=s_{23212320}.
\eeq
Action of $t_{\al_0}$ on $X_1$ is given by Equation \eqref{tjf}:
\begin{equation}\label{tlfB3}
t_{\al_0}(f)=f-\pi(\oc{\tilde{\al}})=f-h_2, \quad f\in X_1,
\end{equation}
where we have used Equation \eqref{plB3} for the last equality.\\
On $\{\al_1, \al_2, \al_3, \al_0\}$ basis of $V^{(1)}$ actions of $t_{\al_0}$ are given by Equation \eqref{tjal}
and $C(B_3^{(1)})$ in Equation \eqref{CarB3a},
\beq\label{tlvalB3}
t_{\al_0}:\{\al_1, \al_2, \al_3, \al_0\}\mapsto
\{\al_1, \al_2+\de, \al_3, \al_0-2\de\}.
\eeq
If we are using Proposition \ref{tbfv}
to get an expression for $t_1=t_{\al_1+\de}$ it is easier to consider first its inverse
$t_1^{-1}=t_{-\al_1+\de}$ and observe that 
\beq
s_{232}(\al_0)=s_{232}(\de-\tilde{\al})=-\al_1+\de,
\eeq
that is, we have, 
\beq
s_{-\al_1+\de}=s_{s_{232}(\al_0)}=s_{232}s_0s_{232}.
\eeq
Then by Proposition \ref{tbfv}, let $\be=-\al_1+\de$, that is
$\al=-\al_1$ and $m=1$, we have
\begin{align}\nonumber
t_1^{-1}&=t_{-\al_1+\de}=t_\be=s_{\al}s_{\be},\\\label{tn1B3}
&=s_{-\al_1}s_{-\al_1+\de},\\\nonumber
&=s_1s_{2320232}=s_{12320232}.
\end{align}
Finally, we have
\begin{equation}
t_1=\left(s_{-\al_1}s_{-\al_1+\de}\right)^{-1}=s_{23202321}.
\end{equation}
Actions of $t_1$ on $X_1$ and on $V^{(1)}$ are given by Equations \eqref{tjf} and \eqref{tjal}, respectively,  for $j=1$,
\begin{equation}\label{t1fB3}
t_1(f)=f+\pi(\oc\al_1)=f+2h_1-h_2,\quad f\in X_1
\end{equation}
and
\beq\label{t1valB3}
t_{1}:\{\al_1, \al_2, \al_3, \al_0\}\mapsto
\{\al_1-2\de, \al_2+\de, \al_3, \al_0\},
\eeq
where we have used $C(B_3^{(1)})$ given in Equation \eqref{CarB3a}.
\item{Translations associated with short roots of $\Phi(B_3^{(1)})$.}
We look at two such elements, $t_{\tilde{\al}_s+\de}$ and $t_3$.

First observe that $t_{\tilde{\al}_s+\de}=t_{-\tilde{\al}_s+\de}^{-1}$ and
\begin{equation}
-\tilde{\al}_s+\de=-\al_{123}+\al_{012233}=\al_{023}.
\end{equation} 
Moreover, we have
\beq
s_{12}(\al_3)=\al_{123},\quad\mbox{and}\quad s_{02}(\al_3)=\al_{023},
\eeq
that is, 
\beq
s_{\al_{123}}=s_{12}s_3s_{21},\quad\mbox{and}\quad s_{\al_{023}}=s_{02}s_3s_{20}.
\eeq

Then by Proposition \ref{tbfv}, on letting $\be= -\tilde{\al}_s+\de=\al_{023}$,
that is $\al=-\tilde{\al}_s=-\al_{123}$ and $m=1$,
we have
\begin{align}\nonumber
t_{-\tilde{\al}_s+\de}&=t_\be=s_{\al}s_{\be}=s_{-\tilde{\al}_s}s_{-\tilde{\al}_s+\de}\\
&=s_{\al_{123}}s_{\al_{023}}=s_{12321}s_{02320}.
\end{align}
Finally, we have
\begin{equation}
t_{\tilde{\al}_s+\de}=t_{-\tilde{\al}_s+\de}^{-1}=\left(s_{-\tilde{\al}_s}s_{-\tilde{\al}_s+\de}\right)^{-1}
=\left(s_{\al_{123}}s_{\al_{023}}\right)^{-1}
=s_{\al_{023}}s_{\al_{123}}=s_{02320}s_{12321}.
\end{equation}
Actions of $t_{\tilde{\al}_s+\de}$ on $X_1$ are given by Equation \eqref{tbf}:
\begin{equation}\label{tsfB3}
t_{\tilde{\al}_s+\de}(f)=f+\pi(\oc{\tilde{\al}_s})=f+2h_1,\quad f\in X_1,
\end{equation}
where we have used Equation \eqref{psB3} for the last expression. 
On $V^{(1)}$ $t_{\tilde{\al}_s+\de}$ acts by Equation \eqref{tbal}:
\beq\label{tsvalB3}
t_{\tilde{\al}_s+\de}:\{\al_1, \al_2, \al_3, \al_0\}\mapsto
\{\al_1-2\de, \al_2, \al_3, \al_0+2\de\}.
\eeq
To obtain an expression for $t_3$, we make use of Equation \eqref{Fwh}.
Recall from Example \ref{B3} that $\tilde{\al}_s=\al_{123}$ and $\al_3$ are both short roots
of $\Phi(B_3^{(1)})$ hence belong to the same $W(B_3^{(1)})$-orbit:
\beq
s_{21}(\al_{s})=s_{21}(\al_{123})=\al_3
\eeq
We have,
\begin{equation}
t_3=t_{\al_3+\de}=t_{s_{21}(\tilde{\al}_s+\de)}=s_{21}t_{\tilde{\al}_s+\de}s_{12}
=s_{21}s_{02320}s_{12321}s_{12}
=s_{20123}s_{20123},
\end{equation}
where we have used the defining relations of $W(B_3^{(1)})$ given in Equation \eqref{funWB3a}
to simplify the last expression.
Actions of $t_3$  on $X_1$ and $V^{(1)}$ are given by Equations \eqref{tjf} and \eqref{tjal} for $j=3$:
\begin{equation}\label{t3fB3}
t_3(f)=f+\pi(\oc\al_3)=f-2h_2+2h_3,\quad f\in X_1,
\end{equation}
and
\beq\label{t3valB3}
t_{3}:\{\al_1, \al_2, \al_3, \al_0\}\mapsto
\{\al_1, \al_2+2\de, \al_3-2\de, \al_0\}.
\eeq
\end{enumerate}

\subsection{Translations in $W(C_3^{(1)})$}\label{TaC3}
Recall that in $\De(C_3^{(1)})$, $\al_0$ and $\al_3$ are long, $\al_1$ and $\al_2$ are short.
Moreover,  
$\de=\al_0+\tilde{\al}=\al_{011223}$ 
and $\tilde{\al}_s=\al_{1223}$.

\begin{enumerate}
\item{Translations associated to long roots in $\Phi(C_3^{(1)})$.}
We write done explicit expressions for
$t_0=t_{\al_0}$ and $t_1=t_{\al_3+\de}$.
By Equation \eqref{t0},
we have $t_{\al_0}=s_{\tilde{\al}}s_{0}$.
From Example
\ref{C3} we know that 
\beq
s_{12}(\al_3)=\tilde{\al}=\al_{11223}.
\eeq
That is,
\beq
s_{\tilde{\al}}=s_{s_{12}(\al_3)}=s_{12}s_3s_{21}=s_{12321},
\eeq
so we have
\beq\label{t0C3}
t_{\al_0}=s_{\tilde{\al}}s_{0}=s_{123210}.
\eeq
Action of $t_{\al_0}$ on $X_1$ is given by Equation \eqref{tjf}:
\begin{equation}\label{tlfC3}
t_{\al_0}(f)=f+\pi(\oc\al_0)=f-h_1, \quad f\in X_1,
\end{equation}
where we have used Equation \eqref{hsrC3} for the last expression. On $\{\al_1, \al_2, \al_3, \al_0\}$ basis of $V^{(1)}$ its action is given by Equation \eqref{tjal}:
\beq\label{tlvalC3}
t_{\al_0}:\{\al_1, \al_2, \al_3, \al_0\}\mapsto
\{\al_1+\de, \al_2, \al_3, \al_0-2\de\},
\eeq
with $C(C_3^{(1)})$ given in Equation \eqref{CarC3a}.

To write done an expression for $t_3=t_{\al_3+\de}$ it is easier to consider first its inverse
$t_3^{-1}=t_{-\al_3+\de}$ and observe that 
\beq
s_{21}(\al_0)=\al_{01122}=-\al_3+\de.
\eeq
That is, 
\beq
s_{-\al_3+\de}=s_{s_{21}(\al_0)}=s_{21}s_0s_{12}.
\eeq
Then by Proposition \ref{tbfv},
letting $\be= -\tilde{\al}_3+\de=\al_{01122}$,
that is $\al=-\al_{3}$ and $m=1$, we have
\begin{align}\nonumber
t_3^{-1}=t_{-\al_3+\de}=t_\be&=s_{\al}s_{\be},\\\label{tn1C3}
&=s_{-\al_3}s_{-\al_3+\de},\\\nonumber
&=s_3s_{21012}=s_{321012}.
\end{align}
That is
\begin{equation}
t_3=\left(s_{-\al_3}s_{-\al_3+\de}\right)^{-1}=s_{210123}.
\end{equation}
Actions of $t_3$ on $X_1$ are given by Equation \eqref{tjf} for $j=3$:
\begin{equation}\label{tfC3}
t_3(f)=f+\pi(\oc\al_3)=f-h_2+2h_3,\quad f\in X_1
\end{equation}
and on $\{\al_1, \al_2, \al_3, \al_0\}$ basis of $V^{(1)}$ by Equation \eqref{tjal},
\beq\label{t3valC3}
t_{3}:\{\al_1, \al_2, \al_3, \al_0\}\mapsto
\{\al_1, \al_2+\de, \al_3-2\de, \al_0\}.
\eeq
Any other translations associated with a long root of $\Phi(C_3)$ can be obtained
from these expressions of $t_3$ or $t_0$ by using Equation \eqref{Fwh}.
\item{Translations associated to short roots of $\Phi(C_3^{(1)})$.}
First, observe that 
\begin{equation}
-\tilde{\al}_s+\de=-\al_{1223}+\al_{011223}=\al_{01},
\end{equation}
and from Equation \eqref{rsC3} from Example \ref{C3} we had $s_{232}(\tilde{\al}_s)=\al_1$, then
\beq
s_{232}(\al_{1})=\al_{1223}=\tilde{\al}_s,\quad\mbox{and}\quad s_0(\al_1)=\al_{01},
\eeq so
\beq
s_{\tilde{\al}_s}=s_{2321232},\quad\mbox{and}\quad s_{\al_{01}}=s_{010}.
\eeq
In Proposition \ref{tbfv}, on letting $\be= -\tilde{\al}_s+\de=\al_{01}$,
that is $\al=-\tilde{\al}_s=-\al_{1223}$ and $m=1$,
we have
\begin{align}\nonumber
t_{-\tilde{\al}_s+\de}&=t_\be=s_{\al}s_{\be}=s_{-\tilde{\al}_s}s_{-\tilde{\al}_s+\de},\\
&=s_{\al_{1223}}s_{\al_{01}}=s_{2321232}s_{010}.
\end{align}
Then
\begin{equation}\label{tsC3}
t_{\tilde{\al}_s+\de}=t_{-\tilde{\al}_s+\de}^{-1}=\left(s_{-\tilde{\al}_s}s_{-\tilde{\al}_s+\de}\right)^{-1}
=s_{010}s_{2321232}.
\end{equation}

Action of $t_{\tilde{\al}_s+\de}$ on $X_1$ is given by Equation \eqref{tbf}:
\begin{equation}\label{tsfC3}
t_{\tilde{\al}_s+\de}(f)=f+\pi(\oc{\tilde{\al}_s})=f+h_2,\quad f\in X_1,
\end{equation}
where we have used Equation \eqref{hrC3}. On $\{\al_1, \al_2, \al_3, \al_0\}$ basis of $V^{(1)}$ its action is given by Equation \eqref{tbal}:
\beq\label{tsvalC3}
t_{\tilde{\al}_s+\de}:\{\al_1, \al_2, \al_3, \al_0\}\mapsto
\{\al_1, \al_2-\de, \al_3, \al_0+2\de\}.
\eeq
For $t_1$, using $s_{232}(\tilde{\al}_s)=\al_{1}$ and Equation \eqref{Fwh} we haves
\begin{equation}\label{t1C3}
t_1=t_{\al_1+\de}=t_{s_{232}(\tilde{\al}_s+\de)}=s_{232}t_{\tilde{\al}_s+\de}s_{232}
=s_{232}s_{010}s_{2321232}s_{232}=s_{232}s_{010}s_{2321}.
\end{equation}
The actions of $t_1$  on $X_1$ and $V^{(1)}$ are given by Equations \eqref{tjf} and \eqref{tjal}, respectively, for $j=1$:
\begin{equation}\label{t1fC3}
t_1(f)=f+\pi(\oc\al_1)=f+2h_1-h_2,\quad f\in X_1,
\end{equation}
and
\beq\label{t1valC3}
t_{1}:\{\al_1, \al_2, \al_3, \al_0\}\mapsto
\{\al_1-2\de, \al_2+\de, \al_3, \al_0\}.
\eeq
\end{enumerate}
Translation $t_2=t_{\al_2+\de}$ can be similarly obtained using $s_{132}(\tilde{\al}_s)=\al_2$ from Equation \eqref{rsC3} in Example \ref{C3} and we have,
\begin{equation}\label{t2C3}
t_2=t_{\al_2+\de}=t_{s_{132}(\tilde{\al}_s+\de)}=s_{132}t_{\tilde{\al}_s+\de}s_{231},
\end{equation}
with $t_{\tilde{\al}}$ given in Equation \eqref{tsC3}.

\subsection{Translations in $W(F_4^{(1)})$.}\label{TaF4}
Recall that in $\De(F_4^{(1)})$, $\al_0$, $\al_1$ and $\al_2$ are long while $\al_3$ and
$\al_4$ are short. Recall that we have
$\de=\al_0+\tilde{\al}=\al_0+2\al_1+3\al_2+4\al_3+2\al_4$ 
and $\tilde{\al}_s=\al_1+2\al_2+3\al_3+2\al_4$.                                                  
\begin{enumerate}
\item{Translations associated with long roots of $\Phi(F_4^{(1)})$.}
We look at two such elements, $t_0=t_{\al_0}$ and $t_2$.

From Equation \eqref{t0}
we have $t_{\al_0}=s_{\tilde{\al}}s_{0}$.
It can be checked that 
\beq
s_{1232143}(\al_2)=\tilde{\al}.
\eeq That is,
\beq
s_{\tilde{\al}}=s_{s_{1232143}(\al_2)}=s_{1232143}s_2s_{3412321},
\eeq
where we have used the relation given in Equation \eqref{sr}. Finally,
\beq\label{t0F4}
t_{\al_0}=s_{\tilde{\al}}s_{0}=s_{1232143}s_2s_{3412321}s_0.
\eeq
Action of $t_{\al_0}$ on $X_1$ is given by Equations \eqref{tjf}:
\begin{equation}\label{tlfF4}
t_{\al_0}(f)=f-\pi(\oc{\tilde{\al}})=f-h_1, \quad f\in X_1,
\end{equation}
where we have used  Equation \eqref{plF4} for the last equality.
That is, $t_{\al_0}$ translate on $X_1$ by $-h_1$.\\
On $\{\al_1, \al_2, \al_3, \al_4, \al_0\}$ basis of $V^{(1)}$ actions of $t_{\al_0}$ are given by Equation \eqref{tjal}
with $C(F_4^{(1)})$ from Equation \eqref{CarF4a},
\beq\label{tlvalF4}
t_{\al_0}:\{\al_1, \al_2, \al_3, \al_4, \al_0\}\mapsto
\{\al_1+\de, \al_2, \al_3, \al_4, \al_0-2\de\}.
\eeq
Before using Proposition \ref{tbfv}
to get an expression for $t_2=t_{\al_2+\de}$ it is easier to consider first its inverse
$t_2^{-1}=t_{-\al_2+\de}$ and observe that 
\beq
s_{3412321}(\al_0)=s_{3412321}(\de-\tilde{\al})=-\al_2+\de,
\eeq
that is, we have, 
\beq
s_{-\al_2+\de}=s_{s_{3412321}(\al_0)}=s_{3412321}s_0s_{232143}.
\eeq
Then by Proposition \ref{tbfv}, let $\be=-\al_2+\de$, that is
$\al=-\al_2$ and $m=1$, we have
\begin{align}\nonumber
t_2^{-1}&=t_{-\al_2+\de}=t_\be=s_{\al}s_{\be}\\\label{tn1F4}
&=s_{-\al_2}s_{-\al_2+\de}\\\nonumber
&=s_2s_{3412321}s_0s_{232143}.
\end{align}
Hence we have
\begin{equation}
t_2=\left(s_{-\al_2}s_{-\al_2+\de}\right)^{-1}=s_{2341232}s_{01}s_{2321432}.
\end{equation}
Actions of $t_2$ on $X_1$ and on $V^{(1)}$ are given by Equations \eqref{tjf} and \eqref{tjal}, respectively,  for $j=2$,
\begin{equation}\label{t1fF4}
t_2(f)=f+\pi(\oc\al_2)=f-h_1+2h_2-h_3,\quad f\in X_1
\end{equation}
and
\beq\label{t1valF4}
t_{2}:\{\al_1, \al_2, \al_3, \al_4, \al_0\}\mapsto
\{\al_1+\de, \al_2-2\de, \al_3-\de, \al_4, \al_0\},
\eeq
where we have used $C(F_4^{(1)})$ given in Equation \eqref{CarF4a}.
\item{Translations associated with short roots of $\Phi(F_4^{(1)})$.}
We look at two such elements, $t_{\tilde{\al}_s+\de}$ and $t_4$.

First observe that $t_{\tilde{\al}_s+\de}=t_{-\tilde{\al}_s+\de}^{-1}$ and
\begin{equation}
-\tilde{\al}_s+\de=-\al_{12233344}+\al_{011222333344}=\al_{0123}.
\end{equation} 
It can be checked that, 
\begin{align}\label{F4a4as}
    s_{4321323}(\al_4)&=\tilde{\al}_s,\\[-2.5em]\nonumber
 \intertext{and}
    s_{012}(\al_3)&=\al_{0123},
\end{align} hence we have
\begin{align}
    s_{\tilde{\al}_s}&=s_{4321323}s_4s_{3231234},\\[-2.5em]\nonumber
 \intertext{and}
    s_{\al_{0123}}&=s_{012}s_3s_{210}.
\end{align}

Then by Proposition \ref{tbfv}, on letting $\be= -\tilde{\al}_s+\de=\al_{0123}$,
that is $\al=-\tilde{\al}_s$ and $m=1$,
we have
\begin{align}\nonumber
t_{-\tilde{\al}_s+\de}&=t_\be=s_{\al}s_{\be}
=s_{-\tilde{\al}_s}s_{-\tilde{\al}_s+\de},\\
&=s_{\tilde{\al}_s}s_{\al_{0123}}
=s_{4321323}s_4s_{3231234}s_{012}s_3s_{210}.
\end{align}
Finally, we have
\begin{align}\nonumber
  t_{\tilde{\al}_s+\de}&=t_{-\tilde{\al}_s+\de}^{-1}=\left(s_{-\tilde{\al}_s}s_{-\tilde{\al}_s+\de}\right)^{-1}
=\left(s_{\tilde{\al}_s}s_{\al_{0123}}\right)^{-1}
=s_{\al_{0123}}s_{\tilde{\al}_s},\\\label{thrsF4}
&=s_{0123210}s_{4321323}s_4s_{3231234}.
\end{align}
Actions of $t_{\tilde{\al}_s+\de}$ on $X_1$ are given by Equation \eqref{tbf}:
\begin{equation}\label{tsfF4}
t_{\tilde{\al}_s+\de}(f)=f+\pi(\oc{\tilde{\al}_s})=f+h_4,\quad f\in X_1,
\end{equation}
where we have used Equation \eqref{psF4} for the last equality.
That is, $t_{\tilde{\al}_s+\de}$ translates on $X_1$ by $h_4$.
On $V^{(1)}$ $t_{\tilde{\al}_s+\de}$ acts by Equation \eqref{tbal}:
\beq\label{tsvalF4}
t_{\tilde{\al}_s+\de}:\{\al_1, \al_2, \al_3,\al_4, \al_0\}\mapsto
\{\al_1, \al_2, \al_3,\al_4-\de, \al_0+2\de\}.
\eeq
To obtain an expression for $t_4$, we make use of Equation \eqref{Fwh}.
By Equation \eqref{F4a4as}, we have 
\beq
s_{4321323}(\al_4)=\tilde{\al}_s\quad\mbox{or}\quad
\al_4=s_{3231234}(\tilde{\al}_s).
\eeq
Then,
\begin{equation}
t_4=t_{\al_4+\de}=t_{s_{3231234}(\tilde{\al}_s+\de)}
=s_{3231234}t_{\tilde{\al}_s+\de}s_{4321323},
\end{equation}
with $t_{\tilde{\al}_s+\de}$ given by Equation \eqref{thrsF4}.
Action of $t_4$  on $X_1$ and $V^{(1)}$ are given by Equations \eqref{tjf} and \eqref{tjal} for $j=4$:
\begin{equation}\label{t4fF4}
t_4(f)=f+\pi(\oc\al_4)=f-h_3+2h_4,\quad f\in X_1,
\end{equation} where we have used Equation \eqref{hpaF4}
and
\beq\label{t4valF4}
t_{4}:\{\al_1, \al_2, \al_3,\al_4, \al_0\}\mapsto
\{\al_1, \al_2, \al_3+\de,\al_4-2\de, \al_0\},
\eeq
where we have used $C(F_4^{(1)})$ given in Equation \eqref{CarF4a}.
\end{enumerate}

\subsection{Translations in $W(G_2^{(1)})$.}\label{TaG2}
Recall that in $\De(G_2^{(1)})$, $\al_0$ and $\al_2$ are long while $\al_1$ is short, moreover we have
$\de=\al_0+\tilde{\al}=\al_0+3\al_1+2\al_2$ 
and $\tilde{\al}_s=2\al_1+\al_2$.                                                  
\begin{enumerate}
\item{Translations associated with long roots of $\Phi(G_2^{(1)})$.}
We look at two such elements, $t_0=t_{\al_0}$ and $t_2$.

From Equation \eqref{t0}
we have $t_{\al_0}=s_{\tilde{\al}}s_{0}$.
From Equation \eqref{rsG2} in Example \ref{G2} we know that 
\beq\label{G2lra2}
s_{12}(\tilde{\al})=\al_2,\quad\mbox{or}\quad \tilde{\al}=s_{21}(\al_2) .
\eeq That is,
\beq
s_{\tilde{\al}}=s_{s_{21}(\al_2)}=s_{21}s_2s_{12},
\eeq
where we have used Equation \eqref{sr}. Finally,
\beq\label{t0G2}
t_{\al_0}=s_{\tilde{\al}}s_{0}=s_{21212}s_0.
\eeq
Action of $t_{\al_0}$ on $X_1$ is given by Equation \eqref{tjf}:
\begin{equation}\label{tlfG2}
t_{\al_0}(f)=f-\pi(\oc{\tilde{\al}})=f-h_2, \quad f\in X_1,
\end{equation}
where we have used Equation \eqref{hrG2} for the last equality.
That is, $t_{\al_0}$ translate on $X_1$ by $-h_2$.\\
On $\{\al_1, \al_2, \al_0\}$ basis of $V^{(1)}$ actions of $t_{\al_0}$ are given by Equation \eqref{tjal}
with $C(G_2^{(1)})$ from Equation \eqref{CarG2a},
\beq\label{tlvalG2}
t_{\al_0}:\{\al_1, \al_2, \al_0\}\mapsto
\{\al_1, \al_2+\de,\al_0-2\de\}.
\eeq
Instead of using Proposition \ref{tbfv}
to get an expression for $t_2=t_{\al_2+\de}$ we make use of Equations \eqref{Fwh}
and \eqref{G2lra2}.
Observe that
\beq
t_{\tilde{\al}+\de}=\left(t_{-\tilde{\al}+\de}\right)^{-1}=t_0^{-1},
\eeq and
\beq
t_2=t_{\al_2+\de}=t_{s_{12}(\tilde{\al}+\de)}=s_{12}t_0^{-1}s_{21}.
\eeq
Then by Equation \eqref{t0G2} we have,
\beq
t_2=s_{12}t_0^{-1}s_{21}=s_{12}s_{021212}s_{21}=s_{12021},
\eeq
where we have used $s_1^2=s_2^2=1$ to simplify the last expression.

Actions of $t_2$ on $X_1$ and on $V^{(1)}$ are given by Equations \eqref{tjf} and \eqref{tjal}, respectively,  for $j=2$,
\begin{equation}\label{t2fG2}
t_2(f)=f+\pi(\oc\al_2)=f-h_1+2h_2,\quad f\in X_1
\end{equation}
and
\beq\label{t2valG2}
t_{2}:\{\al_1, \al_2,\al_0\}\mapsto
\{\al_1+\de, \al_2-2\de, \al_0-\de\},
\eeq
where we have used $C(G_2^{(1)})$ from Equation \eqref{CarG2a}.
\item{Translations associated with short roots of $\Phi(G_2^{(1)})$.}
We look at two such elements, $t_{\tilde{\al}_s+\de}$ and $t_1$.

First observe that $t_{\tilde{\al}_s+\de}=t_{-\tilde{\al}_s+\de}^{-1}$ and
\begin{equation}
-\tilde{\al}_s+\de=-\al_{112}+\al_{011122}=\al_{012}.
\end{equation} 
Moreover, we have 
\begin{align}
    s_{12}(\al_1)&=\tilde{\al}_s,\\[-2.5em]\nonumber
 \intertext{and}
    s_{02}(\al_1)&=\al_{012},
\end{align} hence
\begin{align}
    s_{\tilde{\al}_s}&=s_{12}s_1s_{21},\\[-2.5em]\nonumber
 \intertext{and}
    s_{\al_{012}}&=s_{02}s_1s_{20}.
\end{align}

Then by Proposition \ref{tbfv}, on letting $\be= -\tilde{\al}_s+\de=\al_{012}$,
that is $\al=-\tilde{\al}_s$ and $m=1$,
we have
\begin{align}\nonumber
t_{-\tilde{\al}_s+\de}&=t_\be=s_{\al}s_{\be}
=s_{-\tilde{\al}_s}s_{-\tilde{\al}_s+\de}\\
&=s_{\tilde{\al}_s}s_{\al_{012}}
=s_{12121}s_{02120}.
\end{align}
Finally, we have
\begin{equation}\label{thrsG2}
t_{\tilde{\al}_s+\de}=t_{-\tilde{\al}_s+\de}^{-1}=\left(s_{-\tilde{\al}_s}s_{-\tilde{\al}_s+\de}\right)^{-1}
=\left(s_{\tilde{\al}_s}s_{\al_{012}}\right)^{-1}
=s_{\al_{012}}s_{\tilde{\al}_s}=s_{02120}s_{12121}.
\end{equation}
Actions of $t_{\tilde{\al}_s+\de}$ on $X_1$ are given by Equation \eqref{tbf}:
\begin{equation}\label{tsfG2}
t_{\tilde{\al}_s+\de}(f)=f+\pi(\oc{\tilde{\al}_s})=f+h_1,\quad f\in X_1,
\end{equation}
where we have used Equation \eqref{hsrG2} for the last equality.
That is, $t_{\tilde{\al}_s+\de}$ translates on $X_1$ by $h_1$. 
On $V^{(1)}$ $t_{\tilde{\al}_s+\de}$ acts by Equation \eqref{tbal}:
\beq\label{tsvalG2}
t_{\tilde{\al}_s+\de}:\{\al_1, \al_2, \al_0\}\mapsto
\{\al_1-\de, \al_2, \al_0+3\de\}.
\eeq
For an expression of $t_1=t_{\al_1+\de}$ we make use of Equation \eqref{Fwh}.
Recall that,
\beq
\al_1=s_{21}(\tilde{\al}_s).
\eeq
Then we have,
\beq
t_1=t_{\al_1+\de}=t_{s_{21}(\tilde{\al}_s+\de)}=
s_{21}t_{\tilde{\al}_s+\de}s_{21}=s_{21}s_{02120}s_{12121}s_{12}
=s_{2102120}s_{121},
\eeq
where we had the expression of $t_{\tilde{\al}_s+\de}$ given by Equation \eqref{thrsG2}, and
$s_1^2=s_2^2=1$ was used to simplify the last equality.

Actions of $t_1$  on $X_1$ and $V^{(1)}$ are given by Equations \eqref{tjf} and \eqref{tjal} for $j=1$:
\begin{equation}\label{t1fG2}
t_1(f)=f+\pi(\oc\al_1)=f+2h_1-3h_2,\quad f\in X_1,
\end{equation} where we have used Equation \eqref{hpaG2}
and
\beq\label{t4valG2}
t_{1}:\{\al_1, \al_2, \al_0\}\mapsto
\{\al_1-2\de, \al_2+3\de, \al_0\},
\eeq
with $C(G_2^{(1)})$ given by Equation \eqref{CarG2a}.
\end{enumerate}
\section{Extended affine Weyl groups }\label{EAW}
Previously in Section \ref{AW}, we have seen how $W^{(1)}$ decomposes into a semi-direct product of
the finite Weyl group $W$ and an abelian group of translations on the root lattice 
$Q$, $W^{(1)}=W\ltimes Q$. 
Generators of $Q$, $t_j$ are associated with
translation on $X_1$ by $\pi(\oc\al_j)$ ($1\leq j \leq n$).
In order to describe translations on the weight lattice $P$ we construct certain extensions of ${W}^{(1)}$, $\widetilde{W}^{(1)}=W\ltimes P$.

For affine Weyl groups of type $A_n$,$B_n$, $C_n$, $D_n$, $E_6$ and $E_7$,
$P/Q$ amounts to some group $A$ of diagram automorphsims of
the affine Dynkin diagram ${\Ga}^{(1)}$. The group $A$ preserves the simple system $\De^{(1)}$ and hence
normalises ${W}^{(1)}$. That is, we have $\widetilde{W}^{(1)}=W\ltimes P=A\ltimes{W}^{(1)}$.

For types $E_8$,
$F_4$ and $G_2$, $P$ is isomorphic to $Q$ since the affine Dynkin diagrams for these types have no non-trivial diagram automorphisms.
{\Def\label{uj}
For each $1\leq j \leq n$, define a linear transformation\\
$U_j=R(u_j)=F_{h_j}:V^{(1)*}\to V^{(1)*}$ by,   
\begin{equation}\label{RFhj}
\left(R(u_j)\right)(f)=F_{h_j}(f)=f+\lan\de,f\ran h_j\quad\mbox{for each}\quad h_j\in X_0\quad 1\leq j \leq n,
\end{equation}
for all $f\in V^{(1)*}$. Furthermore, let $U$ be 
a free abelian group generated by $\{u_j\, \mid 1\leq j\leq n\}$, 
\beq\label{PtoUd}
U=\lan u_j\, \mid 1\leq j\leq n\ran.
\eeq
}
The weight lattice $P$, which  is the integer span of $\{h_j| 1\leq j\leq n\}$, is isomorphic to $U$ by the map: 
\beq\label{PtoU}
\sum_{i=1}^n n_ih_i\mapsto \prod_{i=1}^n u_i^{n_i},\quad n_i\in\mathbb{Z}.
\eeq
We define an action of ${W}^{(1)}$ on $U$ by requiring 
\begin{equation}\label{Fwuj}
R(wu_jw^{-1})=F_{w(h_j)}=F_{\sum_{i=1}^n n_{ij}h_i}=\prod_{i=1}^n U_i^{n_{ij}}=R\left(\prod_{i=1}^n u_i^{n_{ij}}\right)
\end{equation}
for $1\leq j\leq n$, and $w\in W^{(1)}$.
That is, if we have $w(h_j)=\sum_{i=1}^n n_{ij}h_i$ for some $n_{ij}\in\mathbb{Z}$ then,
\begin{equation}\label{ujnj}
wu_jw^{-1}=\prod_{i=1}^n u_i^{n_{ij}}.
\end{equation}
{\Def
Let the extended affine Weyl group $\widetilde{W}^{(1)}$ be the semi-direct product
of $U$ by $W$,
\beq
\widetilde{W}^{(1)}=W\ltimes U=W\ltimes\lan u_j\, \mid 1\leq j\leq n\ran =W\ltimes P,
\eeq where $R$ is extended to a representation of $\widetilde{W}^{(1)}$ on $V^{(1)*}$.
}
Equation \eqref{RFhj}
tells us how $u_j$ acts on $V^{(1)*}$.

{\prop\label{ujf}
Action of $u_j\in \widetilde{W}^{(1)}$ on $V^{(1)*}$ is given by
\begin{equation}\label{ujfe}
u_j(f)=f+\lan\de,f\ran h_j,\quad 1\leq j \leq n,
\end{equation} for all $f\in V^{(1)*}$.
In particular, actions of $u_j$ on $\{h_1, h_2, ..., h_n, h_{\de}\}$ basis
of $V^{(1)*}$ are particularly simple. We have
\beq\label{ujhi}
u_j(h_i)=h_i,\quad 1\leq i \leq n\quad\mbox{and}\quad u_j(h_\de)=h_\de+h_j,\quad
\quad 1\leq j \leq n.
\eeq
}
\begin{proof}
For Equation \eqref{ujhi} we have used the facts $\lan \de,h_i\ran=0$ ($1\leq i \leq n$)
and $\lan \de,h_\de\ran=1$.
\end{proof}

It remains to
determine, via the contragredient actions of $R$, how the generators $u_j$ of $U$ act on $V^{(1)}$.
{\prop\label{ujv}
Action of $u_j\in \widetilde{W}^{(1)}$ on $v\in V^{(1)}$ is given by
\begin{equation}\label{ujve}
u_j(v)=v-\lan v,h_j\ran\de,\quad 1\leq j \leq n.
\end{equation}
In particular, we have for $1\leq i,j\leq n$,
\beq\label{ujai}
u_j(\al_i)=
 \begin{cases}
 \al_i, &\text{for}\quad i\neq j,\\
 \al_j-\de, &\text{for}\quad i=j,\\
 \end{cases}
 \eeq
 and
 \beq\label{uja0}
u_j(\al_0)=\al_0+c_j\de,
 \eeq
where $c_j$ is the coefficient of $\al_j$ in $\de$.
}
\begin{proof} For each $j$ and $v\in V^{(1)}$ we have,
\begin{align*}
    \lan u_j^{-1}(v),f\ran&=\lan v,R(u_j)(f)\ran,\\
    &=\lan v,F_{h_j}(f)\ran,\\
    &=\lan v,f+\lan\de,f\ran h_j\ran,\\
    &=\lan v,f\ran+\lan\de,f\ran \lan v,h_j\ran,\\
    &=\lan v+\lan v,h_j\ran\de,f\ran,
\end{align*}
that is, $u_j^{-1}(v)=v+\lan v,h_j\ran\de$, or
\beq
u_j(v)=v-\lan v,h_j\ran\de.
\eeq
Actions of $u_j$ on $\{\al_1, \al_2, ..., \al_0\}$ basis
of $V^{(1)}$ are given by,
\begin{equation}\label{ujal}
u_{j}(\al_i)=\al_i-\lan \al_i,h_j\ran\de, \quad\mbox{for all}\quad 0\leq i\leq n,\; 1\leq j\leq n.
\end{equation}
For Equations \eqref{ujai} and \eqref{uja0} we have used the facts $\lan \al_i,h_j\ran=0$
for $i\neq j$, $\lan \al_j,h_j\ran=1$, and $\lan \al_0,h_j\ran=-c_j$
for $1\leq i,j\leq n$.
\end{proof}

{\prop 
$P$ is preserved by ${W}^{(1)}$. That is
\beq\label{wh}
w(h_j)=\sum\limits_{i=1}^n n_{ij}h_i,\quad w\in {W}^{(1)}, \quad\mbox{for some}\quad n_{ij}\in\mathbb{Z}.
\eeq

In particular we have for $1\leq i,j\leq n$,
\beq\label{sihjn}
s_i(h_j)=
 \begin{cases}
 h_j, &\text{for}\quad i\neq j,\\
 h_j-\sum\limits_{k=1}^n a_{ki}h_k, &\text{for}\quad i=j,\\
 \end{cases}
 \eeq
 and
 \beq\label{s0hj}
s_0(h_j)=h_j+c_j\sum\limits_{k=1}^n a_{k0}h_k,
 \eeq
where $c_i$'s are the coefficients of $\al_i$ in $\de$, and $a_{ij}$ are the entries of $C(\Ga^{(1)})$.
}
\begin{proof}
From Equation \eqref{sbf} for $1\leq j\leq n$, $0\leq i\leq n$  we have,
\beq\label{sihj}
s_i(h_j)=h_j-\lan \al_i,h_j\ran\pi(\oc \al_i),
\eeq
with $\pi(\oc \al_i)$ expressed in terms of $h_k$ ($1\leq k\leq n$) given by Proposition \ref{pa0h}.
For Equations \eqref{sihjn} and \eqref{s0hj}
 we have used $\lan \al_i,h_j\ran=0$ for $i\neq j$;
$\lan \al_j,h_j\ran=1$, and $\lan \al_0,h_j\ran=-c_j$ for for $1\leq j\leq n$.
The coefficients $c_i$ and $a_{ij}$ are all integers. That is, the generators 
$s_i\;(0\leq i \leq n)$
and hence ${W}^{(1)}$ preserve $P$.
\end{proof}
Earlier in Section \ref{TX1}, for discussions on $t_j\in {W}^{(1)}$, that is elements of translations on $X_1$ by 
$\pi(\oc \al_j)$,
we used the fact that all long (short) roots belong to
the same $W$-orbit of $\tilde{\al}$ ($\tilde{\al}_s$). 
For discussions on $u_j\in \widetilde{W}^{(1)}$, that is elements of translations on $X_1$ 
by $h_j$
we compute the orbit of $h_j$ under
the finite Weyl group $W$
in the following proposition.

{\prop\label{orbWP}
Each $h_j$ forms a single $W$-orbit, $Wh_j$, for $1\leq j \leq n$.
In particular,

\begin{equation}\label{orbWe}
|Wh_j|=\frac{|W|}{|W_{h_j}|},
\end{equation}
where $W_{h_j}=\lan s_{\al_i}\mid\al_i\in\De, \mbox{for}\;i\neq j\ran$ is the stabilizer of $h_j$ in $W$. 
}

\begin{proof}
First, let us recall the coset theorem of a group.

A group $G$ acts transitively on a set $X$. For any $x\in X$, let $Gx$ be the 
orbit of $x$ under $G$, and $G_x$ denotes the stabilizer of $x$ in $G$, that is
$G_x=\{w\in G\mid wx=x, x\in G\}$. We have
\begin{equation}\label{orbG}
|G|=|Gx||G_x|,
\end{equation}
where $|G|$ denotes the order of the group $G$.

Now let $G=W$, $x=h_j$ and we have $|W|=|Wh_j||W_{h_j}|$,
where $W_{h_j}$ is the stabilizer of $h_j$ in $W$ and
$Wh_j$ is the $W$-orbit of $h_j$.
Computations of $W_{h_j}$ are particularly easy
since by Equation \eqref{sihjn}, $h_j$ is stabilised by all the simple reflections
associated with nodes that are not connected to the node $j$ in the Dynkin diagram. Then different $h_j$ can not have the same stabilizer
 hence  each $h_j$ belongs to a different $W$-orbit, $Wh_j$, for $1\leq j \leq n$.
\end{proof}

Since $W$ preserves length,
all vectors in the same weight orbit have the same length. The other direction
of this statement is not true, that is, two vectors of the same length may or may not be 
in the same weight orbit. Recall that by Proposition \ref{blh}, squared lengths of $h_j$'s
are given by the diagonal entries of $C(\Ga^{(1)})$, which are listed in Table \ref{basic} of Appendix \ref{list}.
For example, we see that the two fundamental weights of $W(E_6)$, $h_2$ and $h_6$ both have squared lengths of $4/3$ 
but they belong to different orbits under the actions of $W(E_6)$.

{\Rem{Shortest weights and basic translations.}\label{sw}
Consider the shortest weight/s given in Table \ref{basic} of Appendix \ref{list}. We see that for type $A_n$,
they are $h_1$ and $h_n$; for type $B_n$,
it is $h_1$; for type $C_n$, it is $h_n$;
for type $D_n$,
they are $h_{n-1}$ and $h_n$; for type $E_6$,
they are $h_2$ and $h_6$;for type $E_7$,
it is $h_7$; for type $E_8$,
it is $h_8$; for type $F_4$,
it is $h_1$; and  for type $G_2$,
it is $h_2$. 
}

{\Def\label{swtran}
Let $h_j$ (for some $j$ or $j$'s) be the shortest of the fundamental weights of a type $\Ga^{(1)}$ system,
the corresponding element of translation is $u_j\in \widetilde{W}(\Ga^{(1)})$. we call
a translation $T\in \widetilde{W}(\Ga^{(1)})$ {\it basic} if 
\begin{equation}\label{Tfh}
T(f)=f+h,\quad f\in X_1,\; h\in Wh_j.
\end{equation}
A translation is basic if it translates by an element of the $W$-orbit of $h_j$.
Moreover, since $h\in Wh_j$ there is some $w\in W$, such that $w(h_j)=h$ and
we have $T=wu_jw^{-1}$.
}

Translation, $T$, is ``basic'' in the sense that any other translations on the weight and root lattices 
of type $\Ga^{(1)}$ system can be
obtained from $T$ by conjugation and compositions of elements of $\widetilde{W}(\Ga^{(1)})$.
All mappings that give rise to discrete \Pa equations
in Sakai's classification \cite{sak:01}
correspond to some basic translations. We discuss such a system of $E_8^{(1)}$ type in Section \ref{E8e}.

Now we give an example of how weight orbits in a Weyl group can be computed and used.

{\eg\label{BOrbh12}
Previously in Remark \ref{Bnr}, we stated that 
there are $2n(n-1)$ long roots and $2n$ short roots in the
root system $\Phi$ of $B_n$ type.
Now we make use of Proposition \ref{orbWP} to show this.

For long roots, first observe that the orbit of $\pi(\oc{\tilde{\al}})=h_2$ (see Table \ref{basic} of Appendix \ref{list})
under $W(B_n)$ has the same carnality as the orbit of $\tilde{\al}$ under $W(B_n)$ (that
is, the number of long roots in $\Phi$). We compute the orbit of $h_2$ under $W(B_n)$
using Proposition \ref{orbWP}. First, the stabilizer of $h_2$ in $W(B_n)$ is given by

\begin{align}\nonumber
W(B_n)_{h_2}&=\lan s_{\al_i}\mid\al_i\in\De,\;\mbox{for}\;i\neq 2\ran,\\\label{stbh2}
&\cong W(A_1)\times W(B_{n-2}),
\end{align}
where $\De$ is the simple system of $B_n$.
Then,
\begin{align}\nonumber
  |W(B_n)h_2|&=\frac{|W(B_n)|}{|W(A_1)\times W(B_{n-2})|}, \\
  &=\frac{2^n n!}{2*2^{n-2}(n-2)!},\\\nonumber
  &=2n(n-1),
\end{align}
where we have used the orders of Weyl groups of type $A_n$ and $B_n$ given in Table \ref{basic} of Appendix \ref{list}. That is, there are $2n(n-1)$ long roots in $\Phi$ of $B_n$ type.

For short roots in $\Phi$, recall that $\pi(\oc{\tilde{\al}_s})=2h_1$. So one just needs to find
the size of the orbit of $h_1$ under $W(B_n)$.
The stabilizer of $h_1$ in $W(B_n)$ is given by

\begin{align}\nonumber
W(B_n)_{h_1}&=\lan s_{\al_i}\mid\al_i\in\De, \mbox{for}\;i\neq 2\ran\\\label{stbh1B}
&\cong W(B_{n-1}),
\end{align}
so we have
\begin{align}\nonumber
  |W(B_n)h_1|&=\frac{|W(B_n)|}{|W(B_{n-1})|}, \\
  &=\frac{2^n n!}{2^{n-1}(n-1)!},\\\nonumber
  &=2n.
\end{align}
 That is, there are $2n$ long roots in $\Phi$ of $B_n$ type.
}
\subsection{A normal subgroup of translations, $\widetilde{W}^{(1)}=W\ltimes P=A\ltimes{W}^{(1)}$}
It can be shown that $\widetilde{W}^{(1)}$ is
decomposed as a semidirect product of ${W}^{(1)}$ by the group of Dynkin diagram automorphisms
$A$, that is
$\widetilde{W}^{(1)}=A\ltimes{W}^{(1)}$.
That is, elements of $\widetilde{W}^{(1)}$ can be written in the form $aw$ with
$a\in A$ and $w\in {W}^{(1)}$. To explicitly write down 
translations $u_j\in \widetilde{W}^{(1)}$ on the weight lattice $P$,
one needs to work out the elements of $A$. Definition \ref{Nuaw} can be extended to the case of $\widetilde{W}^{(1)}$. 

{\Def\label{Nueaw}
For each $u\in\widetilde{W}^{(1)}$, define 
\beq
N(u)=\{\al\in \Phi^{(1)}_{+}\mid u(\al)\in \Phi^{(1)}_{-}\}.
\eeq
That is, $N(u)$ is the set of positive roots that
$u$ takes to some negative roots.
}
The group $A$ preserves the
simple system $\De^{(1)}$, acting on the simple roots as permutations.
Hence elements of $A\subset\widetilde{W}^{(1)}$ are characterised by 
having the property, that $N(a)$ is the empty set for any $a\in A$.

Definition \ref{Nueaw} and Equation \eqref{lfun} provide us with a strategy for writing any element $u\in\widetilde{W}^{(1)}$
in the form
\beq\label{uaw}
u=aw=as_{l_k}...s_{l_1},\quad l_1, ..., l_k\in\{0,1,...,n\},\quad a\in A,
\quad w\in{W}^{(1)},
 \eeq
where we have $l(w)=k$ and $s_j$ are the simple reflections of ${W}^{(1)}$.
We now illustrate properties of $\widetilde{W}^{(1)}$ discussed in Section \ref{EAW}
for $B_3$ and $C_3$ type systems. Recall that for types $F_4$ and $G_2$,
the two lattices $P$ and $Q$ are 
isomorphic so discussions on translations have been done in Sections \ref{TaF4} and \ref{TaG2},
respectively.
$\widetilde{W}^{(1)}=W\ltimes P=A\ltimes{W}^{(1)}$
\subsection{Translations on the weight lattice of $\widetilde{W}(B_3^{(1)})
=W(B_3)\ltimes P=A\ltimes{W}(B_3^{(1)})$.}\label{B3h}
\begin{figure}[ht]
\centering
\begin{tikzpicture}[scale=1]
\node  (a1) {};
\node [right=of a1](a2){$\circ$};
\node [right=of a2](a5){$\circ$};

\node [above=of a1](a10) {$\circ$};
\node [below=of a1](a11) {$\circ$};
\node [left=of a1](an){};
\draw (a5) node [anchor=north] {$3$};
\draw (a2) node [anchor=north] {$2$} ;
\draw (a10) node [anchor=east] {$0$} ;
\draw (a11) node [anchor=east] {$1$} ;
\draw (a1) node [anchor=east] {$\sigma$} ;
\draw (a1) node {$\curvearrowupdown$};
\draw[double distance=1.5pt] (a2) -- node {} (a5);
\draw[-] (a2) --  (a10);
\draw[-] (a2) -- (a11);
\node (b) at ($(a2)!0.5!(a5)$) {$>$};.
\end{tikzpicture}
%
\caption{Dynkin diagram of affine $B_3$ type with the diagram automorphism, $\widetilde\Ga(B_3^{(1)})$.}\label{rsB3ae}
\end{figure}
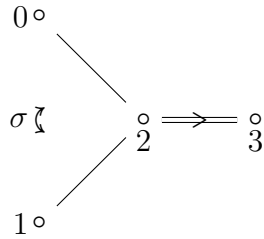

Here we find the group $A$ and write down explicit expressions for 
$u_j\, (1\leq j\leq 3)$ in terms of the generators of $\widetilde{W}(B_3^{(1)})$. 
Recall that from Equation \eqref{hijB3} in Example \ref{pB3}, we have
$|h_j|^2$ for $1\leq j\leq 3$ are $1$, $2$ and $3$, respectively. Moreover, by Equation \eqref{SymB3d},
the two squared lengths for the coroots are $2$ and $4$.

It is useful to recall that $\de=\al_{012233}$, that is $c_1=1$, $c_2=2$, $c_3=2$ for the applications of Proposition \ref{ujv} in this section.
\begin{enumerate}
\item Action of $u_1$  on $X_1$ is given by Proposition \ref{ujf},
\begin{equation}\label{u1fB3}
u_1(f)=f+h_1,\quad f\in X_1.
\end{equation}

By Proposition \ref{ujv},
$u_1$ acts on $\{\al_1, \al_2, \al_3, \al_0\}$ basis of $V^{(1)}$ by,
\beq\label{u1valB3}
u_{1}:\{\al_1, \al_2, \al_3, \al_0\}\mapsto
\{\al_1-\de, \al_2, \al_3, \al_0+\de\}.
\eeq
We see that $u_1(\al_1)=\al_1-\de=-\al_{02233}\in \Phi^{(1)}_{-}$, hence $l(u_1s_1)=l(u_1)-1$ by 
Equation \eqref{lfun}. For $u_1s_1$, we have
\beq
u_{1}s_1:\{\al_1, \al_2, \al_3, \al_0\}\mapsto
\{\al_{02233}, -\al_{0233}, \al_3, \al_0+\de\},
\eeq
where the actions of $s_1$ are given by Equation \eqref{sij0} with $a_{ij}$
being the $(i,j)$-entry of $C(B_3^{(1)})$
given by Equation \eqref{CarB3a}. Here we see that $u_1s_1(\al_2)=-\al_{0233}\in \Phi^{(1)}_{-}$, hence $l(u_1s_1s_2)=l(u_1s_1)-1$ by 
Equation \eqref{lfun}. Continuing in this way we have,
\beq
u_{1}s_1s_2:\{\al_1, \al_2, \al_3, \al_0\}\mapsto
\{\al_{2}, \al_{0233},-\al_{023}, \al_{012}\},
\eeq
\beq
u_{1}s_1s_2s_3:\{\al_1, \al_2, \al_3, \al_0\}\mapsto
\{\al_{2}, -\al_{02},\al_{023}, \al_{012}\},
\eeq
\beq
u_{1}s_1s_2s_3s_2:\{\al_1, \al_2, \al_3, \al_0\}\mapsto
\{-\al_{0}, \al_{02},\al_{3}, \al_{1}\},
\eeq
\beq
u_{1}s_1s_2s_3s_2s_1=\sigma:\{\al_1, \al_2, \al_3, \al_0\}\mapsto
\{\al_{0}, \al_{2},\al_{3}, \al_{1}\}.
\eeq
The element $\sigma=u_{1}s_1s_2s_3s_2s_1$ fixes $\al_2$ and $\al_3$, while switches $\al_0$ and
$\al_1$ (see Figure \ref{rsB3ae}), hence $\sigma^2=1$, and it is an element of $A$.
It turns out that it is the generator of $A$. We have,
\beq\label{u1B3}
u_1=\sigma s_{12321},
\eeq
where we have used the fundamental relations of ${W}(B_3^{(1)})$ given in Equation \eqref{funWB3a}.
\item Action of $u_2$  on $X_1$ is given by Proposition \ref{ujf},
\begin{equation}\label{u2fB3}
u_2(f)=f+h_2,\quad f\in X_1.
\end{equation}
By Proposition \ref{ujv},
$u_2$ acts on $\{\al_1, \al_2, \al_3, \al_0\}$ basis of $V^{(1)}$ by,
\beq\label{u2valB3}
u_{2}:\{\al_1, \al_2, \al_3, \al_0\}\mapsto
\{\al_1, \al_2-\de, \al_3, \al_0+2\de\}.
\eeq
We see that $u_2(\al_2)=\al_2-\de=-\al_{01233}\in \Phi^{(1)}_{-}$, hence $l(u_2s_2)=l(u_2)-1$ by 
Equation \eqref{lfun}. However, before launching into the length-reducing procedure we recall
the following fact about $h_2$.
By Equations \eqref{plB3} and \eqref{pa0ah} in Example \ref{pB3}
we have,
\beq
h_2
=\pi(\oc\al_1)+2\pi(\oc\al_2)+\pi(\oc\al_3)=\pi(\oc{\tilde{\al}})=-\pi(\oc\al_0),
\eeq
Which means that $u_2$ is a translation on the root lattice $Q$, in particular by Equation \eqref{t0B3}
we have,
\beq\label{u2B3}
u_2=t_{\al_0}^{-1}=t_{0}^{-1}=s_{02321232}.
\eeq
\item Action of $u_3$  on $X_1$ is given by,
\begin{equation}\label{u3fB3}
u_3(f)=f+h_3,\quad f\in X_1,
\end{equation}
and by Proposition \ref{ujv},
$u_3$ acts on $\{\al_1, \al_2, \al_3, \al_0\}$ basis of $V^{(1)}$ by,
\beq\label{u3valB3}
u_{3}:\{\al_1, \al_2, \al_3, \al_0\}\mapsto
\{\al_1, \al_2, \al_3-\de, \al_0+2\de\}.
\eeq
Applying the same length-reducing procedure we find that,
\beq
u_{3}s_3s_2s_1s_3s_2s_0s_3s_2s_1=\sigma:\{\al_1, \al_2, \al_3, \al_0\}\mapsto
\{\al_{0}, \al_{2},\al_{3}, \al_{1}\}.
\eeq
That is,
\beq\label{u3B3}
u_3=\sigma s_{123023123}.
\eeq
\end{enumerate}
Finally, we have 
\beq
\widetilde{W}(B_3^{(1)})=W(B_3)\ltimes P=W(B_3)\ltimes \lan u_j\mid 1\leq j \leq 3 \ran=A\ltimes{W}(B_3^{(1)})
=\lan\s \;|\; \sigma^2=1\ran\ltimes{W}(B_3^{(1)}).
\eeq

\subsection{Translations on the weight lattice $\widetilde{W}(C_3^{(1)})
=W(C_3)\ltimes P=A\ltimes{W}(C_3^{(1)})$.}\label{C3h}
Here we find the group $A$ and write down explicit expressions for 
$u_j\, (1\leq j\leq 3)$ in terms of the generators of $\widetilde{W}(C_3^{(1)})$. 
Recall that from Equation \eqref{hijC3} in Example \ref{pC3}, we have
$|h_j|^2$ for $1\leq j\leq 3$ are $1$, $2$ and $\frac{3}{4}$, respectively. Moreover, by Equation \eqref{SymC3d},
the two squared lengths for the coroots are $2$ and $1$.

It is useful to recall that $\de=\al_{011223}$, that is  we have $c_1=2$, $c_2=2$, $c_3=1$ for the applications of Proposition \ref{ujv}.

\begin{figure}[ht]
\centering
\begin{tikzpicture}[scale=1]
			\centering
			\node  (a1) {$\circ$};
			\node  [left=of a1](a0) {$\circ$};
			\node [right=of a1](a2) {$\circ$} ;
			\node [right=of a2](a4) {$\circ$} ;
			\draw (a1) node [anchor=north] {$1$} ;
			\draw (a2) node [anchor=north] {$2$} ;
			\draw (a4) node [anchor=north] {$3$} ;
			\draw (a0) node [anchor=north] {$0$} ;
			\draw[-] (a1) -- node {} (a2);
			\draw[double distance=1.5pt] (a2) -- node {} (a4);
			\draw[double distance=1.5pt] (a1) -- node {} (a0);
			\node (b) at ($(a4)!0.5!(a2)$) {$<$};
			\node (b) at ($(a0)!0.5!(a1)$) {$>$};
			\node (b) at (.7,.5) [rotate=-90]{$\curvearrowupdown$};
			\draw (b) node [anchor=north] {$\sigma$};
			\path[use as bounding box] (-1.5,0) rectangle (0,0); .
			\end{tikzpicture}
		\caption{Dynkin diagram of affine $C_3$ type
			with the diagram automorphism, $\widetilde\Ga(C_3^{(1)})$.}\label{rsC3ae}
\end{figure}
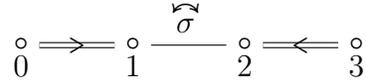
\begin{enumerate}
\item Action of $u_1$  on $X_1$ is given by Proposition \ref{ujf},
\begin{equation}\label{u1fC3}
u_1(f)=f+h_1,\quad f\in X_1.
\end{equation}

By Proposition \ref{ujv},
$u_1$ acts on $\{\al_1, \al_2, \al_3, \al_0\}$ basis of $V^{(1)}$ by,
\beq\label{u1valC3}
u_{1}:\{\al_1, \al_2, \al_3, \al_0\}\mapsto
\{\al_1-\de, \al_2, \al_3, \al_0+2\de\}.
\eeq
By Equations \eqref{hsrC3} and \eqref{pa0ah} in Example \ref{pC3} we have
\beq
h_1
=\pi(\oc\al_1)+\pi(\oc\al_2)+\pi(\oc\al_3)=\pi(\oc{\tilde{\al}})=-\pi(\oc\al_0),
\eeq
Which means that $u_1$ is a translation on the root lattice $Q$, in particular by Equation \eqref{t0C3} in Example \ref{TaC3}
we have,
\beq\label{u1C3}
u_1=t_{\al_0}^{-1}=t_{0}^{-1}=(s_{123210})^{-1}=s_{012321}.
\eeq
\item Action of $u_2$  on $X_1$ is given by Proposition \ref{ujf},
\begin{equation}\label{u2fC3}
u_2(f)=f+h_2,\quad f\in X_1.
\end{equation}

By Proposition \ref{ujv},
$u_2$ acts on $\{\al_1, \al_2, \al_3, \al_0\}$ basis of $V^{(1)}$ by,
\beq\label{u2valC3}
u_{2}:\{\al_1, \al_2, \al_3, \al_0\}\mapsto
\{\al_1, \al_2-\de, \al_3, \al_0+2\de\}.
\eeq
By Equation \eqref{hrC3}we have
\beq
h_2
=\pi(\oc\al_1)+2\pi(\oc\al_2)+2\pi(\oc\al_3)=\pi(\oc{\tilde{\al}_s}),
\eeq
Which means that $u_2$ is a translation on the root lattice $Q$, in particular by Equation \eqref{tsC3} in Section \ref{TaC3}
we have,
\beq\label{u2C3}
u_2=t_{\tilde{\al}_s+\de}=s_{010}s_{2321232}.
\eeq
\item Action of $u_3$  on $X_1$ is given by Proposition \ref{ujf},
\begin{equation}\label{u3fC3}
u_3(f)=f+h_3,\quad f\in X_1.
\end{equation}

By Proposition \ref{ujv},
$u_3$ acts on $\{\al_1, \al_2, \al_3, \al_0\}$ basis of $V^{(1)}$ by,
\beq\label{u3valC3}
u_{3}:\{\al_1, \al_2, \al_3, \al_0\}\mapsto
\{\al_1, \al_2, \al_3-\de, \al_0+\de\}.
\eeq
Applying the length-reducing procedure using Equation \eqref{lfun} we find that
\beq
u_{3}s_3s_2s_3s_1s_2s_3=\sigma:\{\al_1, \al_2, \al_3, \al_0\}\mapsto
\{\al_{2}, \al_{1},\al_{0}, \al_{1}\}.
\eeq
That is, the element $\sigma=u_{3}s_3s_2s_3s_1s_2s_3$ switches $\al_0$ and 
$\al_3$; $\al_1$ and 
$\al_2$ (see Figure \ref{rsC3ae}), hence $\sigma^2=1$, and it is the generator of $A$, and
\beq\label{u3C3}
u_3=\sigma s_{321323}.
\eeq
\end{enumerate}
Finally, we have 
\begin{align}\label{CPA}
    \widetilde{W}(C_3^{(1)})&=W(C_3)\ltimes P=W(C_3)\ltimes \lan u_j\mid 1\leq j \leq 3 \ran,\\\nonumber
    &=A\ltimes{W}(C_3^{(1)})
=\lan\s \;|\; \sigma^2=1\ran\ltimes{W}(C_3^{(1)}).
\end{align}
\section{Applications}\label{Dis}
Here, we illustrate the power of the formulas developed in the earlier sections of the paper in 
studying different integrable systems recently appeared in the literature. 

\subsection{Two discrete \Pa systems of type $E_8^{(1)}$.} \label{E8e}
In \cite{NN_elliptic}, an elliptic \Pa equation associated with a translation $T_{J,1}\in W(E_8^{(1)})$ was constructed.
$T_{J,1}$ acts on the $\{\al_j\mid 0\leq j\leq 8\}$ basis of $V^{(1)}$ by
\beq\label{TJ1}
T_{J,1}:\{\al_1, \al_6\}\mapsto
\{\al_1-2\de, \al_6+\de\}.
\eeq
On the other hand, the element $T_1\in W(E_8^{(1)})$ which gives
Sakai's $e$-{\bf P}$(E_8^{(1)})$ equation \eqref{ePE8} acts
on $\{\al_j\mid 0\leq j\leq 8\}$ by
\beq\label{TJ2}
T_{1}:\{\al_1, \al_3\}\mapsto
\{\al_1-2\de, \al_3+\de\}.
\eeq
In Equation \eqref{T1form}, we have shown that $T_1$ as defined in Equation \eqref{T1decomp} is indeed a translation by $\al_1$
in $W(E_8^{(1)})$.

Here we show that $\al_1$, a vector of squared length $2$ is in the $W(E_8)$-orbit of $h_8$, the shortest of fundamental weights of $E_8$ type, 
hence $T_{1}$ and any of its conjugations by $W(E_8)$ correspond
to a translation by one of the 240 roots of the $W(E_8)$ root system and hence are basic (by Definition \ref{swtran}). 
Then we identify the vector associated with the translation $T_{J,1}$
and express it as a vector sum of $\al_1$ and another root from the $W(E_8)$ root system.
This allows us to express $T_{J,1}$ as a composition of two basic translations in $W(E_8^{(1)})$. However, before doing so we need to lay down some basic properties of the group
$W(E_8^{(1)})$.

The Dynkin diagram of type $E_8^{(1)}$, $\Ga(E_8^{(1)})$ is given in Figure \ref{claW}.
The corresponding $E_8^{(1)}$ simple system, $\De^{(1)}=\{\al_j\mid 0\leq j\leq 8\}$ forms a basis for an $9$-dimensional real vector space
$V^{(1)}$ equipped with a semidefinite symmetric positive bilinear form given by Equation
\eqref{alaij0} using the generalised Cartan matrix of type $E_8^{(1)}$,
\beq\label{CarE8a}
C(E_8^{(1)})=(a_{ij})_{1\leq i,j\leq 8,0}=(\al_i\cdot\oc\al_j)_{1\leq i,j\leq 8,0}=\left(
\begin{array}{ccccccccc}
 2 & 0 & -1 & 0 & 0 & 0 & 0 & 0 & 0 \\
 0 & 2 & 0 & -1 & 0 & 0 & 0 & 0 & 0 \\
 -1 & 0 & 2 & -1 & 0 & 0 & 0 & 0 & 0 \\
 0 & -1 & -1 & 2 & -1 & 0 & 0 & 0 & 0 \\
 0 & 0 & 0 & -1 & 2 & -1 & 0 & 0 & 0 \\
 0 & 0 & 0 & 0 & -1 & 2 & -1 & 0 & 0 \\
 0 & 0 & 0 & 0 & 0 & -1 & 2 & -1 & 0 \\
 0 & 0 & 0 & 0 & 0 & 0 & -1 & 2 & -1 \\
 0 & 0 & 0 & 0 & 0 & 0 & 0 & -1 & 2 \\
\end{array}
\right).
\eeq
The defining relations of $W(E_8^{(1)})=\lan s_i\mid 0\leq i \leq 8\ran$ 
can be read off from $\Ga(E_8^{(1)})$ with the rules
given in Table \ref{CD}.
Generators $s_j\in W(E_8^{(1)})$ act on $V^{(1)}$ by Equation \eqref{sij0}, where $a_{ij}$ is the $(i,j)$-entry
of $C(E_8^{(1)})$ from Equation \eqref{CarE8a}. 
We have the null root and the highest root given by,
\begin{equation}\label{deE8}
\de=\al_0+\tilde{\al}=\al_0+\sum_{i=1}^{8}c_i\al_i
=\al_0+2\al_1+3\al_2+4\al_3+6\al_4+5\al_5+4\al_6+3\al_7+2\al_8.
\end{equation}
where $c_i$ $(1\leq i \leq 8)$ of $E_8^{(1)}$ type can be found in Table \ref{basic} of Appendix \ref{list}.
A dual space $V^{(1)\ast}$ with basis $\{ h_1, \ldots, h_8, h_\de\}$ is given by Definition \ref{ds},
in which we consider the subspace $X_0$ with basis $\{\pi(\oc\al_j)\mid1\leq j\leq n\}$ and the hyperplane $X_1=h_\de+X_0$ (see Definition \ref{Xk}).
Since $\Ga(E_8^{(1)})$ is simply-laced, and $C(E_8)$ is symmetric,
we can identify the simple coroots $\pi(\oc\al_j)$ with the simple roots $\al_j$ $(1\leq j\leq 8)$.
Then, by Equation \eqref{pah} we have
\beq\label{pahE8}
\al_j
=\sum_{k=1}^{8}\left(C(E_8)\right)_{kj}h_k
=\sum_{k=1}^{8}a_{kj}h_k, \quad 1\leq j\leq 8.
\eeq
By Proposition \ref{blh},
the bilinear form $({}\,,{})$ in $\{h_j\mid1\leq j\leq 8\}$ basis of $X_0$ is
\beq\label{hijE8}
\left((h_i, h_j)\right)_{1\leq i,j\leq8}
=C(E_8)^{-1}=
\left(
\begin{array}{cccccccc}
 4 & 5 & 7 & 10 & 8 & 6 & 4 & 2 \\
 5 & 8 & 10 & 15 & 12 & 9 & 6 & 3 \\
 7 & 10 & 14 & 20 & 16 & 12 & 8 & 4 \\
 10 & 15 & 20 & 30 & 24 & 18 & 12 & 6 \\
 8 & 12 & 16 & 24 & 20 & 15 & 10 & 5 \\
 6 & 9 & 12 & 18 & 15 & 12 & 8 & 4 \\
 4 & 6 & 8 & 12 & 10 & 8 & 6 & 3 \\
 2 & 3 & 4 & 6 & 5 & 4 & 3 & 2 \\
\end{array}
\right).
\eeq
Moreover, by Equation \eqref{hpa} we have
\beq\label{hpaE8}
h_i=\sum_{k=1}^8C(E_8)^{-1}_{ik}\al_k\quad
\mbox{for}\quad1\leq i\leq n,
\eeq
with $C(E_8)^{-1}$ given in Equation \eqref{hijE8}.

As it was discussed earlier, the weight lattice and the root lattice of $W(E_8^{(1)})$ are isomorphic.
The weight lattice is generated by translations of the fundamental weights
$\{u_j\mid1\leq j\leq 8\}$ (given by Propositions \ref{ujf} and \ref{ujv})  or equivalently by translations of the simple coroots
$\{t_j\mid1\leq j\leq 8\}$ (given by Proposition \ref{tj}).
Moreover, it is enough to know an element of translation by a shortest weight (a basic translation) - any other translations on the weight lattice can be obtained by conjugations or compositions of the basic translations.
We see from Equation \eqref{hijE8}
that the shortest weight is $h_8$ with $|h_8|^2=2$. 
By Equation \eqref{hpaE8}, entries of the last row of $C(E_8)^{-1}$ in Equation \eqref{hijE8} gives the coefficients of
$\al_k$ ($1\leq k \leq 8$) in $h_8$. Compare these coefficients with those of
$\al_k$ ($1\leq k \leq 8$) in $\tilde{\al}$ given in Equation \eqref{deE8} we see that $h_8=\tilde{\al}$.
That is the $W(E_8)$-orbit of $h_8$
coincides with $\Phi$, the finite $E_8$ root system.

Let us compute the size of orbit of $h_8$ under $W(E_8)$
using Proposition \ref{orbWP}. First, the stabilizer of $h_8$ in $W(E_8)$ is given by
\begin{align}\nonumber
W(E_8)_{h_8}&=\lan s_{\al_i}\mid\al_i\in\De, \quad\mbox{for}\;i\neq 8\ran,\\\label{stbh8}
&\cong W(E_{7}).
\end{align}
Then the $W(E_8)$-orbit of $h_8$ is given by,
\begin{align}\nonumber
  |W(E_8)h_8|&=\frac{|W(E_8)|}{|W(E_{7})|}, \\\nonumber
  &=\frac{2^{14}3^55^27}{2^{10}3^45\cdot7},\\\label{E8h8}
  &=2^4\cdot3\cdot5=240,
\end{align}
where we have used the orders of Weyl groups of type $E_7$ and $E_8$ given in Table \ref{basic}. That is, there are $240$ roots in $\Phi$ of $E_8$ type, as we know from again Table \ref{basic}.
Looking at the diagonal entries of the matrix in Equation \eqref{hijE8}, we see that
the weight of the next weight length is $h_1$ with $|h_1|^2=4$.
As it will be relevant to the problem we are discussing, we compute also the size of $W$-orbit of $h_1$.
First, the stabilizer of $h_1$ in $W(E_8)$ is given by

\begin{align}\nonumber
W(E_8)_{h_1}&=\lan s_{\al_i}\mid\al_i\in\De, \;\mbox{for}\;i\neq 1\ran,\\\label{stbh1}
&\cong W(D_{7}).
\end{align}
Then the $W(E_8)$-orbit of $h_1$ is given by,
\begin{align}\nonumber
  |W(E_8)h_1|&=\frac{|W(E_8)|}{|W(D_{7})|}, \\\nonumber
  &=\frac{2^{14}3^55^27}{2^{6}7!},\\\label{E8h1}
  &=2160,
\end{align}
where we have used the orders of Weyl groups of type $D_7$ and $E_8$ given in Table \ref{basic}. That is, there are $2160$ weight vectors in the  $W(E_8)$-orbit of $h_1$.

 
Now we are ready to consider the relationship between the two mappings 
given by
$T_{J,1}$ and $T_{1}$.
By Proposition \ref{ujv} and the actions of $T_{J,1}$ and $T_{1}$ given in Equations \eqref{TJ1} and
\eqref{TJ2},
we recognise that $T_{J,1}$ and $T_{1}$ are translations by $2h_1-h_6$ and $2h_1-h_3$ on $X_1$, 
respectively.

Next, we compute the squared lengths of $2h_1-h_6$ and $2h_1-h_3$ using the symmetric bilinear form given in Equation \eqref{hijE8}.
We have $|2h_1-h_6|^2=4$ and $|2h_1-h_3|^2=2$. 

By Proposition \ref{orbWP} we see that $2h_1-h_3$ must be in the $W(E_8)$-orbit
of the shortest weight $h_8$, or equivalently $\Phi$. Using Equation \eqref{pahE8} and an inspection of the Cartan matrix in Equation \eqref{CarE8a}
tells us that $\al_1=2h_1-h_3$, that is $T_{1}$ is the translation by the simple root 
$\al_1$, 
\begin{equation}
    T_{1}=t_1=t_{\al_1+\de},
\end{equation}
with $t_1$ given by $j=1$ in Proposition \ref{tj}. Moreover, $T_{1}$ and
any of its $240$ conjugations under the actions of $W(E_8)$ are all basic translations.

\begin{figure}
 \centering
   \begin{tikzpicture}[scale=1.5]
  \coordinate  (C) at (-.5,-1.cm);
  \coordinate  (A) at (1.5cm,-1.0cm);
  \coordinate  (B) at (1.5cm,1.0cm);
  \draw (C) -- node[above, rotate=45] {$\al_{11233445}$} (B) -- node[right] {$\al_{1233445}$} (A) -- node[below] {$\al_1$} (C);
 \draw (A) node {$>$};
 \draw (B) node[rotate=45] {$>$};
 \draw (B) node [anchor=east, rotate=90] {$>$};
  \draw (1.25cm,-1.0cm) rectangle (1.5cm,-0.75cm);
 \draw pic[draw, "$45^\circ$", -, angle eccentricity=1.4,angle radius = .8cm] {angle = A--C--B}; 
 \draw pic[draw, "$45^\circ$", -, angle eccentricity=1.4,angle radius = .8cm] {angle = C--B--A};
\end{tikzpicture}
    \caption{Addition of the roots in $\Phi$ that give rise to the vector 
    $2h_1-h_6=\al_{11233445}$
    associated with $T_{J,1}$.}
    \label{Py}
\end{figure}
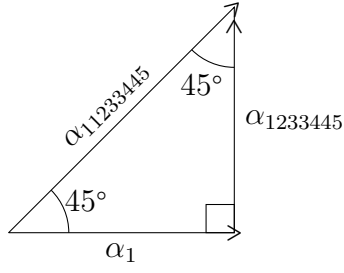

Again by Proposition \ref{orbWP}, the vector $2h_1-h_6$ (of squared length 4) is in the $W(E_8)$-orbit
of $h_1$ (see Equation \eqref{E8h1}). Moreover, it must be a vector sum of two orthogonal roots in $\Phi$, since
by Pythagoras theorem we have $(\sqrt{2})^2+(\sqrt{2})^2=2^2=4$ (See Figure \ref{Py}).

Using Equation \eqref{hpaE8} and $C(E_8)^{-1}$ in Equation \eqref{hijE8} we first
write
$2h_1-h_6$ as a sum of simple roots, and then as a sum of two orthogonal roots as follows,
\beq
2h_1-h_6=\al_{11233445}=\al_1+\al_{1233445}=2h_1-h_3+\al_{1233445}.
\eeq
The orthogonality between $\al_1$ and $\al_{1233445}$ can be checked using the bilinear form given 
in Equation \eqref{CarE8a}.

The vector $\al_{1233445}$ is a root, and hence is in $\Phi$, the $W(E_8)$-orbit of $\al_1$. We found the element in $W(E_8)$
that takes $\al_1$ to $\al_{1233445}$,
\beq\label{wE8}
w(\al_1)=s_{345243}(\al_1)=\al_{1233445}, 
\eeq
where we have used the convention of writing roots and products of simple reflections introduced in Remark \ref{rsconv}.
Then by Equation \eqref{Fwtj}, the element of translation associated to the root $\al_{1233445}$ 
is given by $wT_{1}w^{-1}$.
Finally, the element of translation associated to $\al_{11233445}=2h_1-h_6$ is given by a composition of $T_{1}$ and $wT_{1}w^{-1}$,
\beq
T_{J,1}
=T_{1}wT_{1}w^{-1},
\eeq
with $w$ given in Equation \eqref{wE8}. Moreover, the \Pa equation given by $T_{J,1}$ in \cite{NN_elliptic}
is equivalent to those given by any of its $2160$ conjugates (size of the orbit of $h_1$, computed in Equation \eqref{E8h1}) under the actions of $W(E_8)$.

\subsection{A subsystem of type $F_4^{(1)}$ in $E_8^{(1)}$.}\label{f4inE8e}
Let $\De^{(1)}=\{\al_j\mid 0\leq j\leq 8\}$
be the $E_8^{(1)}$ simple system with the numbering on $\Ga(E_8^{(1)})$ given
in Figure \ref{claW},
and $h_j$ ($1\leq j\leq 8$)
the fundamental weights.
We have
$V^{(1)}=\mbox{Span}(\De^{(1)})$, the real vector space
on which $W(E_8^{(1)})=\lan s_i\mid 0\leq i \leq 8\ran$ acts as a group of reflections.

A subsystem of type $F_4^{(1)}$, Equation  \eqref{RCG}, was found for Sakai's 
$e$-{\bf P}$(E_8^{(1)})$ equation \eqref{ePE8} in \cite{ahjn:16}.
It was verified (using MAGMA \cite{magma}) that  
the element $\varphi_a$ (given in Equation \eqref{psidecomp1}),
that gives rise to the discrete evolution in Equation  \eqref{RCG},
is an element
of a $F_4^{(1)}$ type subgroup of $W(E_8^{(1)})$.

Now we explain how such a subgroup arises from $W(E_8^{(1)})$ 
using the normalizer theory of Coxeter groups \cite{BH}. 
We shall see that although the generators of this subgroup, which are
the involutions in general, satisfy the defining relations of a Weyl group of type
$F_4^{(1)}$ they can not be realised as reflections on $V^{(1)}$.
We find a subspace in $V^{(1)}$
on which they can be realised as simple reflections of $F_4^{(1)}$ type, hence allowing the construction of
translational type elements. Finally we show that $\varphi_a$ is a translation in this $F_4^{(1)}$ subgroup, and an
element of
quasi-translation in $W(E_8^{(1)})$.

{\prop\label{f4inE8}
Take $J=\{\al_2, \al_5, \al_7, \al_0\}\subset \De^{(1)}$.
The normalizer of ${W}_J=\lan s_2, s_5, s_7, s_0\ran$ in ${W}(E_8^{(1)})$
is given by 
\begin{equation}\label{NWJf4e8}
N(W_J)=N_J\ltimes W_J=\lan b_i \mid 0\leq i \leq 4 \ran\ltimes
\lan s_2, s_5, s_7, s_0\ran
\cong W(F_4^{(1)})\ltimes {W}(4A_1),
\end{equation}  
with
\begin{align}\label{bF4}
    b_0&=s_{8708},\quad
    b_1=s_{6576},\quad
    b_2=s_{4254},\quad
    b_3=s_3,\quad
    b_4=s_1.
\end{align}
}
\begin{proof}
 Recall from Section \ref{Norm} that $N(W_J)=N_J\ltimes W_J$,
where $N_J$ generated by the R- and M-elements,
is the group of all elements of $W(E_8^{(1)})$
that act permutatively on the set $J$. Here we have 
 $J=\{\al_2, \al_5, \al_7, \al_0\}\cong 4A_1$, and
$W_J=\lan s_2, s_5, s_7, s_0\ran\cong {W}(4A_1)$. 
As $J$ is not conjugated to any other type $4A_1$ subsets of 
$\De^{(1)}$ under the actions of $W(E_8^{(1)})$, 
there is no M-elements, and we have $N_J=\lan b_i \mid 0\leq i \leq 4 \ran$ with
\begin{align}\nonumber
    b_0&=v[\al_8,J]=w_{J\cup \{\al_8\}}w_{J}=w_{A_3+2A_1}w_{4A_1}=w_{A_3}w_{2A_1}=w_{\{7,8,0\}}w_{\{7,0\}}
    =v[\al_8,\{\al_0,\al_7\}]=s_{8708},\\\nonumber
    b_1&=v[\al_6,J]=w_{J\cup \{\al_6\}}w_{J}=w_{\{5,6,7\}}w_{\{5,7\}}=v[\al_6,\{\al_5,\al_7\}]=s_{6576},\\\label{bvF4} 
    b_2&=v[\al_4,J]=w_{J\cup \{\al_4\}}w_{J}=w_{\{2,4,5\}}w_{\{2,5\}}=v[\al_4,\{\al_2,\al_5\}]=s_{4254},\\\nonumber
    b_3&=v[\al_3,J]=s_3,\quad
    b_4=v[\al_1,J]=s_1.
\end{align}
For the evaluations of $b_i$ ($0\leq i\leq 2$)
we have used the fact that when two of the simple roots in $J$ are orthogonal to the $a$ in $v[a, J]$ we have $v[a, J]=w_{J\cup {a}}w_{J}=w_{A_3+2A_1}w_{4A_1}=w_{A_3}w_{2A_1}$. 
Then we can use the expression for $w_{A_3}w_{2A_1}$
we have found from
Example \ref{2AinA3}, from which
we also know that $b_0, b_1,b_2$ are involutions and their actions on $J$.
That is, written as permutations on the index set of $J$ we have,
\begin{equation}\label{bepermJ}
    b_0=(70),
    \quad
    b_1=(57),
    \quad
    b_2=(25).
\end{equation}
That is, all the generators of $N_J$ are involutions,
$b_i^2=1$ ($0\leq i\leq 4$), with
$b_3=s_3$ and $b_4=s_1$ being also reflections. Elements $b_i$ $(0\leq i \leq 2)$ cannot be realised as  reflections (as defined in Equation \eqref{sbv})
in $V^{(1)}$, as they do not fix point wisely a co-dimension one hyperplane in $V^{(1)}$.
We now proceed to find the orders of the pairwise products of the generators of $N_J$
to identify its Coxeter type. 

First of all, from $\Ga(E_8^{(1)})$
it is obvious that
$(b_4b_3)^3=(s_1s_3)^3=1$.
Moreover, $(b_4b_2)^2=(b_0b_2)^2=1$ and $(b_ib_j)^2=1$ for $i=3,4$ and $j=1,0$,
since elements that correspond to disjoint nodes (or products of nodes
which are disjoint)
in the Dynkin diagram commute. For the order of $b_3b_2$, we
consider the subset $\{\al_3, \al_4, \al_2, \al_5\}\cong D_4$, noticing
that
\begin{equation}
  v[\al_3,\{\al_2,\al_5\}]=s_3=b_3,\quad\mbox{and}\quad v[\al_4,\{\al_2,\al_5\}]=s_{4254}=b_2.  
\end{equation}
We have from Equation \eqref{lwLJ},
\begin{equation}
 l(v[\{\al_3, \al_4\},\{\al_2, \al_5\}])=l(w_{D_4})-l(w_{2A_1})=
\mid\Phi^{+}_{D_4}\mid-2\mid\Phi^{+}_{A_1}\mid
=12-2=10,   
\end{equation}
where the value of $\mid\Phi^{+}_{D_4}\mid$ can be found in Table \ref{basic}.
By Equation \eqref{vab}, the two standard expressions of $v[\{\al_3, \al_4\},\{\al_2, \al_5\}]$ are given  by
\begin{align*}
&v[\al_3,\{\al_2,\al_5\}]v[\al_4,\{\al_2,\al_5\}]v[\al_3,\{\al_2,\al_5\}]v[\al_4,\{\al_2,\al_5\}]\\
&=v[\al_4,\{\al_2,\al_5\}]v[\al_3,\{\al_2,\al_5\}]v[\al_4,\{\al_2,\al_5\}]v[\al_3,\{\al_2,\al_5\}],
\end{align*}
since
\begin{align*}
&l(v[\al_3,\{\al_2,\al_5\}])+l(v[\al_4,\{\al_2,\al_5\}])+l(v[\al_3,\{\al_2,\al_5\}])\\\nonumber
&\hspace{10cm}+l(v[\al_4,\{\al_2,\al_5\}])\\\nonumber
&=l(v[\al_4,\{\al_2,\al_5\}])+l(v[\al_3,\{\al_2,\al_5\}])+l(v[\al_4,\{\al_2,\al_5\}])\\\nonumber
&\hspace{10cm}+l(v[\al_3,\{\al_2,\al_5\}]),\\
\mbox{or}\quad &1+4+1+4=4+1+4+1,\quad\mbox{that is we have}\quad(b_3b_2)^4=1.
\end{align*}

To find the order of $b_2b_1$ we consider the subset $\{\al_4,\al_6,\al_2,\al_5,\al_7\}\cong A_5$, noticing that
\begin{equation}
 v[\al_6,\{\al_2,\al_5,\al_7\}]=s_{6576}=b_1,\quad\mbox{and}\quad 
 v[\al_4,\{\al_2,\al_5,\al_7\}]=s_{4254}=b_2.   
\end{equation}
We have 
\begin{equation}
 l(v[\{\al_4,\al_6\},\{\al_2,\al_5,\al_7\}])
=l(w_{A_5})-l(w_{3A_1})
=\mid\Phi^{+}_{A_5}\mid-3\mid\Phi^{+}_{A_1}\mid
=15-3=12.   
\end{equation}

The two standard expressions of $v[\{\al_4,\al_6\},\{\al_2,\al_5,\al_7\}]$ are
given by,
\begin{align}\nonumber
&v[\al_6,\{\al_2,\al_5,\al_7\}]v[\al_4,\{\al_2,\al_5,\al_7\}]v[\al_6,\{\al_2,\al_5,\al_7\}]\\\nonumber
&=v[\al_4,\{\al_2,\al_5,\al_7\}]v[\al_6,\{\al_2,\al_5,\al_7\}]v[\al_4,\{\al_2,\al_5,\al_7\}],
\end{align}
since
\begin{align*}
 &l(v[\al_6,\{\al_2,\al_5,\al_7\}])l(v[\al_4,\{\al_2,\al_5,\al_7\}])l(v[\al_6,\{\al_2,\al_5,\al_7\}])\\\nonumber
&=l(v[\al_4,\{\al_2,\al_5,\al_7\}])l(v[\al_6,\{\al_2,\al_5,\al_7\}])l(v[\al_4,\{\al_2,\al_5,\al_7\}]),\\\nonumber
\mbox{or}\quad&4+4+4=4+4+4,\quad\mbox{that is we have}\quad(b_2b_1)^3=1.
\end{align*}
Similarly, by considering the subset $\{\al_5,\al_6,\al_7,\al_8,\al_0\}\cong A_5$
one can show that $(b_1b_0)^3=1$.
Orders for the ${5 \choose 2}$ pairwise products 
of the generators of $N_J$
thus obtained are summarized by the Dynkin diagram in Figure \ref{bF4f}, using the rules given in Table \ref{CD}.
That is, we found that the generators of $N_J=\lan b_i \mid 0\leq i \leq 4 \ran=\langle s_{8708}, s_{6576},
s_{4254}, s_3, s_1\rangle$ 
satisfy the defining relations (given in Equaiton \eqref{funWF4a}) for a Weyl group of type $F_4^{(1)}$.
\end{proof}
\subsubsection{Sub-root system and translations for
$N_J\cong\widetilde{W}( F_{4}^{(1)} )$}
\label{Tf4A}
Here we find the under-laying root system of type $F_4^{(1)}$ for $N_J$.
It spans a subspace in $V^{(1)}$ on which $b_i\in N_J$
$(0\leq i \leq 4)$ can be realised as simple reflections of $F_4^{(1)}$
type.
{\prop\label{brsF4}
Recall that $V_J=\mbox{Span}\left(J\right)$, and $V_J^\perp$ is the orthogonal
complement of $V_J$ in $V^{(1)}$, that is, $V^{(1)}=V_J\bigoplus
V_J^\perp$.
The root system for $N_J$ given in Equation \eqref{bF4} is generated by
$\be=\{\be_i\mid 0\leq i \leq 4\}$ with
\begin{align}\label{brsF4A}
    \be_0&=\al_{7880},\quad
    \be_1=\al_{5667},\quad
    \be_2=\al_{2445},\quad
    \be_3=\al_3,\quad
    \be_4=\al_1,
\end{align}
is of $F_4^{(1)}$ type, and $V_J^\perp=\mbox{Span}\left(\be\right)$.
The group $N_J=\lan b_i \mid 0\leq i \leq 4 \ran$ with the 
$b_i$'s given by Equation \eqref{bvF4} can be realised as 
a reflection group of type $F_4^{(1)}$ on $V_J^\perp$, with $b_i$ actings as the 
reflection along the root $\be_i$.
}
\begin{proof}
Let $\be=\{\be_i\mid 0\leq i \leq 4\}$ with the $\be_i$'s
as given in Equation \eqref{brsF4A}. 
The bilinear form on $V^{(1)}$ in the $\De^{(1)}$ basis, given by Equation \eqref{alaij0} with $a_{ij}$ being the $(i,j)$-entry of $C(E_8^{(1)})$ from Equation \eqref{CarE8a}, is used to check that the $\be$-system
is orthogonal to $V_J$. That is, $V_J^\perp=\mbox{Span}\left(\be\right)$.
Moreover, we find 
$|\be_i|^2$ for $0\leq i \leq 4$ to be $4, 4, 4, 2$ and $2$, respectively.

Compute $(\be_i\cdot\oc\be_j)$ again using this bilinear form and we have,
\begin{equation}\label{CarF4a5}
 C(F_4^{(1)})=(\be_i\cdot\oc\be_j)_{1\leq i,j\leq 4, 0}=\bp
2&-1&0&0&-1\\
-1&2&-2&0&0\\
0&-1&2&-1&0\\
0&0&-1&2&0\\
-1&0&0&0&2
\ep,
\end{equation}
which is the generalized Cartan matrix of type $F_4^{(1)}$ (see
Equation \eqref{CarF4a} in Example \ref{pF4}).
That is, $\be$ forms a simple system of $F_4^{(1)}$ type. This is summarised by the Dynkin diagram in Figure \ref{bF4f}.
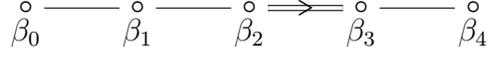
\begin{figure}[h]
\centering
 \begin{tikzpicture}[scale=1]
			\centering
			\node  (a1) {$\circ$};
			\node [left=of a1](a0) {$\circ$} ;
			\node [right=of a1](a2) {$\circ$} ;
			\node [right=of a2](a3) {$\circ$} ;
			\node [right=of a3](a4) {$\circ$} ;
			\node [left=of a0](an){};
			\draw (a1) node [anchor=north] {$\be_1$} ;
			\draw (a2) node [anchor=north] {$\be_2$} ;
			\draw (a3) node [anchor=north] {$\be_3$} ;
			\draw (a4) node [anchor=north] {$\be_4$} ;
			\draw (a0) node [anchor=north] {$\be_0$} ;
			\draw[-] (a1) -- node {} (a2);
			\draw[-] (a3) --node {} (a4);
			\draw[double distance=1.5pt] (a2) -- node {} (a3);
			\node (b) at ($(a2)!0.5!(a3)$) {$>$};
			\draw[-] (a1) -- node {} (a0);
			\path[use as bounding box] (-1.5,0) rectangle (0,0); 
			\end{tikzpicture}
    \caption{Dynkin diagram for the $\be=\{\be_i|\;0\leq i \leq 4\}$ system, where $|\be_i|^2=4$ for $0\leq i \leq 2$
   and  $|\be_i|^2=2$ for i=3, 4.}
    \label{bF4f}
\end{figure}
The null root of this $F_4^{(1)}$ subsystem is given by,
\begin{align}\label{deF4E81}
\de_{F_4}&=\be_0+\tilde{\be}_{F_4}=\be_0+2\be_1+3\be_2+4\be_3+2\be_4.\\\nonumber
&=\al_{7880}+2\al_{5667}+3\al_{2445}+4\al_{3}+2\al_1.\\\nonumber
&=\al_0+2\al_1+3\al_2+4\al_3+6\al_4+5\al_5+4\al_6+3\al_7+2\al_8,\\\nonumber
&=\al_0+\tilde{\al}_{E_8}=\de_{E_8}=\de,
\end{align}
where $\tilde{\be}_{F_4}$ and $\tilde{\al}_{E_8}$ denote
the highest roots of root systems of type $F_4$ and $E_8$, respectively.
The coefficients of simple roots in $\tilde{\be}_{F_4}$ and $\tilde{\al}_{E_8}$ can be found in Table \ref{basic}.
We see that $\de_{F_4}=\de_{E_8}=\de$.

Now, let us we look at the actions of $b_i$ $(0\leq i \leq 4)$ on the 
$\be \cup J=\{\be_1, \be_2, \be_3, \be_4, \be_0, \al_2, \al_5, \al_7, \al_0\}$ basis
of $V^{(1)}$. These are computed using 
Figure \ref{bF4} by
composing the actions of $s_j\in W(E_8^{(1)})$ 
on $V^{(1)}$ given by Equation \eqref{sij0} and $C(E_8^{(1)})$. We have:
\begin{align}\label{bact}\nonumber
    b_1:&\{\be_1, \be_2, \be_3, \be_4, \be_0, \al_2, \al_5, \al_7, \al_0\}\mapsto\\\nonumber
        &\{-\be_1, \be_2+\be_1, \be_3, \be_4, \be_0+\be_1, \al_2, \al_7, \al_5, \al_0\},\\\nonumber
    b_2:&\{\be_1, \be_2, \be_3, \be_4, \be_0, \al_2, \al_5, \al_7, \al_0\}\mapsto\\\nonumber
        &\{\be_1+\be_2, -\be_2, \be_3+\be_2, \be_4, \be_0, \al_5, \al_2, \al_7, \al_0\},\\
    b_3:&\{\be_1, \be_2, \be_3, \be_4, \be_0, \al_2, \al_5, \al_7, \al_0\}\mapsto\\\nonumber
    &\{\be_1, \be_2+2\be_3, -\be_3, \be_4+\be_3, \be_0, \al_2, \al_5, \al_7, \al_0\},\\\nonumber
    b_4:&\{\be_1, \be_2, \be_3, \be_4, \be_0, \al_2, \al_5, \al_7, \al_0\}\mapsto\\\nonumber
    &\{\be_1, \be_2, \be_3+\be_4, -\be_4, \be_0, \al_2, \al_5, \al_7, \al_0\},\\\nonumber
    b_0:&\{\be_1, \be_2, \be_3, \be_4, \be_0, \al_2, \al_5, \al_7, \al_0\}\mapsto\\\nonumber
    &\{\be_1+\be_0, \be_2, \be_3, \be_4, -\be_0, \al_2, \al_5, \al_0, \al_7\}.
\end{align}
In the $\be\cup J$ basis it is easy to see that $b_i$ act on the $\be$-system
exactly as the reflection along the root $\be_i$ of $F_4^{(1)}$ type (for $0\leq i \leq 4$), while acting permutatively on $J$ as
given in Equation \eqref{bepermJ}.
\end{proof} 
Recall that  
the element $\varphi_a$ 
which gives rise to the discrete evolution in Equation \eqref{RCG},
is given as $\varphi_a=\varphi_s^2$ 
with $\varphi_s$ expressed as a product of simple reflections of $W(E_8^{(1)})$
in Equation \eqref{psidecomp1}.
The action of $\varphi_s$ on the simple system $\De^{(1)}$ of $W(E_8^{(1)})$
can be computed by composing the actions of $s_j\in W(E_8^{(1)})$ 
on $V^{(1)}$ given by Equation \eqref{sij0} and $C(E_8^{(1)})$,
\begin{align}\label{AHJNphis}
    \varphi_s&:\{\al_1, \al_2, \al_3, \al_4, \al_5, \al_6, \al_7,\al_8, \al_0\}\\\nonumber
    &\mapsto
  \{-\al_{1233445}+\de, -\al_7, -\al_{1234455667}, -\al_8, -\al_0, 
  \al_{13456780}, \al_2,\al_4, \al_5\}, 
\end{align}
while its action on the 
$\be \cup J=\{\be_1, \be_2, \be_3, \be_4, \be_0, \al_2, \al_5, \al_7, \al_0\}$ basis
is given by,
\begin{align}\label{psact}\nonumber
    \varphi_s:&\{\be_1, \be_2, \be_3, \be_4, \be_0, \al_2, \al_5, \al_7, \al_0\}\\\nonumber
      \mapsto  &\{\be_{1233440}, -\be_0, -\be_{1234}, \be_{11223345}, \be_2, -\al_7, -\al_0, \al_2, \al_5\}.
\end{align}
We can now
use Equation \eqref{lfun} on the $\be$-$J$-system to write $\varphi_s$ as a product of 
generators of the normalizer $N(W_J)=N_J\ltimes W_J$ given in Equation \eqref{NWJf4e8},
and we have,
\begin{equation}\label{psbs}
\varphi_s=b_{1}b_{0}b_{2}b_{1}b_{4}b_{3}b_{2}b_{3}b_{2}s_{5}s_{2}.    
\end{equation}
That is, $\varphi_s$ is an element of $N(W_J)$, and so is $\varphi_a$.

Recall that by Proposition \ref{Tnormsub},
an affine Weyl group has
the following decomposition 
\begin{equation}\label{eawF4Q}
{W}(F_4^{(1)})={W}(F_4)\ltimes Q
=\langle b_{j} \mid 1\leq j\leq 4\rangle\ltimes\langle t_{j} \mid 1\leq j\leq 4\rangle, \end{equation}
where $t_j$ given by Proposition \ref{tj},
is the translation by simple coroot $\pi(\oc\be_j)$ ($1\leq j\leq 4$)
and $Q=\langle t_{j} \mid 1\leq j\leq 4\rangle$ is the root lattice of the
$\be$-system.
Since for $F_4^{(1)}$ type Weyl group the weight lattice is isomorphic to 
the root lattice we have also,
\begin{equation}\label{eawF4P}
{W}(F_4^{(1)})={W}(F_4)\ltimes P
=\langle b_{j} \mid 1\leq j\leq 4\rangle\ltimes\langle U_{j} \mid 1\leq j\leq 4\rangle, \end{equation}
where $U_j$ given in Definition \ref{uj}, is the translation by the fundamental weight $H_j$
($1\leq j\leq 4$) of the $\be$-system, and $P=\lan U_j\mid\,1\leq j\leq 4\ran$ is the weight lattice.

From Equation \eqref{bact}, we see that $b_i$ $(0\leq i \leq 2)$ can not be reflections on the
whole of $V^{(1)}$ since they act permutatively on $\{\al_2, \al_5, \al_7, \al_0\}$.
For this reason, we do not expect an element of translation in this $F_4^{(1)}$ type subsystem to be also a translation in the original $W(E_8^{(1)})$ group in general, but
rather a quasi-translation. However, first, we need to take a closer look
at the coroots and weights of the $\be$-system.

The Dynkin diagram for the dual root system $\{\pi(\oc\be_i)\mid \;1\leq i \leq 4\}$, given in Figure \ref{bF4fd}, 
is obtained by reversing the direction of the arrow in the $F_4^{(1)}$ Dynkin diagram of Figure \ref{bF4f}. The lengths of the simple coroots are,
\begin{equation}
    |\pi(\oc\be_1)|^2=|\pi(\oc\be_1)|^2=1,\quad
   \mbox{and}\quad  |\pi(\oc\be_3)|^2=|\pi(\oc\be_4)|^2=2
\end{equation}  
computed by Equation \eqref{picl} using  
   $|\be_1|^2=|\be_2|^2=4$ 
   and  $|\be_3|^2=|\be_4|^2=2$ found earlier.
   
\begin{figure}[ht]
\centering
%
\begin{tikzpicture}
		
			\node  (a1) {$\circ$};
			\node [right=of a1](a2) {$\circ$} ;
			\node [right=of a2](a3) {$\circ$} ;
			\node [right=of a3](a4) {$\circ$} ;
			\node [left=of a0](an){};
			\draw (a1) node [anchor=north] {$\pi(\oc\be_1)$} ;
			\draw (a2) node [anchor=north] {$\pi(\oc\be_2)$} ;
			\draw (a3) node [anchor=north] {$\pi(\oc\be_3)$} ;
			\draw (a4) node [anchor=north] {$\pi(\oc\be_4)$} ;
			\draw[-] (a1) -- node {} (a2);
			\draw[-] (a3) --node {} (a4);
			\draw[double distance=1.5pt] (a2) -- node {} (a3);
			\node (b) at ($(a2)!0.5!(a3)$) {$<$};
			\path[use as bounding box] (-1.5,0) rectangle (0,0); 
\end{tikzpicture}
\caption{Dynkin diagram for the dual $\{\pi(\oc\be_i)\mid \,0\leq i \leq 4\}$ system, where we have $|\pi(\oc\be_1)|^2=|\pi(\oc\be_1)|^2=1$
   and  $|\pi(\oc\be_3)|^2=|\pi(\oc\be_4)|^2=2$.}\label{bF4fd}
\end{figure}

From Equations \eqref{pahF4} and \eqref{hpaF4} we have 
\beq\label{pahbF4}
\bp
\pi(\oc\be_1)\\
\pi(\oc\be_2)\\
\pi(\oc\be_3)\\
\pi(\oc\be_4)
\ep=C(F_4)^T\bp
H_1\\
H_2\\
H_3\\
H_4
\ep=\bp
2&-1&0&0\\
-1&2&-1&0\\
0&-2&2&-1\\
0&0&-1&2
\ep\bp
H_1\\
H_2\\
H_3\\
H_4
\ep,
\eeq
and
\beq\label{hpabF4}
\bp
H_1\\
H_2\\
H_3\\
H_4
\ep=
\left(C(F_4)^T\right)^{-1}
\bp
\pi(\oc\be_1)\\
\pi(\oc\be_2)\\
\pi(\oc\be_3)\\
\pi(\oc\be_4)
\ep=\bp
2 & 3 & 2 & 1 \\
 3 & 6 & 4 & 2 \\
 4 & 8 & 6 & 3 \\
 2 & 4 & 3 & 2
\ep\bp
\pi(\oc\be_1)\\
\pi(\oc\be_2)\\
\pi(\oc\be_3)\\
\pi(\oc\be_4)
\ep,
\eeq
respectively.
The matrix of symmetric bilinear form $(\;,\;)$ in $\{\pi(\oc\be_i)\mid \,0\leq i \leq 4\}$ basis is given by Equation \eqref{paipj}: 
\begin{align}\label{SymbF4d}\nonumber
&\left(\left(\pi(\oc\be_i),\pi(\oc\be_j)\right)\right)_{1\leq i,j\leq4}
=\left(\frac{2}{|\be_i|^2}a_{ij}\right)_{1\leq i,j\leq4},\\
&=\bp
1/2&0&0&0\\
0&1/2&0&0\\
0&0&1&0\\
0&0&0&1
\ep\bp
2&-1&0&0\\
-1&2&-2&0\\
0&-1&2&-1\\
0&0&-1&2
\ep
=\left(
\begin{array}{cccc}
 1 & -\frac{1}{2} & 0 & 0 \\
 -\frac{1}{2} & 1 & -1 & 0 \\
 0 & -1 & 2 & -1 \\
 0 & 0 & -1 & 2 \\
\end{array}
\right),    
\end{align}
where we have used the fact that $|\be_i|^2$ for $1\leq i \leq 4$ are $4, 4, 2$ and $2$.

The bilinear form in $\{H_j \mid 1\leq j\leq 4\}$ basis is given by Equation
\eqref{hij}:
\begin{align}\label{hijbF4}\nonumber
&\left((H_i, H_j)\right)_{1\leq i,j\leq4}
=\left(C(F_4)^T\right)^{-1}\left(\frac{2}{|\be_k|^2}\de_{kj}\right)_{1\leq k,j\leq4},\\
&=\left(C(F_4)^T\right)^{-1}
\bp
1/2&0&0&0\\
0&1/2&0&0\\
0&0&1&0\\
0&0&0&1
\ep=\bp
2 & 3 & 2 & 1 \\
 3 & 6 & 4 & 2 \\
 4 & 8 & 6 & 3 \\
 2 & 4 & 3 & 2
\ep\bp
1/2&0&0&0\\
0&1/2&0&0\\
0&0&1&0\\
0&0&0&1
\ep
=\left(
\begin{array}{cccc}
 1 & \frac{3}{2} & 2 & 1 \\
 \frac{3}{2} & 3 & 4 & 2 \\
 2 & 4 & 6 & 3 \\
 1 & 2 & 3 & 2 \\
\end{array}
\right).    
\end{align}
The diagonal entries of the last matrix in Equation \eqref{hijbF4}
tell us that $|H_j|^2$ for $1\leq j\leq 4$ are 1, 3, 6, and 2, respectively. 
That is, $H_1$ (with $|H_1|^2=1$) is the shortest of the fundamental weights of this $F_4^{(1)}$ type $\be$-system, which
means that $U_1$ is a basic translation of the $\be$-system.
Recall that for non-simply-laced systems we 
can identify $\pi(\oc\be_i)$ with $\frac{2\be_i}{|\be_i|^2}$,
so that we can express the simple coroots of the $F_4^{(1)}$ system
in terms of the simple roots of the $E_8^{(1)}$ system,

\begin{equation}\label{pbal}
    \pi(\oc\be_0)=\frac{\al_{70}}{2}+\al_8,\quad
   \pi(\oc \be_1)=\frac{\al_{57}}{2}+\al_6,\quad
   \pi(\oc \be_2)=\frac{\al_{25}}{2}+\al_4,\quad
  \pi(\oc  \be_3)=\al_3,\quad
   \pi(\oc \be_4)=\al_1. 
\end{equation}

Moreover, using Equations \eqref{hpabF4} and \eqref{pbal} we have
\begin{align}\label{H1F4}
    H_1&=2\pi(\oc\be_1)+3 \pi(\oc\be_2)+ 2\pi(\oc\be_3)+\pi(\oc\be_4),\\\nonumber
    &=2\left(\frac{\al_{57}}{2}+\al_6\right)+3\left(\frac{\al_{25}}{2}+\al_4\right)+
    2\al_3+\al_1,\\\nonumber
    &=\al_1+\frac{3}{2}\al_2+2\al_3+3\al_4+\frac{5}{2}\al_5+2\al_6+\al_7,
\end{align}
where we see that $H_1$ of the $F_4^{(1)}$ type subsystem is neither a root nor weight of the $E_8^{(1)}$ system.
On the other hand, we have,
\begin{align}\label{Htoh}\nonumber
   2H_1&=2\al_1+3\al_2+4\al_3+6\al_4+5\al_4+4\al_6+2\al_7,\\
   &=h_6-2h_8,
\end{align}
where we have used Equation \eqref{pahE8}
for the expression in terms of the fundamental weights $h_j$ ($1\leq j\leq 8$), of the original $E_8^{(1)}$ system.

Earlier in Section \ref{TaF4}, we discussed some elements of translations
by coroots in $W(F_4^{(1)})$. In particular 
from Equation \eqref{t0F4} we have the translation by $\pi(\oc\be_0)$ is given by,
\begin{equation}
t_{\be_0}=b_{1232143}b_2b_{3412321}b_0.    
\end{equation}
Moreover, from Equation \ref{p0Fh} in Example \ref{pF4} we found $\pi(\oc\be_0)=-H_1$. 
That is, $t_{\be_0}$ is the translation by $-H_1$, then we have
\beq\label{t0F4b}
U_1=t_{\be_0}^{-1}=b_0b_{1232143}b_2b_{3412321}.
\eeq
$U_1$ acts on the $\De^{(1)}$ basis of $V^{(1)}$ by
\begin{equation}\label{U1alF4b}
 U_1:\{\al_1, \al_2, \al_3, \al_4, \al_5, \al_6, \al_7,\al_8, \al_0\}\mapsto
\{\al_1, \al_5, \al_3, \al_4, \al_2, \al_{1233444555666778}-\de, \al_0,\al_{8}+\de, \al_7\}, \end{equation}
moreover,
\begin{equation}\label{U12alF4b}
 U_1^2:\{\al_1, \al_2, \al_3, \al_4, \al_5, \al_6, \al_7,\al_8, \al_0\}\mapsto
\{\al_1, \al_2, \al_3, \al_4, \al_5, \al_6-\de, \al_7,\al_8+2\de, \al_0\}.
\end{equation}
By Proposition \ref{ujv} we recognise that
$U_1^2$ is a translation in $W(E_8^{(1)})$
by $h_6-2h_8$,
and that
$|h_6-2h_8|^2=|2H_1|^2=4|H_1|^2=4$. 
The nature of $U_1$ becomes more evident in the 
$\be\cup J$ basis
of $V^{(1)}$, we have
\begin{equation}\label{U1balF4b}
 U_1:\{\be_1, \be_2, \be_3, \be_4, \be_0, \al_2, \al_5, \al_7, \al_0\}\mapsto
\{\be_1-\de, \be_2, \be_3, \be_4, \be_0+2\de, \al_5, \al_2, \al_0, \al_7\},  
\end{equation}
where we see that although $U_1$ acts on the $\be$-system
like a translation, it also acts on the subset $J$ as a permutation of order two, $(25)(07)$. We call the element $U_1$, given by Equation \eqref{t0F4b}, a quasi-translation
of order two.

In \cite{ahjn:16}, the discrete dynamics of equation \eqref{RCG}
is given by $\varphi_a\in W(E_8^{(1)})$ (see Equation \eqref{psidecomp1}), whose
actions on $\De^{(1)}$ are,
\begin{align}\label{AHJNphiA}
    \varphi_a&:\{\al_1, \al_2, \al_3, \al_4, \al_5, \al_6, \al_7,\al_8, \al_0\}\\\nonumber
    &\mapsto
  \{\al_1+\de, -\al_2, \al_{23445}-\de, -\al_4, -\al_5, -\al_{12334456}+\de, -\al_7,-\al_8, -\al_0\}, 
    \end{align}
moreover, we have,
\begin{align}\label{AHJNphi2A}
    \varphi_a^2&:\{\al_1, \al_2, \al_3, \al_4, \al_5, \al_6, \al_7,\al_8, \al_0\}\\\nonumber
    &\mapsto
  \{\al_1+2\de, \al_2, \al_3-2\de, \al_4, \al_5, \al_6+\de, \al_7,\al_8, \al_0\}.
    \end{align}
That is, $\varphi_a^2$ is a translation
by $-2h_1+2h_3-h_6$ and
$\varphi_a$ a quasi-translation
of order two. It can be computed using 
the symmetric bilinear form given in Equation \eqref{hijE8} that $|-2h_1+2h_3-h_6|^2=4$.  That is, $U_1$ and $\varphi_a$ are both quasi-translations
of order two for some translations by vectors of squared length 4.
{\prop\label{conT}
$U_1^2$ is related to $\varphi^2$ by a conjugation of an element of
${W}\left(F_4\right)=\langle b_{j} \mid 1\leq j\leq 4\rangle$,
 \begin{equation}\label{conTe}
    XU_1^2X^{-1}=\varphi_a^2,\quad\mbox{where}\quad X=b_{124321}.
\end{equation}
}
\begin{proof}
First we know that
\begin{equation}
 |h_6-2h_8|^2=|2H_1|^2=4|H_1|^2=4=|-2h_1+2h_3-h_6|^2. 
\end{equation}
That is $-2h_1+2h_3-h_6$ must be in the same ${W}\left(F_4\right)$-orbit
of $2H_1$. Let us compute the size of orbit of $H_1$ under $W(F_4)$, $|W(F_4)H_1|$,
using Proposition \ref{orbWP}.
The stabilizer of $H_1$ in $W(F_4)$ is given by
\begin{align}\nonumber
W(F_4)_{H_1}&=\lan b_2, b_3, b_4\ran,\\\label{stbH1}
&\cong W(C_3).
\end{align}
Then by Proposition \ref{orbWP} we have,
\begin{align}\nonumber
  |W(F_4)H_1|&=\frac{|W(F_4)|}{|W(C_3)|}, \\\nonumber
  &=\frac{2^73}{2^3 3!},\\\label{F4H1}
  &=2^33=24,
\end{align}
where we have used the orders of Weyl groups of type $C_3$ and $F_4$ given in Table \ref{basic}.
We find that,
\begin{equation}\label{X1}
    X(2H_1)=X(h_6-2h_8)=b_{412321}(h_6-2h_8)=-2h_1+2h_3-h_6.
\end{equation}
Then by Equation \eqref{ujnj} we have Equation \eqref{conTe}.
\end{proof}
{\prop Any element of the form $yXU_1X^{-1}$ $(y\in \lan b_2, b_0\ran\times 
\lan s_2s_5, s_7s_0\ran)$ has the property that it iterated twice is
a translation by $-2h_1+2h_3-h_6$. In particular, we have
\begin{equation}
    s_{0752}b_0b_2XU_1X^{-1}=\varphi_a.
\end{equation}
}
\begin{proof}
Let $y$ be an involution in $N(W_J)$ that commutes with $XU_1X^{-1}$. That is,
\begin{equation}
    yXU_1X^{-1}=XU_1X^{-1}y,\quad \mbox{and}\quad y^2=1,
\end{equation}
then we have,
\begin{align}\nonumber
&XU_1^2X^{-1}\\\nonumber
&=XU_1X^{-1}XU_1X^{-1},\\
&=XU_1X^{-1}y^2XU_1X^{-1},\\\nonumber
&=yXU_1X^{-1}yXU_1X^{-1},\\\nonumber
&=(yXU_1X^{-1})^2,\\\nonumber
&=\varphi_a^2.
\end{align}
Now it is left to find all involutions in $N(W_J)$ that commute with $XU_1X^{-1}$.
From the actions of $XU_1X^{-1}$ on $\be\cup J$,
\begin{equation}\label{XU1balF4b}
 XU_1X^{-1}:\{\be_1, \be_2, \be_3, \be_4, \be_0, \al_2, \al_5, \al_7, \al_0\}\mapsto
\{\be_1+\de, \be_2, \be_3-\de, \be_4+\de, \be_0, \al_5, \al_2, \al_0, \al_7\},  
\end{equation}
we see that $XU_1X^{-1}$ commutes with $b_2, b_0$. Moreover, we have
\begin{equation}
    XU_1X^{-1}s_2=s_5XU_1X^{-1},\quad\mbox{and}\quad XU_1X^{-1}s_7=s_0XU_1X^{-1},
\end{equation}
hence elements $s_2s_5$ and $s_7s_0$ and their products are
all involutions that commute with $XU_1X^{-1}$. In particular, we find that
\begin{equation}
 s_{0752}b_0b_2XU_1X^{-1}=\varphi_a.   
\end{equation}
That is, $\varphi_a$ is an element of quasi-translation of order two
in a normalizer of ${W}(4A_1)$ in $W(E_8^{(1)})$ given in Proposition \ref{f4inE8}.
\end{proof}

\section{Conclusion}\label{Con}
In this work, we reviewed some formulations and properties of the affine Weyl group
and demonstrated how they can be used in studying discrete integrable systems with affine Weyl symmetries. We exploited the fact that a Weyl group
defined abstractly as a Coxeter system,
(that is, a generating set with defining relations) has, on the one hand, a birational 
representation (symmetries for a nonlinear equation), while on the other, classical linear representations
known for studying the many remarkable properties of the group.
In particular, we showed that some behaviours of integrable equations are
manifestations of certain properties of the Weyl group, such as translation
and quasi-translation,
hence can be 
dealt with effectively using a well-chosen linear representation.
Computations that may be difficult in the context of
the integrable system, such as finding a birational transformation between two 
nonlinear equations
amounted to relating two vectors in a linear representation of the affine Weyl group,
as shown for our two examples of systems of type $E_8^{(1)}$ and $F_4^{(1)}$. 
This approach is useful for understanding and clarifying the nature of the plethora of 
discrete integrable equations that appear in the literature, 
since unless by construction,
equations rarely come with the full symmetries of their types (as listed in Figure \ref{Sakai}) and are even less likely to be in the canonical form. 

Finally, properties discussed here for the affine Weyl group have
their generalisations in
the theory of Coxeter groups, which is a rich and well-developed
area of mathematics, with major
breakthroughs still being made \cite{EW_14}.
It occupies a special place in the theory of groups in that
certain questions which are undecidable for
general groups such as the normalizer problem \cite{BH} can 
be answered for Coxeter groups. Integrable system occupies a similar position in the theory of equations.
It would be interesting and profitable to
explore further implications of the many
remarkable properties of the Coxeter groups in the context of the integrable system.

\vskip5mm
\section*{Acknowledgement}
The author would like to express her gratitude to R. B. Howlett 
for the enlightening discussions related to the Coxeter groups.

\newpage
\appendix
\section{}\label{list}
Below, we collect some data from the Weyl groups relevant for our
discussions. 
Except when specifically stated, the index $i$ runs from $1$ to $n$.
We have:
$|\Phi_+|$ is the number of positive roots in a finite root system;
$c_j$ are coefficients of $\al_j$ in $\tilde{\al}$;
$|W|$ is the order of the group;
$k_j$ are coefficients of $\pi(\oc\al_j)$
in $\pi(\oc{\tilde{\al}})$;
the $-$ indicates that the $k_j$'s are the same as the $c_j$'s of that row;
$|h_i|^2$ are the squared lengths
of the fundamental weights. Finally, we write the coroots for the highest long and short root
in terms of the fundamental weights.(Lengths of $\pi(\oc{\tilde{\al}})$
and $\pi(\oc{\tilde{\al}_s})$ are given in Equation \eqref{picl}).

\begin{table}[ht]
\centering
\begin{tabular}{|c|c|c|c|c|c|c|c|} 
 \hline
 Type &$|\Phi_+|$&$c_j$ & $|W|$ &$k_j$&$|h_i|^2$&$\pi(\oc{\tilde{\al}})$ &$\pi(\oc{\tilde{\al}_s})$\\ [0.5ex] 
 \hline
 $A_n$& $\frac{n(n+1)}{2}$&$1, 1, ..., 1$ &$(n+1)!$ & -&$\frac{i(n+1-i)}{(n+1)}$&$h_1+h_n$&-\\ 
 $B_n$&$n^2$&$1, 2, 2,..., 2$ &$2^n n!$&$1, 2, ..., 2, 1$  & $i$&$h_2$&$2h_1$ \\ 
$C_n$&$n^2$&$2, 2, ..., 2, 1$ &$2^n n!$&$1, 1,..., 1$& $i\; (1\leq i\leq n-1), \frac{n}{4}$&$2h_1$&$h_2$\\ 
$D_n$&$n(n-1)$&$1, 2, ..., 2, 1, 1$ &$2^{n-1}n!$&-&$ i\; (1\leq i\leq n-2), \frac{n}{4}, \frac{n}{4}$&$h_2$&-  \\ 
$E_6$&$36$&$1, 2, 3, 2, 1, 2$ &$2^73^45$&-  &$2, \frac{4}{3}, \frac{10}{3}, 6, \frac{10}{3}, \frac{4}{3}$&$h_1$&-\\
$E_7$&$63$&$2,2, 3, 4, 3, 2, 1 $ &$2^{10}3^45\,7$&-& $2, \frac{7}{2}, 6, 12, \frac{15}{2}, 4, \frac{3}{2}$&$h_1$&-\\
$E_8$&$120$&$2, 3, 4, 6, 5, 4, 3, 2$ &$2^{14}3^55^27$&-& $4, 8, 14, 30, 20, 12, 6, 2$&$h_8$&- \\
$F_4$&$24$&$2, 3, 4, 2$ &$2^73^2$& $2, 4, 3, 2$&$2, 6, 12, 4$&$h_1$&$h_4$\\
$G_2$&$6$&$3, 2$ &$12$& $2, 3$&$2, \frac{2}{3}$&$h_2$&$h_1$\\
  \hline
 \end{tabular}
  \vspace{1em}
\caption{Some useful data of the Weyl groups.}\label{basic}
 \end{table}
 \section{}
 \subsection{A birational representation of $W(E_8^{(1)})$}\label{biratE8}
Transformations on the variables $t$, $f$, $g$, and
parameters $b_i (1\leq i \leq 8)$
of Sakai's $e$-{\bf P}$(E_8^{(1)})$ equation \eqref{ePE8}
given in Equation \eqref{msye8rep} satisfy
the defining relations of $W(E_8^{(1)})=\lan s_i\mid 0\leq i \leq 8\ran$ \cite{MSY:13}.
That is, the transformations generate a birational representation of $W(E_8^{(1)})$.
\begin{subequations}\label{msye8rep}
\begin{align}
s_2&:\left(\begin{array}{l}
b_1, b_2, b_3, b_4\\
b_5, b_6, b_7, b_8\end{array} t, f, g\right)\\\nonumber &\hspace{1em}\mapsto\left(\begin{array}{l}
b_1-3 \frac{2 t+b_1+b_2}{4} , b_2-3 \frac{2 t+b_1+b_2}{4} , b_3+\frac{2 t+b_1+b_2}{4} , b_4+\frac{2 t+b_1+b_2}{4} \\
b_5+\frac{2 t+b_1+b_2}{4} , b_6+\frac{2 t+b_1+b_2}{4} , b_7+\frac{2 t+b_1+b_2}{4} , b_8+\frac{2 t+b_1+b_2}{4}, t-\frac{2 t+b_1+b_2}{4}
\end{array},t, f, \widetilde{g}\right),    
\end{align}
\begin{equation}
 \begin{gathered}
s_1: \quad(t, f, g) \mapsto(-t, g, f), \quad s_i: \quad\left(b_{i-1}, b_i\right) \mapsto\left(b_i, b_{i-1}\right) \quad(i=3, \ldots, 7), \\
s_8: \quad\left(b_1, b_2\right) \mapsto\left(b_2, b_1\right), \quad s_0: \quad\left(b_7, b_8\right) \mapsto\left(b_8, b_7\right),
\end{gathered}   
\end{equation}
where $\tilde{g}$ is given by

 \begin{align}\label{s2g}
& \frac{\widetilde{g}-\wp\left(2 t-\left(b_1-b_2\right) / 2\right)}{\widetilde{g}-\wp\left(2 t-\left(-b_1+b_2\right) / 2\right)} \\\nonumber
& \quad=\frac{f-\wp\left(b_2+t\right)}{f-\wp\left(b_1+t\right)} \frac{\wp\left(t-\left(b_1+b_2\right) / 2\right)-\wp\left(2 t-\left(b_1-b_2\right) / 2\right)}{\wp\left(t-\left(b_1+b_2\right) / 2\right)-\wp\left(2 t-\left(-b_1+b_2\right) / 2\right)} \frac{\wp(2 t)-\wp\left(t-b_2\right)}{\wp(2 t)-\wp\left(t-b_1\right)} \frac{g-\wp\left(t-b_1\right)}{g-\wp\left(t-b_2\right)} .
\end{align}
\end{subequations}

The actions of $T_1\in W(E_8^{(1)})$ on $t, f, g$ and $b_i\, (1\leq i \leq 8)$,  given in Equations \eqref{ePE8}
and \eqref{T1act1}, can be computed using the expression of $T_1$
given in Equation \eqref{T1decomp1}
by composing the actions of the $s_i$'s given in Equation \eqref{msye8rep} from left to right.
\subsection{Discrete \Pa equations}\label{E8}
\begin{figure}[ht]
\centering
\begin{tikzpicture}[scale = 1.2]
\begin{scope}
\coordinate (Y01s) at (0,1);
\coordinate (Y21s) at (0,-2);
\coordinate (Y21e) at ($(Y21s)+(0.4,0)$);
\coordinate (Y22s) at ($(Y21e)+(0.8,0)$);
\coordinate (Y22e) at ($(Y22s)+(0.4,0)$);
\coordinate (Y23s) at ($(Y22e)+(0.8,0)$);
\coordinate (P11s) at (0,0);
\coordinate (P11e) at ($(P11s)+(0.4,0)$);
\coordinate (P12s) at ($(P11e)+(0.8,0)$);
\coordinate (P12e) at ($(P12s)+(0.4,0)$);
\coordinate (P13s) at ($(P12e)+(0.8,0)$);
\coordinate (P13e) at ($(P13s)+(0.4,0)$);
\coordinate (P14s) at ($(P13e)+(0.8,0)$);
\coordinate (P14e) at ($(P14s)+(0.4,0)$);
\coordinate (P15s) at ($(P14e)+(0.8,0)$);
\coordinate (P15e) at ($(P15s)+(0.4,0)$);
\coordinate (P16s) at ($(P15e)+(1.8,0)$);
\coordinate (P16e) at ($(P16s)+(0.4,0)$);
\coordinate (P17s) at ($(P16e)+(1.8,0)$);
\coordinate (P17e) at ($(P17s)+(0.4,0)$);
\coordinate (P18s) at ($(P17e)+(0.8,0)$);
\coordinate (P18e) at ($(P18s)+(0.4,0)$);
\coordinate (P19s) at ($(P18e)+(0.8,0)$);
\coordinate (P19e) at ($(P19s)+(0.4,0)$);
\coordinate (P21s) at (0,-0.8);
\coordinate (P21e) at ($(P21s)+(0.4,0)$);
\coordinate (P22s) at ($(P21e)+(0.8,0)$);
\coordinate (P22e) at ($(P22s)+(0.4,0)$);
\coordinate (P23s) at ($(P22e)+(0.8,0)$);
\coordinate (P23e) at ($(P23s)+(0.4,0)$);
\coordinate (P24s) at ($(P23e)+(0.8,0)$);
\coordinate (P24e) at ($(P24s)+(0.4,0)$);
\coordinate (P25s) at ($(P24e)+(0.8,0)$);
\coordinate (P25e) at ($(P25s)+(0.4,0)$);
\coordinate (P26s) at ($(P25e)+(1.8,0)$);
\coordinate (P26e) at ($(P26s)+(0.4,0)$);
\coordinate (P27s) at ($(P26e)+(1.8,0)$);
\coordinate (P27e) at ($(P27s)+(0.4,0)$);
\coordinate (P28s) at ($(P27e)+(0.8,0)$);
\coordinate (P28e) at ($(P28s)+(0.4,0)$);
\coordinate (P29s) at ($(P28e)+(0.8,0)$);
\coordinate (P29e) at ($(P29s)+(0.4,0)$);
\coordinate (P31s) at (0,-2);
\coordinate (P31e) at ($(P31s)+(0.4,0)$);
\coordinate (P32s) at ($(P31e)+(0.8,0)$);
\coordinate (P32e) at ($(P32s)+(0.4,0)$);
\coordinate (P33s) at ($(P32e)+(0.8,0)$);
\coordinate (P33e) at ($(P33s)+(0.4,0)$);
\coordinate (P34s) at ($(P33e)+(0.8,0)$);
\coordinate (P34e) at ($(P34s)+(0.4,0)$);
\coordinate (P35s) at ($(P34e)+(0.8,0)$);
\coordinate (P35e) at ($(P35s)+(0.4,0)$);
\coordinate (P36s) at ($(P35e)+(1.0,0)$);
\coordinate (P36e) at ($(P36s)+(1.2,0)$);
\coordinate (P37s) at ($(P36e)+(1.4,0)$);
\coordinate (P37e) at ($(P37s)+(0.8,0)$);
\coordinate (P38s) at ($(P37e)+(0.8,0)$);
\coordinate (P38e) at ($(P38s)+(0.4,0)$);
\coordinate (P39s) at ($(P38e)+(0.8,0)$);
\coordinate (P39e) at ($(P39s)+(0.4,0)$);
\coordinate (P41s) at (0,-3);
\coordinate (P41e) at ($(P41s)+(0.4,0)$);
\coordinate (P42s) at ($(P41e)+(0.8,0)$);
\coordinate (P42e) at ($(P42s)+(0.4,0)$);
\coordinate (P43s) at ($(P42e)+(0.8,0)$);
\coordinate (P43e) at ($(P43s)+(0.4,0)$);
\coordinate (P44s) at ($(P43e)+(0.8,0)$);
\coordinate (P44e) at ($(P44s)+(0.4,0)$);
\coordinate (P45s) at ($(P44e)+(0.8,0)$);
\coordinate (P45e) at ($(P45s)+(0.4,0)$);
\coordinate (P46s) at ($(P45e)+(1.8,0)$);
\coordinate (P46e) at ($(P46s)+(0.4,0)$);
\coordinate (P47s) at ($(P46e)+(1.0,0)$);
\coordinate (P47e) at ($(P47s)+(1.2,0)$);
\coordinate (P48s) at ($(P47e)+(0.8,0)$);
\coordinate (P48e) at ($(P48s)+(0.4,0)$);
\coordinate (P49s) at ($(P48e)+(0.8,0)$);
\coordinate (P49e) at ($(P49s)+(0.4,0)$);
\node at ($(Y01s)-(0.4,0)$){$E_8^{(1)}$};
\node at ($(Y01s)-(0.9,0)$){$e$:\;};
\node at ($(Y21s)-(0.4,0)$){$E_8^{(1)}$};
\node at ($(Y21s)-(0.9,0)$){$d$:\;};
\node at ($(P21s)-(0.9,0)$){$q$:\;};
\node at ($(Y22s)-(0.4,0)$){$E_7^{(1)}$};
\node at ($(Y23s)-(0.4,0)$){$E_6^{(1)}$};

\node at ($(P18s)-(0.4,0)$){$A_1^{(1)'}$};
\node at ($(P21s)-(0.4,0)$){$E_8^{(1)}$};
\node at ($(P22s)-(0.4,0)$){$E_7^{(1)}$};
\node at ($(P23s)-(0.4,0)$){$E_6^{(1)}$};
\node at ($(P24s)-(0.4,0)$){$D_5^{(1)}$};
\node at ($(P25s)-(0.4,0)$){{$A_4^{(1)}$}};
\node at ($(P26s)-(0.9,0)$){$(A_2+A_1)^{(1)}$};
\node at ($(P27s)-(0.9,0)$){{$(A_1+A'_1)^{(1)}$}};
\node at ($(P28s)-(0.4,0)$){$A_1^{(1)}$};
\node at ($(P29s)-(0.4,0)$){$A_0^{(1)}$};
\node at ($(P35s)-(0.4,0)$){$D_4^{(1)}$};
\node at ($(P35s)-(0.4,0.5)$){$(P_\text{VI})$};
\node at ($(P36s)-(0.4,0)$){$A_3^{(1)}$};
\node at ($(P36s)-(0.4,0.5)$){$(P_\text{V})$};
\node at ($(P37s)-(0.6,0)$){$(2A_1)^{(1)}$};
\node at ($(P37s)-(0.5,0.5)$){$(P_\text{III})$};
\node at ($(P38s)-(0.4,0)$){$A_1^{(1)'}$};
\node at ($(P39s)-(0.4,0)$){$A_0^{(1)}$};
\node at ($(P47s)-(0.4,0)$){$A_2^{(1)}$};
\node at ($(P47s)-(0.4,0.5)$){$(P_\text{IV})$};
\node at ($(P48s)-(0.4,0)$){$A_1^{(1)}$};
\node at ($(P48s)-(0.4,0.5)$){$(P_\text{II})$};
\node at ($(P49s)-(0.4,0)$){$A_0^{(1)}$};
\node at ($(P49s)-(0.4,0.5)$){$(P_\text{I})$};
\draw [->, thick] (Y21s)--(Y21e);
\draw [->, thick] (Y22s)--(Y22e);
\draw [->, thick] (Y23s)--(P34s);
\draw [->, thick] ($(P11s)-(0.4,-0.4)$)--($(P21s)-(0.4,-0.4)$);
\draw [->, thick] ($(P21s)-(0.4,0.3)$)--($(Y21s)-(0.4,-0.4)$);
\draw [->, thick] ($(P22s)-(0.4,0.3)$)--($(Y22s)-(0.4,-0.4)$);
\draw [->, thick] ($(P23s)-(0.4,0.3)$)--($(Y23s)-(0.4,-0.4)$);
\draw [->, thick] (P21s)--(P21e);
\draw [->, thick] (P22s)--(P22e);
\draw [->, thick] (P23s)--(P23e);
\draw [->, thick] (P24s)--(P24e);
\draw [->, thick] (P25s)--(P25e);
\draw [->, thick] (P26s)--(P26e);
\draw [->, thick] (P27s)--(P27e);
\draw [->, thick] (P35s)--($(P35e)+(0.2,0)$);
\draw [->, thick] (P36s)--($(P36e)+(0.15,0)$);
\draw [->, thick] (P37s)--(P37e);
\draw [->, thick] (P38s)--(P38e);
\draw [->, thick] (P47s)--(P47e);
\draw [->, thick] (P48s)--(P48e);
\draw [->, thick] ($(P27s)+(0,0.2)$)--($(P17e)-(0,0.2)$);
\draw [->, thick] ($(P18s)-(0.1,0.2)$)--($(P28e)+(0,0.2)$);
\draw [->, thick] ($(P24s)-(0,0.2)$)--($(P34e)+(0,0.2)$);
\draw [->, thick] ($(P25s)-(0,0.2)$)--($(P35e)+(0.2,0.2)$);
\draw [->, thick] ($(P26s)-(0,0.2)$)--($(P36e)+(0.15,0.2)$);
\draw [->, thick] ($(P27s)-(0,0.2)$)--($(P37e)+(0,0.2)$);
\draw [->, thick] ($(P28s)-(0,0.2)$)--($(P38e)+(0,0.2)$);
\draw [->, thick] ($(P36s)-(0,0.2)$)--($(P46e)+(0.2,0.1)$);
\draw [->, thick] ($(P37s)-(0,0.2)$)--($(P47e)+(0,0.2)$);
\draw [->, thick] ($(P38e)+(0.3,-0.3)$)--($(P48e)+(0.3,0.3)$);
\draw [->, thick] ($(P26s)-(0.2,0.3)$)--($(P46e)+(0.25,0.25)$);
\draw[->, thick] ($(P28s)-(0.2,0.2)$) .. controls ($(P38s)+(0.3,0)$) .. ($(P48s)+(-0.1,0.2)$);
\draw[->, thick] ($(P29s)-(0.2,0.2)$) .. controls ($(P39s)+(0.1,0)$) .. ($(P49s)+(-0.1,0.2)$);

\end{scope}
\end{tikzpicture}
\caption{Sakai's classification of 22 types of discrete \Pa equations listed by symmetry types.}\label{Sakai}
\end{figure}
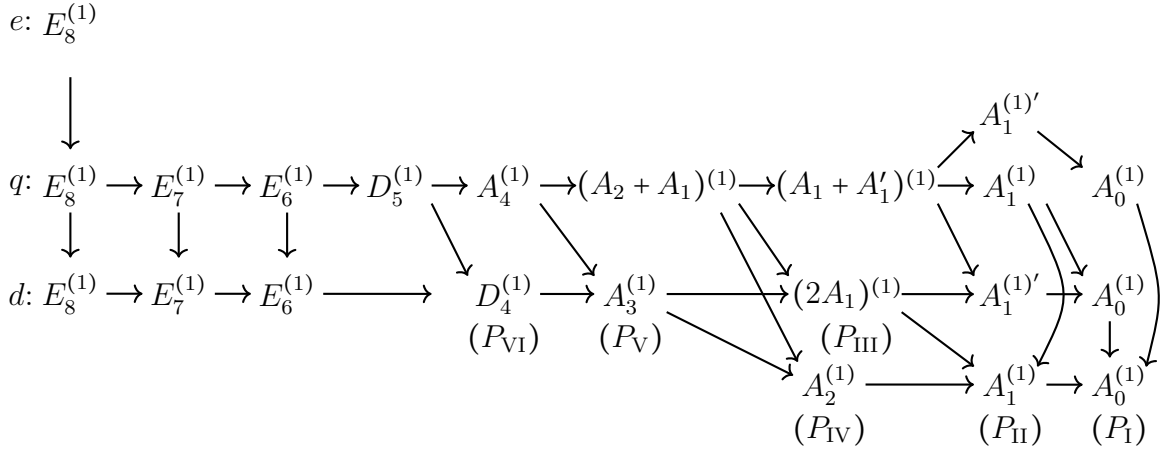

Sakai's classification of 22 types of discrete \Pa equations \cite{sak:01}
listed by their symmetry types are shown in Figure \ref{Sakai}.
They arise naturally as three classes of difference equations: 
elliptic-difference ($e-$), q-difference ($q-$) and
additive difference ($d-$).

\end{document}